\documentclass[12pt, twoside]{article}
\let\counterwithin\relax
\usepackage{chngcntr}

\usepackage[T1]{fontenc}
\usepackage[latin9]{inputenc}
\usepackage{geometry}
\usepackage{amsmath}
\usepackage{amsthm}
\usepackage{amssymb}
\usepackage{amstext}
\usepackage{bm}
\usepackage{booktabs} 
\usepackage{color}
\usepackage{enumerate} 
\usepackage[shortlabels]{enumitem}
\usepackage{epsfig,epsf,psfrag}
\usepackage{epstopdf}
\usepackage{float}
\usepackage{graphics}
\usepackage{graphicx}
\usepackage{latexsym}
\usepackage{lscape}
\usepackage{mathabx} 
\usepackage{mathtools}
\usepackage{mathrsfs}
\usepackage{multirow}
\usepackage{natbib}
\usepackage{rotate}
\usepackage{setspace}
\usepackage[linesnumbered,ruled,vlined]{algorithm2e}
\usepackage{sectsty}
\usepackage{caption}
\usepackage{subcaption}
\usepackage{xargs}[2008/03/08]
\usepackage{xcolor}

\usepackage{ifthen}
\usepackage{fancyhdr}
\usepackage[hidelinks]{hyperref}
\hypersetup{
	colorlinks,
	linkcolor={red!50!black},
	citecolor={blue!50!black},
	urlcolor={blue!80!black}
}
\usepackage{url}

\renewcommand\thesection{\arabic{section}}
\renewcommand{\thesubsection}{\thesection.\arabic{subsection}}

\sectionfont{\centering\bf\large}
\subsectionfont{\normalsize}

\usepackage{titlesec}
\newcommand{\Appendix}{\appendix
	\def\thesection{Appendix \Alph{section}}
	\titleformat{\section}{\centering\bf\large}{\thesection}{0em}{: }
	\def\thesubsection{\Alph{section}.\arabic{subsection}}
	\renewcommand{\theequation}{\Alph{section}.\arabic{equation}}
	\setcounter{section}{0}
	\counterwithin*{equation}{section}
	\renewcommand{\thethm}{\Alph{section}\arabic{thm}}
	\counterwithin*{thm}{section}
	\renewcommand{\thelem}{\Alph{section}\arabic{lem}}
	\counterwithin*{lem}{section}
	\renewcommand{\thecor}{\Alph{section}\arabic{cor}}
	\counterwithin*{cor}{section}
	
	\renewcommand{\thefigure}{\Alph{section}.\arabic{figure}}
}
\newcommand\unappendix{
	\titleformat{\section}{\centering\bf\large}{}{0em}{}
}

\renewcommand{\baselinestretch}{1.6} 
\newcommand{\single}{\renewcommand{\baselinestretch}{1.2}\normalsize}

\newcommand{\bea}{\begin{eqnarray*}}
	\newcommand{\eea}{\end{eqnarray*}}
\newcommand{\be}{\begin{eqnarray}}
	\newcommand{\ee}{\end{eqnarray}}
\newcommand{\beq}{\begin{equation}}
	\newcommand{\eeq}{\end{equation}}

\newcommand{\bal}{\begin{equation}\aligned}
	\newcommand{\eal}{\endaligned\end{equation}}
\newcommand{\bgt}{\begin{equation}\begin{gathered}}
		\newcommand{\egt}{\end{gathered}\end{equation}}
\newcommand{\ed}{\end{document}}

\newcommand{\btab}{\begin{tabular}}
\newcommand{\etab}{\end{tabular}}

\newcommand{\bi}{\begin{itemize}}
\newcommand{\ei}{\end{itemize}}
\newcommand{\bfi}{\begin{figure}}
\newcommand{\efi}{\end{figure}}
\newcommand{\ben}{\begin{enumerate}}
\newcommand{\een}{\end{enumerate}}
\newcommand{\bay}{\begin{array}}
\newcommand{\eay}{\end{array}}
\newcommand{\bs}{\boldsymbol}

\def\bco{\iffalse}

\newcommand{\no}{\noindent}
\newcommand{\bc}{\begin{center}}
\newcommand{\ec}{\end{center}}

\newcommand{\bpmt}{\begin{pmatrix}}
\newcommand{\epmt}{\end{pmatrix}}

\newcommand{\bsp}{\begin{split}}
\newcommand{\esp}{\end{split}}

\newcommand{\bdes}{\begin{description}}
\newcommand{\edes}{\end{description}}

\newcommand{\bass}{\begin{assumption}}
\newcommand{\eass}{\end{assumption}}
\newcommand{\bthm}{\begin{thm}}
\newcommand{\ethm}{\end{thm}}
\newcommand{\blem}{\begin{lem}}
\newcommand{\elem}{\end{lem}}
\newcommand{\bcor}{\begin{cor}}
\newcommand{\ecor}{\end{cor}}
\newcommand{\bpf}{\begin{proof}}
\newcommand{\epf}{\end{proof}}
\newcommand{\bprop}{\begin{prop}}
\newcommand{\eprop}{\end{prop}}

\def\Cov{{\rm Cov}}

\def\Var{{\rm Var}}
\def\corr{{\rm cor}}
\def\diag{{\rm diag}}
\def\trace{{\rm trace}}

\def\eps{\varepsilon}

\def\wh{\widehat}

\def\lra{\longrightarrow}
\def\ra{\rightarrow}

\def \mbb {\mathbb}
\def \mbf {\mathbf}
\def \mc {\mathcal}

\def \tbf{\textbf}
\def \lime{\underset{\eps \to 0}{\lim}\ }

\def \inve{\frac{1}{\eps}}
\def \invee{\frac{1}{\eps^2}}

\def \Dtri{\triangledown}

\def \whDtri{\wh{\Dtri}}

\def \balpha{\boldsymbol{\alpha}}
\def \bbeta{\boldsymbol{\beta}}
\def \gopA {g_{1\oplus}}
\def \gopB {g_{2\oplus}}
\def \hop {h_\oplus}
\def \sighat{\wh{\sigma}_{kl}(\true)}
\def \Shat{\wh{S}}
\def \Sighat{\wh{\Sigma}(\true)}

\def \bTheta{\bar{\Theta}}
\def \DH {\bm{\mathit{\Delta}}H}

\def \Vn{V_n}
\def \DVn {\bm{\mathit{\Delta}}\Vn}

\def \DDH {\bm{\mathit{\Delta^2}}H}

\def \Vn{V_n}
\def \DDVn {\bm{\mathit{\Delta^2}}\Vn}

\def \toP{\overset{P}{\rightarrow}}
\def \toD{\overset{D}{\rightarrow}}

\def \argminomega{\underset{\omega \ \in \ \Omega}{\argmin}}
\def \argminbtheta{\underset{\bpara \ \in\ \bar{\Theta}}{\argmin}}
\def \argmintheta{\underset{\para \ \in \ \Theta}{\argmin}}

\DeclareMathOperator*{\argmin}{argmin}

\def\expect{\mbb{E}}
\def\prob{P}
\def\real{\mbb{R}}

\def\indicator#1{\mbb I\left(#1\right)}

\def\t{^{\top}}
\def \btrue{{\boldsymbol{\bar{\theta}_0}}}

\def \SIPBT{\tbfX \t {\boldsymbol{\bar{\theta}_0}}}
\def \bttrue{\boldsymbol{\tilde{\widebar{\theta}}}}
\def \bhtrue{\boldsymbol{\wh{\widebar{\theta}}}}
\def \bpara{\boldsymbol{\bar{\theta}}}
\def \mcTbpara{\mathcal{T}_{\bpara}}
\def \mcT{\mathcal{T}}

\def\true{\boldsymbol{\theta_0}}
\def\ttrue{\tilde{\boldsymbol{\theta}}}
\def\para{\boldsymbol{\theta}}

\def\tbfX{\tbf{X}}
\def\tbfx{\tbf{x}}

\def \htrue{\hat{\boldsymbol{\theta}}}

\def \Yl{\tilde{Y}_l}
\def \Xl{\tilde{\tbfX}_l}
\def \tL{\tilde{L}_b}
\def \hL{\hat{L}_n}
\def \tVn{\tilde{V}_n}

\def\td{\boldsymbol{\theta_{\delta}}}
\def\rd{\rho_{n  \delta}}

\newcommand \mop[1]{m_\oplus( #1)}
\newcommand \hmop[1]{\hat{m}_\oplus(#1)}
\newcommand \tmop[1]{\tilde{m}_\oplus(#1)}
\newcommand \loomop[1] {\hat{m}_{(-i)}\left(#1\right)}

\newcommand{\ltwoNorm}[2][]{%
\ifthenelse{ \equal{#1}{} }
{\ensuremath{\|#2\|}}
{\ensuremath{\|#2\|_{#1}}}
} 
\theoremstyle{plain}
\newtheorem{thm}{Theorem}[section]
\newtheorem{lem}{Lemma}
\newtheorem{cor}{Corollary}
\newtheorem{prop}{Proposition}

\newcommand\Title{}
\newcommand\Author{}
\begin{document}
	\single \bc {\bf \sc \Large Single Index Fr\'echet Regression}
	\vspace{0.15in}\\
	Satarupa Bhattacharjee$^\ast$ 
	and  Hans-Georg M\"uller$^\dagger$\footnote{Research supported in part by NSF grant DMS-2014626 and an NIH ECHO grant.}
	\\
	$^\ast$ Department of Statistics, Pennsylvania State University\\ 
	$^\dagger$Department of Statistics, University of California, Davis\\
	\ec \centerline{July, 2023}
	
	{\bc{\bf \sf ABSTRACT}\ec}
	\no Single index models provide an effective dimension reduction tool in regression, especially for high dimensional data,  by projecting a general multivariate predictor onto a direction vector. 
	We propose a novel single-index model  for regression models where metric space-valued random object responses are  coupled with multivariate Euclidean predictors. The  responses in this regression model include complex, non-Euclidean data,  including  covariance matrices, graph Laplacians of networks, and univariate probability distribution functions,  among  other complex objects that lie in abstract metric spaces.  While Fr\'echet regression has proved useful for 
	modeling the conditional mean of such random objects given multivariate Euclidean vectors,  it does not 
	provide for  regression parameters such as slopes or intercepts, since the metric space-valued responses  are not amenable to linear operations. As a consequence, distributional results for Fr\'echet regression have been elusive. We show here that for the case of multivariate Euclidean predictors, the parameters that define a single index and   projection vector can be used to substitute for the inherent absence of parameters in Fr\'echet regression. Specifically, we derive the asymptotic distribution of 
	suitable estimates of these parameters, which then can be utilized to test linear hypotheses for the parameters, subject to an  identifiability condition.  Consistent estimation of the link function of the single index Fr\'echet regression model is obtained through local  linear Fr\'echet regression. 
	We demonstrate the finite sample performance of estimation  and inference for the proposed single index Fr\'echet regression model through simulation studies, including the special cases where responses are probability distributions and graph adjacency matrices. The  method is  illustrated for resting-state functional Magnetic Resonance Imaging (fMRI) data from the ADNI study.\\
	\vspace{0.1em}
	
	\no {KEY WORDS}: Single index, Dimension reduction, Random objects, Non-Euclidean data, Local Fr\'echet regression, M-estimation, FMRI.
	\clearpage
	
\section{Introduction}
\label{sec:intro}
Modeling the regression relationship between a real-valued response $Y$ and a multivariate Euclidean predictor vector $\tbfX$ corresponds to specifying the form of the conditional means  $m(\tbfx)=\expect{(Y|\tbfX = \tbfx)}.$ Higher dimensionality of $\tbfX$ can be problematic when one is interested to go beyond the standard multiple linear models and aims for a nonparametric estimation of $m(\tbfx).$ This provides strong motivation to consider regression models  that provide dimension reduction. Single index models are one of the most popular approaches to achieve this under the assumption that the influence of the predictors on the response can be collapsed to a single index, i.e., a projection on a specific direction, complemented by a nonparametric link function. This reduces the predictors to  a univariate index while still capturing relevant features and since the nonparametric link function  acts only on a one-dimensional index, these models are not subject to the curse of dimensionality.  The single index model generalizes linear regression, where the link function is the identity. 
For a real-valued response, $Y$ and a $p$-dimensional predictor $\tbfX$, the semiparametric single index regression model is given by
\bgt \label{model:real}
\expect{(Y|\tbfX= \tbfx)}  = \expect{(Y|\tbfX\t\btrue= t)}  = m(t,\btrue).
\egt 
In model~\eqref{model:real},
the dependence between $Y$ and $\tbfX,$ characterized by the conditional mean, is summarized by the parameter vector  $\btrue$ and the link function $m$. 

The function $m$ is nonparametric and thus includes  location and level changes, and therefore the vector $\tbfX$ cannot include a constant that would serve as  an intercept. For identifiability reasons, $\btrue$ is often assumed to be a unit vector with a positive first coordinate. A second approach is  to require  one component  to equal one. This presupposes  that the component that is set to equal 1  indeed  has a non-zero coefficient \citep{lin:07,cui:11}.
Model (\ref{model:real})  is only meaningful if  the Euclidean predictor vector $\tbfX$ is  of dimension $2$ or larger. If $\tbfX$ is one-dimensional, the
corresponding special case of the  model is the one-dimensional nonparametric regression $\expect{(Y|X =x)} = m(x)$, which does not feature any parametric component.

The classical single index regression model with Euclidean responses has attracted attention from the scientific community for a long time due to its flexibility and the interpretability of the (linear) coefficients and flexibility, owing to the nonparametric link function, as well as due to its wide applicability in many scientific fields.  The coefficient $\btrue$ that defines the single index $\tbfx \t \btrue$ along with the shape of the nonparametric component $m$  characterizes the  relationship between the response and the predictor. 
The parametric component $\btrue$ is 
of primary interest for inference in this model.  
The problem of recovering the true direction $\btrue$ can be viewed as a  subclass of sufficient dimension reduction (SDR) techniques, where identifying the  central subspace of $\tbfX$ that explains most of the variation in $Y$  has been a prime target \citep{li:89,cook:94, li:07}. 

In addition to sufficient dimension reduction techniques, various related  approaches to estimate $\btrue$ in \eqref{model:real} have been studied.  These include projection pursuit regression (PPR) \citep{frie:81, hall:89},   average derivatives \citep{hard:89a, stok:86}, sliced inverse regression (SIR) \citep{li:91}, 
conditional minimum average variance estimation (MAVE) \citep{xia:09} and various other methods \citep{xia:06, xia:07}.
These approaches have focused on  the nonparametric estimation of the link function to recover the index parameter in \eqref{model:real}   \citep{hard:93, huh:02, hris:01}, partially linear versions \citep{carr:97,yu:02} and  various noise models \citep{chan:10a,wang:10}. Inference for the index parameters has also been well studied \citep{fan:05a,lian:10,gao:97} for the classical single index model.

Various extensions of single index regression have  been considered more recently  \citep{zhao:20,kere:20}, including models with  multiple indices or 
high-dimensional predictors \citep{zhu:09b,zhou:08, 
	kuch:20}, censored data \citep{lope:13}, and longitudinal and functional data as predictors \citep{jian:11, chen:11, ferr:11, novo:19}. However, none of these extensions has covered situations where responses are not in a Euclidean vector space, even though this case is increasingly important for data analysis. Two very recent exceptions are \cite{ying:20} and~\cite{zhan:21}, who considered extending sufficient dimension reduction approaches for the case of random objects.
The overall  lack of available methodology for single-index models with random object responses motivates our approach. 
Non-Euclidean complex data structures arising in areas such as biological or social sciences are becoming increasingly common, due to  technological advances that have made it possible to record and efficiently store  sensor data and images \citep{peyr:09}, shapes \citep{smal:12} or networks \citep{tsoc:04}. 
For example, one might be interested in  functional connectivity,  quantified in the form of  correlation matrices obtained from neuroimaging  studies,  to study the effect of predictors on brain connectivity, an application that we explore in Section~\ref{sec:data:ADNI}.

Other examples of general metric space objects include probability distributions  \citep{deli:17},  
such as age-at-death distributions as observed in demography or network objects, such as internet traffic networks. Such ``object-oriented data'' \citep{marr:alon:14}  or 
``random objects'' \citep{mull:16} can be viewed as random variables taking  values in a separable metric space that is devoid of a vector space structure and where only pairwise distances between the observed data are available. Almost all existing methodology for single-index models as briefly reviewed above assumes that one has Euclidean responses, and these methods rely in a fundamental way on the vector space structure of the space where the responses reside. When there is no linear structure, a new  methodology is needed and this paper contributes to this development. 

A natural measure of location  for random elements of a metric space is the  Fr\'echet mean \citep{frec:48}, which  is a direct generalization of the standard mean and is defined as the element of the metric space for which the expected squared distance to all other elements, known as the  Fr\'echet function, is minimized. Depending on the space and metric, Fr\'echet means may or may not exist as unique minimizers of the Fr\'echet function. Fr\'echet regression is an extension of Fr\'echet means  to the notion of  conditional Fr\'echet means, and local as well as global versions have been  recently studied in several papers \citep{pete:19, pete:19b, scho:19, scho:20a, bhat:22}. 

Global Fr\'echet regression is a generalization of linear regression for random object responses. In analogy to classical linear regression, it features a restrictive structural model assumption. While the local linear version of Fr\'echet regression is  more flexible, it suffers from the curse of dimensionality as the dimension of the predictors increases.  Further, neither version of the Fr\'echet regression incorporates an interpretable inference regime.
In this paper, we introduce (single) Index Fr\'echet Regression (IFR) to facilitate inference in the context of Fr\'echet regression when the response variable is a random object lying in general metric space and the predictor is a $p$-dimensional Euclidean vector $\tbfX$ with $p\geq 1.$
Our goal is to develop an extension of the conventional estimation and inference paradigm for single-index models for this challenging case.  It is assumed that the conditional expectation (Fr\'echet regression) of $Y$ depends on the predictor vector $\tbfX$ only through the projection or index $\tbfX\t\btrue$ for a parameter vector  $\btrue \in \bTheta \in\real^p.$ Since there is no notion of direction or sign in a  general metric space, we interpret the index parameter in the proposed index Fr\'echet regression model (IFR)  as the direction in the predictor space along which the variability of the response is maximized. The semiparametric framework provided by the proposed single index model facilitates stable estimation and  interpretable inference.

It turns out to be useful to cast  the direction estimation problem in the framework of M-estimation for an appropriate objective function and to use empirical process theory to show consistency of the proposed estimate. We derive an asymptotic normality result  for these estimators under mild assumptions on the metric space and the unknown  link function by utilizing  an appropriate version of   recent results of \cite{chen:22} concerning local linear Fr\'echet regression estimators. 
Under suitable regularity assumptions, the asymptotic distribution of the estimated index parameter can then be harnessed to construct a  Wald-type statistic to conduct inference. 
Combining this with an auxiliary result on the asymptotic convergence of the estimated covariance matrix makes it possible to employ a 
bootstrap method to obtain inference in finite sample situations.

When we finalized this work, we became aware  that independently and simultaneously another group also developed an approach for single index Fr\'echet regression \citep{ghos:21}.
We wish to emphasize that this paper was not in any way influenced by this parallel development (with preprints becoming available within days of each other).

The paper is organized as follows: The basic setup is defined in Section~\ref{sec:model:methods} and the theory on the asymptotic behavior of the index parameter is provided in Section~\ref{sec:theory}, with a  focus on results for inference. The index vector is assumed to lie on a hyper-sphere, with a non-negative first element to facilitate identifiability. Then it is natural to quantify the  performance of the proposed estimators  by the geodesic distances between the estimated and true directions.  The results of simulation studies with various types of random objects as  responses are reported  in Section~\ref{sec:simul} with additional results in the Supplement.  In Section~\ref{sec:applications} we apply the methods to infer and analyze the effect of age, sex, total Alzheimer's brain score  and the stage  of  Alzheimer's Disease on  the brain connectivity  of patients with dementia. Brain connectivity is derived  from fMRI signals of  brain regions of interest  \citep{thom:11} and  quantified in the form of  correlation matrix objects.  We present additional illustrations for human mortality data as distributional objects and mood data of unemployed workers as compositional objects, with details in the Supplement.  A  brief discussion follows in Section~\ref{sec:disc}.

\section{Model and Estimation Methods}
\label{sec:model:methods}
In all of the following, $(\Omega,d, P)$ is a totally bounded metric space with metric $d$ and a probability measure $P.$ The random objects $Y$ take values in $\Omega$. 
This is coupled with a $p$-dimensional real-valued predictor $\tbfX.$  Throughout we will use bold letters  to denote multivariate real vectors. The conditional Fr\'echet mean of $Y$ given $\tbfX$ is a generalization of $\expect{(Y|\tbfX = \tbfx)}$ to metric spaces, defined as the argmin of $\expect{(d^2(Y,\omega)|\tbfX = \tbfx)},$ $\omega \in \Omega$~\citep{pete:19}, i.e., 
\begin{align}
	\label{fr:reg}
	\expect_\oplus{(Y|\tbfX = \tbfx)} : = \argminomega \ \expect{(d^2(Y,\omega)|\tbfX= \tbfx)}.
\end{align}
Evaluated at the minimizer, the objective function in~\eqref{fr:reg} is the corresponding generalized measure of dispersion around the conditional Fr\'echet mean and can be viewed as a conditional Fr\'echet function. 

As discussed earlier, obtaining inference for  Fr\'echet regression is an elusive goal, for both the more restrictive  global as well as the more flexible  but the curse of dimensionality afflicted local version of Fr\'echet regression. 
To move towards inference, we propose here a more structured model, inspired by its Euclidean single index equivalent in~\eqref{model:real},  given by 
\begin{align}
	\label{fr:reg:sim}
	\expect_\oplus{(Y|\tbfX = \tbfx)} = m_\oplus(\tbfx \t \btrue,\btrue),
\end{align}
where $\btrue$ is the true direction parameter of interest. Model \eqref{model:real} emerges as a  special case of model \eqref{fr:reg:sim} for a Euclidean response, as the conditional Fr\'echet mean coincides with the conditional expectation $\expect{(Y|\tbfX)}$ for the choice of the absolute Euclidean distance metric for the case $\Omega = \real.$
In other words, the conditional Fr\'echet mean is assumed to be a function of $\btrue$ in such a way that the distribution of $Y$ only depends on $\tbfX$ only through the index $\SIPBT,$ that is, $Y \perp \expect_\oplus{(Y|\tbfX)}|(\tbfX \t \btrue).$ Thus
\begin{align*}
	\expect_\oplus{(Y|\tbfX = \tbfx)} = \expect_\oplus{(Y|\tbfX\t \btrue = t)} = m_\oplus(t,\btrue), 
\end{align*}
and invoking local linear nonparametric Fr\'echet regression for the one-dimensional index promises to overcome  the curse of dimensionality problem.  
For projections  $\tbfX\t\btrue \in \mathcal{T}_{\btrue} \subset \real$, which depend  on $\btrue$, we  consider predictors $\tbfX$ with  bounded norm such that $\mathcal{T}_{\btrue} \subset \mcT$, where $\mcT$ is a compact interval on $\real.$ We note that the link function, for given $\btrue\in \bTheta,$ $m_\oplus : \mathcal{T}_{\btrue} \mapsto (\Omega,d)$ in the true model depends on the multivariate predictor $\tbfX = \tbfx$ only through the single-index $t = \tbfx\t\btrue$, as well as on the direction vector $\btrue$ implicitly.
Thus, explicitly characterizing this dependence, we define the Index Fr\'echet Regression (IFR) model for random object response $Y$ and Euclidean predictor $\tbfX$ as
\begin{gather} \label{model:sim}
	\mop{t,\btrue} := \argminomega \ \expect{(d^2(Y,\omega)|\tbfX\t\btrue = t)}.
\end{gather}
The coefficient $\btrue \in \real^p$ is the quantity of interest for the  single index Fr\'echet model owing to its interpretability by quantifying the contribution of each predictor component. More generally, the quantity in model~\eqref{model:sim} can be evaluated for any direction vector $\bpara \in \bTheta$ by
\begin{align}
	\label{model:bpara}
	\mop{\tbfx \t \bpara,\bpara} = \argminomega \ \expect{(d^2(Y,\omega)|\tbfX\t\para = \tbfx\t \bpara)}.
\end{align}

In the Euclidean case, identifiability conditions for the direction parameter have been widely discussed in the literature \citep{carr:97, lin:07, cui:11, zhu:06}. We assume the parameter space  $\bar\Theta$ to be constrained in order to ensure that $\bpara$ in the representation \eqref{model:bpara} is uniquely defined, where 
\begin{align}
	\label{full:para:sp}
	\bar{\Theta} :=   \{\bpara = (\theta_1, \dots, \theta_p)\t :  \ltwoNorm{\bpara} = 1,\ \theta_1 > 0, \ \bpara \in \real^p\}.
\end{align} 

We first choose an identifiable parametrization that transforms the boundary of a unit ball in $\real^p$ to the interior of a unit ball  in $\real^{(p-1)}$. By  eliminating $\theta_1,$ the parameter space $\bar \Theta$ can be rearranged to $\{( (1- \sum_{r=2}^p \theta_r^2)^{1/2}, \theta_2,\dots, \theta_p)\t : \sum_{r=2}^p \theta_r^2 < 1\}.$ This re-parametrization is the key to analyzing the asymptotic properties of the estimates for $\para$ and also facilitating efficient computation.  
The true parameter is then partitioned into $\bpara = (\theta_1, \para)\t,$ where  $\para = (\theta_2,\dots, \theta_p)\t.$ We estimate the $(p-1)-$ dimensional vector $\para$ in the single-index model and then use  $\theta_1 = (1- \sum_{r=2}^p \theta_r^2)^{1/2}$ to obtain $\hat{\theta}_1.$
\bprop[Identifiability of model \eqref{model:sim}]
\label{lem:iden:sim:obj}
Suppose $\hop({\tbfx}) = \expect_\oplus{(Y|\tbfX = \tbfx)}$, that the support $S$ of $\hop({\cdot})$ is a convex bounded set with at least one interior point and that $\hop({\cdot})$ is a non-constant continuous function on $S.$ If
$$\hop(\tbfx) = \gopA(\balpha\t \tbfx, \balpha) = \gopB(\bbeta \t \tbfx, \bbeta), \text{ for all } \tbfx \in S, $$  for some continuous object-valued link functions  $\gopA$ and $\gopB$, and some $\balpha, \bbeta \in \bTheta,$ where $\bTheta$ is as described in~\eqref{full:para:sp}. 
Then $\balpha = \bbeta$ and $\gopA \equiv \gopB$ on $\{\balpha \t \tbfx| \tbfx \in S\}.$
\eprop

\no The above result can be proved using a similar argument as given in the proof of Theorem $1$ of \cite{lin:07}. 

Scrutinizing the special case of a Euclidean response $Y$ in model~\eqref{model:real},  the variation in $Y$ is seen to result from the variation in $\SIPBT$ as well as from the variation in the error term in the model, denoted by $\eps$ \citep{ichi:93}. On the contour line $\tbfX \t \btrue = c$, the variability in $Y$ only results from the variability in $\eps$. Along  contour lines $\tbfX \t \bpara= c$ for $\bpara\neq \btrue$,  $\SIPBT$ is not constant and therefore  the variability in $Y$ along the contour lines $\tbfX \t \bpara= c,\ \bpara\neq \btrue$ is due to both the variation in $\SIPBT$ and in $\eps$.  Since $\Var\left(Y|\tbfX \t \bpara= c\right)$ measures the variability in $Y$ on a contour line $\tbfX \t \bpara= c,\ \bpara\neq \btrue$, one can characterize  $\btrue$ as  the minimizer of the objective function $H(\bpara)$, where
$H(\bpara) : = \expect{( \Var(Y|\tbfX\t\bpara))} \text{ and } \btrue = \argmin_{\bpara\in\bTheta} H(\bpara).$
The constraint $\bpara \t \bpara = 1,$ with the first element of the index $\theta_{1}>0$,  ensures the identifiability of the objective function. Defining  an equivalence class of the parameter vector $\bar{\Theta}_{\btrue} :=\{\bpara\in\bTheta : m(\tbfx\t\bpara) = m(\tbfx\t\btrue) \text{ a.e. in } \tbfx \text{ for some } m\}$ for $\bpara \notin \bar{\Theta}_{\btrue}$, one has $H(\btrue) < H(\bpara).$

To recover the true direction of the single index from model~\eqref{model:sim}, the conditional variance of $Y$ given $\tbfX = \tbfx$ for a real-valued response can be replaced by the conditional Fr\'echet variance  $d^2(Y,\mop{\tbfx \t \bpara, \bpara})$ for any given unit orientation vector $\bpara.$ Thus, for a general object response $Y\in (\Omega,d),$ $\btrue$ can alternatively be expressed as
\bal
\label{sim:obj}
\btrue = &\argminbtheta\ H(\bpara),  \ \text{where } H(\bpara)= \expect{\left(d^2\left(Y,\mop{\tbfX \t \bpara,\bpara}\right)\right)},\\
\mop{t,\bpara} &= \underset{\omega\in \Omega}{\argmin} \ M(\omega, t,\bpara),  \ \text{with } M(\omega,t,\bpara):= \expect{\left(d^2(Y,\omega) |\tbfX\t\bpara= t\right)}.
\eal
This corresponds to  finding the true parameter through the optimal direction that maximizes the total variability of the responses, an  idea developed in~\cite{ichi:93} for the case of Euclidean responses. Instead of  choosing the parameter minimizing   the expected variance explained by the  single index $\tbfX\t\para$, for object responses  the new goal is to choose the parameter minimizing  the expected Fr\'echet variance.

To recover $\btrue$ from the representation~\eqref{sim:obj}, one needs to also estimate the conditional Fr\'echet mean, as  in the IFR model \eqref{model:sim}, for which we employ the local linear Fr\'echet regression estimate \citep{pete:19}.  The idea is  as specified below. We approximate the conditional Fr\'echet mean $m_\oplus$ in \eqref{sim:obj}  by a locally weighted Fr\'echet mean that we refer to as intermediate weighted Fr\'echet mean. The weights for this intermediate Fr\'echet mean are derived from a  weight function $S(\cdot,\cdot,\cdot)$ that characterizes the effect on the predictors via a chosen kernel function $K(\cdot)$ and a bandwidth parameter $b$ such that $K_b(\cdot) = (1/b)K(\cdot/b).$ For any given unit direction index $\bpara,$ this intermediate localized weighted Fr\'echet mean is 
\bal
\label{intermed:local:fr}
\tmop{t,\bpara} &= \underset{\omega\in \Omega}{\argmin} \ \tL(\omega, t,\bpara),  \ \text{with } \tL(\omega,t,\bpara):= \expect{\left(S(\tbfX\t \bpara,\ t,b)d^2(Y,\omega)\right)},
\eal
where
\bal 
\label{local:Fr:weight}  
&S(\tbfX \t \bpara,\ t,b) = \frac{1}{\sigma_0^2(t,\bpara)}K_b(\tbfX \t \bpara- t ) [\mu_2(t,\bpara) - \mu_1(t,\bpara)(\tbfX \t \bpara- t)],\\
&\mu_l(t,\bpara) = \expect{(K_b( \tbfX \t \bpara - t ) \ ( \tbfX \t \bpara- t )^l)},\ l = 0,1,2,\quad 
\sigma_0^2(t,\bpara) = \mu_2(t,\bpara) \mu_0(t,\bpara)- \mu_1^2(t,\bpara), 
\eal
and $M(\cdot,t,\bpara) = \tL(\cdot,t,\bpara) + O(b)$ for all $t$ and $\bpara$; note that $\tmop{t,\bpara}$ is a non-random population quantity. 

Suppose we observe a random sample of paired observations  $(\tbfX_i,Y_i),\ i=1,\dots,n$, where $\tbfX_i$ is a $p-$dimensional Euclidean predictor and $Y_i$ is an object response situated in a 
metric space $(\Omega,d).$ Using the form of the intermediate target in \eqref{intermed:local:fr} and replacing the auxiliary parameters by their corresponding empirical estimates, the local Fr\'echet regression estimator at a given value $t$ of the single index for a given direction parameter $\bpara\in\bTheta$ is defined as
\bal
\label{est:local:fr}
\hmop{t,\bpara} &= \argminomega \ \hL(\omega, t,\bpara),  \ \text{with } \hL(\omega,t,\bpara):= \frac{1}{n}\sum_{i=1}^n \Shat(\tbfX_i \t \bpara,\ t,b)d^2(Y_i,\omega),
\eal
where
\bal 
\label{est:local:Fr:weight}  
&\Shat(\tbfX_i \t \bpara,\ t,b) = \frac{1}{\hat{\sigma}_{0}^2(t,\bpara)}K_b( \tbfX_i \t \bpara - t) [\hat{\mu}_{2}(t,\bpara) - \hat{\mu}_{1}(t,\bpara)(\tbfX_i \t \bpara -  t)],\\
&\hat{\mu}_{l}(t,\bpara) = \frac{1}{n} \sum_{j=1}^n K_b( \tbfX_i \t \bpara - t ) \ (\tbfX_i \t \bpara - t)^l, \ l= 0,1,2,\ 
\hat{\sigma}_{0}^2(t,\bpara) = \hat{\mu}_{2}(t,\bpara) \hat{\mu}_{0}(t,\bpara)- \hat{\mu}_{1}^2(t,\bpara).
\eal

The following assumption pertains to the  existence and uniqueness of the Fr\'echet means  in~\eqref{sim:obj} and \eqref{est:local:fr}.
\ben[label = (A\arabic*), series = fregStrg, start = 0]
\item \label{ass:fr:exist} The conditional and weighted Fr\'echet means in \eqref{sim:obj}, \eqref{intermed:local:fr}, and \eqref{est:local:fr} are well defined, i.e., they exist and are unique, the latter one almost surely. Further, for all $\bpara \in \bTheta$ such that $\bpara \neq \btrue$,  $\prob{(X \in \real^p: \ \mop{\tbfX\t\bpara, \bpara} \neq \mop{\tbfX\t\btrue, \btrue})} >0.$

\een

Existence and uniqueness of Fr\'echet means depend on the nature of the metric space and the underlying probability measure and will be discussed further after (A4) in section 3. For example, in the case of Euclidean responses, Fr\'echet means coincide with the usual means for random vectors with finite second moments. In the case of Riemannian manifolds, the existence, uniqueness, and convexity of the center of mass are guaranteed under certain conditions \citep{afsa:11,penn:18}. In a space with a negative or zero curvature, or in a Hadamard space  unique Fr\'echet means always exist \citep{bhat:03,bhat:05,patr:15,kloe:10}. The existence of unique Fr\'echet means in assumption~\ref{ass:fr:exist} is satisfied for the space $(\Omega,d_W)$ of univariate probability distributions with the 2-Wasserstein metric and also for the space $(\Omega,d_F)$ of covariance matrices 
with the Frobenius metric $d_F$ \citep{pete:19}.

Assume that for all unit direction vectors $\bpara$ the support $\mcTbpara$ of $T:= \tbfX\t \bpara$ is compact, where  all   $\mcTbpara$ are subsets of a fixed interval.  For the derivation of distributional limit results, one needs to establish sufficiently fast convergence of the estimated means. This challenge can be overcome by partitioning the interval where the 
linear predictor is situated. Specifically, we partition $\mcTbpara$ into $M$ equal-width non-overlapping bins $\{B_1, B_2,\dots, B_M\},$ where  data falling in different bins are independent and identically distributed.
We denote by  $\Xl$ and $\Yl$  the representative data points in the $l-$th bin, $l=1,\dots,M.$ 
The number of bins $M$ depends on the sample size $n$, where the choice of the sequence $M = M(n)$ is discussed in (A4) in  section 3 below. 
The proposed estimator for the true direction $\btrue$ in \eqref{sim:obj} is then given by
\bal 
\label{est:sim:obj}
\bhtrue &= \argminbtheta\ \Vn(\bpara), \text{ where }
\Vn(\bpara) = \frac{1}{M} \sum_{l=1}^{M} d^2\left(\Yl, \hmop{\Xl\t\bpara, \bpara}\right).
\eal
Here $\hmop{\Xl\t\bpara,\bpara}, \ l =1,\dots,M,\ $ is the local linear Fr\'echet regression estimator, constructed based on the sample $(\tbfX_i,Y_i), \ i=1,\dots,n$, and evaluated at each sample point of the binned sample $(\Xl,\Yl), \ l = 1,\dots, M,$ as described in \eqref{est:local:fr} and \eqref{est:local:Fr:weight}.
We also require  an intermediate quantity that corresponds to the empirical version of $H(\cdot)$ in \eqref{sim:obj}, defined  as 
\bal 
\label{intermed:sim:obj}
\bttrue &= \argminbtheta\ \tVn(\bpara), \text{ where }
\tVn(\bpara) = \frac{1}{M} \sum_{l=1}^{M} d^2\left(\Yl, \mop{\Xl\t\bpara,\bpara}\right).
\eal

The bandwidth $b = b(n)$ is a tuning parameter and 
features in the rate of convergence of $\hat{m}_\oplus$ to $m_\oplus$.  We note that another possible estimator for $m_\oplus$ could be obtained by applying  global Fr\'echet regression. 
This alternative estimator for the unknown link function in the IFR model~\eqref{model:sim} does not depend on a  tuning parameter as is needed for locally linear Fr\'echet regression but is considerably less flexible.

\section{Theory}
\label{sec:theory}
The unknown quantities that constitute the Index Fr\'echet Regression (IFR) model consist of the nonparametric link function and the index parameter, and thus the asymptotic properties of the estimate of the true unit direction rely on those of the estimates  of the link function (based on local linear Fr\'echet regression) and the index parameter (through an M-estimator of the criterion function $H$ in \eqref{sim:obj}). The  metric space $(\Omega,d)$ is assumed to be totally bounded with diameter $D$, hence separable. In order to  obtain the right bound on the metric entropy of the space $\Omega$,  the boundedness assumption is crucial. While boundedness imposes a restriction that is not needed in the Euclidean case, it  is a quite feasible assumption in general metric spaces, since, for commonly observed non-Euclidean objects,  the underlying metric space satisfies the total boundedness property. Examples include the Wasserstein-2 space of one-dimensional distributions with compact support and the space of  spheres with the geodesic metric and  positive semi-definite matrices with  Frobenius or  power metric.

We make the following assumption on the objective function $H(\cdot)$ in~\eqref{sim:obj}.
\ben[label = (A\arabic*), series = fregStrg, start = 1]
\item \label{H1}There exist $\eta>0$ and $C>0$ such that whenever $\ltwoNorm{\bpara - \btrue} <\eta$ for $\bpara \in \bTheta$, we have $H(\bpara) - H(\btrue) \geq C\ltwoNorm{\bpara -\btrue}^2.$
\een
The above condition on  the curvature of the objective function $H$ is standard in the empirical process theory literature and controls the behavior of $\tVn -H$ near the minimum in order to obtain rates of convergence.
In addition, with regard to the quantities in \eqref{sim:obj}, \eqref{est:local:fr}, and \eqref{est:sim:obj} we require the following assumptions.
\ben[label = (A\arabic*), series = fregStrg, start = 2]
\item \label{ass:reg:cont} The link function $m_\oplus$ is Lipschitz continuous, that is,  there exists a real constant $L \geq 0$ such that, for all  $\tbfx$ with a bounded norm, and for all $\bpara_1,  \bpara_2 \in \bar{\Theta},$
\[
d\left(\mop{\tbfx \t\bpara_1, \bpara_1}, \mop{\tbfx \t\bpara_2, \bpara_2}\right) \leq
L \ltwoNorm{\bpara_1-\bpara_2}.
\]

\item\label{A3}  For any given direction $\bpara,$ the univariate index variable $T := \tbfX \t \bpara$ is assumed to have a density $f_{T,\bpara}(\cdot)$ with a compact support $\mcTbpara  \subset \mcT$ for some bounded $\mcT \subset \real.$ We denote the space of predictors for which this holds by $\mathcal{X}\subset \real^p.$ 

\item For $\beta_1,\beta_2>1$ that satisfy assumption (U3) 
in the Supplement and any  $\eps>0$,  let \bal
\label{rate:an}
a_n =  \max\{ b^{2/(\beta_1 -1)}, (nb^2)^{-1/(2(\beta_2 -1)+\eps)}, (nb^2(-\log b)^{-1})^{1/(2(\beta_2 -1))}\}. 
\eal
\label{ass:tuning:M} 
\no The number of non-overlapping bins $M=M(n),$ as defined in Section~\ref{sec:model:methods}, is 
such that $M = M(n) \ra \infty$ and $Ma_n \ra 0$ as $n\ra \infty.$
\een
We note that for $\beta_1 = \beta_2 =2,$ which is the most common situation, $a_n$ reduces to
\[
a_n =  \max\{ b^{2}, (nb^2)^{-1/(2 +\eps)}, (nb^2(-\log b)^{-1})^{1/2}\}.
\]	

Assumption~\ref{ass:reg:cont} is a strong form of uniform continuity for the link function. Intuitively, it limits how fast the object $m_\oplus$ can change, introducing a concept of smoothness in the link function for the IFR model \eqref{model:sim}. Lipschitz continuity is a natural choice of morphisms between metric spaces.
This assumption is slightly stronger than the assumption of a strictly monotone link function that is commonly used in classical single index literature to ensure identifiability. Since the domain of the link function is compact, in the Euclidean response case, our assumption would translate to having a strictly monotone continuous link function with a bounded derivative. Essentially, assumption~\ref{ass:reg:cont} is weaker than a derivative condition and stronger than assuming only the strict monotonicity of the link function. 
Assumption~\ref{A3}  is basic.  The predictors needed for the nonparametric Fr\'echet regression are required to be randomly distributed over the domain where the function is to be estimated, and on average, to  become denser as more data are collected. 
Sufficient for this to be satisfied is that there is at least one continuous predictor and the predictors $\bf{X}$ are bounded. 
Assumption~\ref{ass:tuning:M}  is required for the rate of convergence and limit distribution results, for which we involve the binning device, and it connects the uniform rate of convergence $a_n$ for the local linear Fr\'echet regression estimator as given in ~\eqref{rate:an} with the number of bins $M$. 

For most types of random objects, such as those in the Wasserstein-2 space (the space of probability distributions equipped with the  2-Wasserstein distance) or  the space of symmetric positive semidefinite matrices endowed with the Frobenius or power metric, one has   $\beta_1= \beta_2 =2$ in the definition of $a_n$ in assumption~\ref{ass:tuning:M} (see assumptions (U1)-(U3) 
in Section S.2. 
of the Supplement). If one chooses the bandwidth sequence $b$ for  the local linear Fr\'echet regression such that, for a given $\eps>0,$ $b\sim n^{-(\beta_1 -1)/(2\beta_1 + 4\beta_2 - 6 +2\eps)},$ then $a_n$ is of the order $n^{-\frac{1}{(\beta_1 +2\beta_2 -3 +\eps)}}$ \citep{chen:22}. For  $\beta_1 = \beta_2 =2,$ this becomes 
$a_n \sim n^{-\frac{1}{3+\eps}}.$ 
Any sequence $M=M(n)=n^{\gamma}$ with 
$0 <\gamma<\frac{1}{3}$ will then satisfy assumption~\ref{ass:tuning:M}.  	

As an alternative characterization for the true direction parameter $\btrue$, an important property of the objective function $H(\cdot)$ in~\eqref{sim:obj} is as follows.
\bprop
\label{lem:H}
Under assumptions~\ref{ass:fr:exist} and~\ref{ass:reg:cont}, $H(\cdot)$ in model \eqref{sim:obj} is a continuous function of $\bpara \in \bar{\Theta},$ and 
$\btrue = \argminbtheta \ H(\bpara).$
\eprop

Additional assumptions (U1)-(U3) and (R1)-(R2) 
have been used previously in \cite{pete:19} and \cite{chen:22}, though in a slightly weaker form, and can be found in  Section S.2. 
of the Supplement. These are regarding 
They concern the existence, uniqueness, and well separateness of the minimizers, the metric entropy condition in terms of the covering number, and the curvature of the metric space near the minimizers   and are commonly used  for the  asymptotic analysis of M estimators utilizing  empirical process theory \citep{vand:00}, here  specifically  to establish  consistency and uniform rate of convergence for the local Fr\'echet regression estimator in \eqref{est:sim:obj}, uniform across the single-index values and the direction parameter.
Uniformity over the single index value $t$ was already required in~\cite{chen:22} to achieve uniform convergence of  local linear Fr\'echet regression.  In the  single index model framework, there is a new parameter vector $\bpara$, the presence of which requires an additional uniformity requirement over $\bpara$. 
Assumptions (R1)-(R2) 
are commonly used  in the local regression literature \citep{silv:78,fan:96}.

We will make use of the following lemma, which is an appropriately modified version of a known  result (Theorem $1$ of \cite{chen:22}), to deal with the link function when investigating  the  asymptotic convergence rates of the proposed IFR estimator.
\blem
\label{lem:unif:local:fr:rate}
Under assumptions (U1)-(U3), (R1)-(R2) 
(see Supplement) and  if $b\to 0,$ such that $nb^2(-\log b)^{-1}$ $\to \infty$ as $n\to \infty,$  for any $\eps>0,$ 
\bgt
\sup_{\bpara \in \bTheta} \underset{t \in \mcTbpara}{\sup} \ d(\hmop{t,\bpara},\mop{t,\bpara}) = O_P(a_n),
\egt
where $a_n$ is as given in equation \eqref{rate:an} in assumption~\ref{ass:tuning:M}.
\elem

It is worth mentioning here that the binning approach is not required for basic consistency results without rates (Theorem~\ref{thm:probConv} and
Corollary~\ref{cor:probConv:ifr}). One can indeed re-define the criteria functions in~\eqref{est:sim:obj} 
based on the whole sample $(\tbfX_i,Y_i)$ $i =1,\dots,n$ as
\begin{align*}
	\bhtrue &= \argminbtheta\ \Vn(\bpara), \text{ where }
	\Vn(\bpara) = \frac{1}{n} \sum_{i=1}^{n} d^2\left(Y_i, \hmop{X_i\t\bpara, \bpara}\right), 
\end{align*}
and carry on with the same proof techniques to show consistency of $\bhtrue$ to the true unit direction vector $\btrue.$ However, to prove rates of convergence and investigate the asymptotic behavior of the estimated parameter, we need to make use of the uniform convergence rate $a_n$ for  local linear Fr\'echet regression, as given in Lemma~\ref{lem:unif:local:fr:rate}.
The binning step is necessary to reduce the effective sample size from $n$ to $M = M(n)$, the latter being intrinsically tied by assumption~\ref{ass:tuning:M} to the uniform convergence rate $a_n.$ The rate is effectively slower than $n^{-1/3}$, again by virtue of the uniform convergence rate $a_n$ for the local linear Fr\'echet regression estimator. One may alternatively consider  global Fr\'echet regression to estimate the unknown link function $m_\oplus$,  resulting  in a near parametric rate of $n^{-1/2}$. However, the global Fr\'echet model may suffer from model-induced bias, since as a direct generalization of linear regression, it may be overly restrictive for random object responses. For a consistent unambiguous representation, we refer to the minimizers in~\eqref{est:sim:obj} and~\eqref{intermed:sim:obj} based on  the binned samples as our quantities of interest throughout the rest of the manuscript.

For all of the following results, the basic assumptions~\ref{ass:fr:exist}-\ref{A3} are assumed to be satisfied.
We first demonstrate  the consistency of the proposed estimator for the true index direction. All proofs can be found in Section S$.1.$ of the Supplement.
\bthm \label{thm:probConv}
Under the basic assumptions~\ref{ass:fr:exist}-\ref{A3}, and the technical assumptions (U1)-(U3), and (R1)-(R2) listed in Section S.2. of the Supplement, 
$$\bhtrue -\btrue \overset{P}{\lra} 0 \text{ on } \bar{\Theta},$$
where $\bTheta$ is as defined in~\eqref{full:para:sp}.
\ethm
Combining the consistency result for the direction vector in Theorem~\ref{thm:probConv} with the uniform convergence of the local linear Fr\'echet regression estimator in Lemma~\ref{lem:unif:local:fr:rate} leads to the asymptotic consistency of the estimated single index regression (IFR) model.
\bcor
\label{cor:probConv:ifr}
Under the conditions required for Theorem~\ref{thm:probConv}, for any $\tbfx \in \mathcal{X} \subset \real^p,$
$$d\left(\hmop{\tbfx \t \bhtrue, \bhtrue}, \mop{\tbfx \t \btrue, \btrue}\right) = o_P(1).$$
\ecor

Since any $\bpara \in \bar{\Theta}$ can be  decomposed into $(\theta_1,\para)\t,$ where $\theta_1>0$ and $\|\bpara\|=1$ due to the identifiability requirement,  $\bpara$ is  a function of $\para.$ This makes it possible to write  the criteria function and the corresponding minimizers in terms of the sub-vector $\para$ only, 
\bal
\label{para:obj}
\true = \argmintheta \, H(\para),\quad 
\ttrue = \argmintheta \, \tVn(\para),\quad	
\htrue   = \argmintheta \, \Vn(\para), \text{ where }
\eal
\begin{align}
	\label{red:para:sp}
	\Theta := \{\para : \para \in \real^{p-1},\t \para \t \para < 1\}.
\end{align}
We note that $\true,$ $\ttrue,$ and $\htrue$ are the unconstrained minimizers for the criteria functions $H(\cdot),$ $\tVn(\cdot),$ and $\Vn(\cdot)$ respectively, which 
are continuous functions of $\para,$ the latter two almost surely. Similarly the link function $\mop{\tbfx\t\bpara,\bpara}$ can be rewritten as $\mop{\tbfx\t r(\para),r(\para)}$, where $r(\para) = (1-\sqrt{\ltwoNorm{\para}^2},\para)\t$.

To study limit distributions, we impose an additional requirement on the interplay between the metric $d(\cdot,\cdot)$ in the metric space of responses and the true regression function $m_\oplus$, namely  that the second order difference of the function $d^2(\cdot, \mop{z_0})$ is bounded away from zero, for any $z_0\in \mc{T},$ where $\mc{T}\subset \real$ denotes the domain of $m_\oplus.$ Specifically, for $z_0 = \mathbf{z}\t r(\para),$ for some $\mathbf{z} \in \real^{p}$ and $\para \in \Theta,$ we denote $\mop{\mathbf{z}\t r(\para), r(\para)}= \mop{z_0,\para}$ by $\mop{z_0}.$ We assume
\ben[label = (A\arabic*), series = fregStrg, start = 5,topsep=0pt]
\item \label{ass:convexity:d_m} For any $z_0 \in \mc{T}\subset\real$  and $u\in\Omega,$ there exists some $\kappa >0,$ and $a_0>0,$ such that for any sufficiently small $0<a<a_0,$ and $z_0 +2a \in \mc{T},$
$$
\frac{1}{a^2}\left[ d^2(u, \mop{z_0+2a}) -2d^2(u, \mop{z_0+a}) + d^2(u, \mop{z_0})\right] \geq \kappa.
$$
\een
In the Euclidean case, assumption~\ref{ass:convexity:d_m} means that $m_{\oplus}$ can be locally approximated by straight lines and is satisfied for twice differentiable functions $m_\oplus$, a common assumption for classical single index modeling. Beyond the Euclidean special case, assumption~\ref{ass:convexity:d_m} can be shown to be satisfied  for fairly general metric spaces.  
An example for this are CAT(0) spaces (see \cite{bura:01}), 
where the regression function between two distinct points $\mop{z_0}$ and $\mop{z_0+a},$ for some small $a>0,$ can be approximated arbitrarily closely by the geodesic path connecting them. Further details on this are provided in Appendix~\ref{append:curvature} and~\ref{append:curvature:assump}. 

The geometric assumption~\ref{ass:convexity:d_m} is crucial to show that the intermediate objective function $\tVn(\cdot)$ has non-negative curvature near its minimizer $\ttrue$ with high probability. This is necessary  to bound the rate of the convergence of the discrepancy between the intermediate index parameter $\ttrue$ and the  estimated version $\htrue.$ 
We proceed to define partial derivatives  of the criteria functions with respect to the components of $\para.$ 
For any $\tbfx \in \real^p$ with bounded norm and $y\in (\Omega,d)$, define the function $f_{\tbfx,y}:\real^{p-1} \mapsto \real$ such that
\bal
\label{finite:diff:fn}
f_{\tbfx,y}(\para) = f_{\tbfx,y}(\theta_2, \dots,\theta_p) = d^2\left(y, \mop{\tbfx\t(\theta_1,\dots,\theta_r,\dots,\theta_s,\dots,\theta_p)}\right), \ r,s =2,\dots,p.
\eal
The first and second ordered forward finite differences of $f_{\tbfx,y}$ are given as follows
\bal
\label{diff_defn1}
\Dtri_a(\tbfx,y,\theta_r) &= f_{\tbfx,y}(\theta_2,\dots,\theta_r +a,\dots,\theta_p) - f_{\tbfx,y}(\theta_2,\dots,\theta_r ,\dots,\theta_p),\\
\Dtri^2_a(\tbfx,y,\theta_r,\theta_s) &= f_{\tbfx,y}(\theta_2,\dots,\theta_r +a,\dots, \theta_s +a,\dots \theta_p) - f_{\tbfx,y}(\theta_2,\dots,\theta_r +a,\dots,\theta_s,\dots,\theta_p) \\
&\quad - f_{\tbfx,y}(\theta_2,\dots,\theta_r ,\dots,\theta_s+a,\dots, \theta_p) +f_{\tbfx,y}(\theta_2,\dots,\theta_r,\dots,\theta_s,\dots, \theta_p).
\eal
Define
\begin{align*}
	\label{diff_defn_H}
	\begin{cases} \DH(\para) := \left(\frac{\partial H(\para)}{\partial \theta_2} ,\dots \frac{\partial H(\para)}{\partial \theta_p} \right)\t, \ 
		&\frac{\partial H(\para)}{\partial \theta_r}  := \lime \inve \expect{(\Dtri_{\eps}(\tbfX,Y,\theta_r))}, \ r =2,\dots,p,\\
		\DDH(\para) := \left(\left(\frac{\partial^2 H(\para)}{\partial \theta_r \partial \theta_s}\right)\right)_{r,s =2,\dots,p},\ 
		& \frac{\partial^2 H(\para)}{\partial \theta_r \partial \theta_s} := \lime \invee \expect{(\Dtri^2_\eps(\tbfX,Y,\theta_r,\theta_s))},  \ r,s =2,\dots,p.
	\end{cases}
\end{align*}

We note that $H(\cdot),$ $\tVn(\cdot),$ and $\Vn(\cdot)$ are all real-valued functions with domain in a constrained subset of $\real^p.$ The appropriate limits for defining the partial derivatives can be shown to exist under \ref{ass:reg:cont} and the assumed total boundedness of the metric space $\Omega.$
The estimated versions of the finite difference derivatives are, for $r,s = 2,\dots,p$,
\begin{align*}
	\begin{cases}
		\DVn(\para) := \left(\frac{\partial \Vn(\para)}{\partial \theta_2} ,\dots \frac{\partial \Vn(\para)}{\partial \theta_p} \right)\t,\
		&\frac{\partial \Vn(\para)}{\partial \theta_r}  := \frac{1}{hM}  \sum_{l=1}^M \whDtri_{h}(\Xl,\Yl,\theta_r),\\
		\DDVn(\para):= \left(\left(\frac{\partial^2 \Vn(\para)}{\partial \theta_r \partial \theta_s} \right) \right)_{r,s =2,\dots,p}, \
		& \frac{\partial^2 \Vn(\para)}{\partial \theta_r \partial \theta_s}  = \frac{1}{h^2M}  \sum_{l=1}^M \wh{\Dtri^2}_{h}(\Xl,\Yl,\theta_r,\theta_s),
	\end{cases}
\end{align*}
\bal
\label{diff_defn2}
\whDtri_{h}(\tbfx,y,\theta_r) &= \hat{f}_{\tbfx,y}(\theta_2,\dots,\theta_r +h,\dots,\theta_p) - \hat{f}_{\tbfx,y}(\theta_2,\dots,\theta_r ,\dots,\theta_p),\\
\wh{\Dtri^2}_{h}(\tbfx,y,\theta_r,\theta_s) &= \hat{f}_{\tbfx,y}(\theta_2,\dots,\theta_r +h,\dots, \theta_s +h,\dots \theta_p) - \hat{f}_{\tbfx,y}(\theta_2,\dots,\theta_r +h,\dots,\theta_s,\dots,\theta_p) \\
&\quad - \hat{f}_{\tbfx,y}(\theta_2,\dots,\theta_r ,\dots,\theta_s+h,\dots, \theta_p) +\hat{f}_{\tbfx,y}(\theta_2,\dots,\theta_r,\dots,\theta_s,\dots, \theta_p),
\eal
with
\bal
\label{finite:diff:fn:est}
\hat{f}_{\tbfx,y}(\para) = \hat{f}_{\tbfx,y}(\theta_2, \dots,\theta_p) = d^2\left(y, \hmop{\tbfx\t(\theta_1,\dots,\theta_r,\dots,\theta_s,\theta_p)}\right), \ r,s =2,\dots,p.
\eal
\no Here $h = h(n)$ is a tuning parameter depending on $n,$ for which we assume that  
\ben[label = (A\arabic*), series = fregStrg, start = 6]
\item \label{ass:tuning:h} $h = h(n) \to 0$ and $M h^2(n) \to \infty,$ as $n \to \infty.$ 
\een
Assumptions~\ref{ass:tuning:M} and~\ref{ass:tuning:h} together imply that furthermore 
$a_n/h^2 \to 0,$ as $n\to \infty.$

Observe that the true and estimated index directions can be framed as  M-estimators of their respective criteria functions. This suggests utilizing empirical process-based approaches  to obtain distributional convergence of $\htrue,$ specifically  
to adopt  a linearization approach \citep{vand:00}. Specifically, 
we show that $\sqrt{M}(\htrue - \ttrue) =o_P(1)$ and $\sqrt{M}(\ttrue - \true) \toD Z,$ where $Z$ is a Gaussian random variable.
Combining these results, it follows that
\bthm
\label{thm:normality}
Under assumptions~\ref{H1}-\ref{ass:tuning:h}, and assumptions (U1)-(U3), and (R1)-(R2) listed in the Supplement Section S.2., 
$$\sqrt{M}(\htrue - \true ) \toD N_{p-1}\left(0,\Lambda(\true)\right),$$ 
where $M$ and $a_n$ are as defined in assumption~\ref{ass:tuning:M}, 
$\Lambda(\true) := \left(\DDH(\true)\right)^{-1}\Sigma(\true) \left(\DDH(\true)\right)^{-1},$ and
$\Sigma(\true) = ((\sigma_{rs}(\true)))_{r,s = 2,\dots}$ with \\$\sigma_{rs}(\true) = \begin{cases}
	\lime \Var \left(\inve \Dtri_{\eps}(\tbfX, Y,\theta_{0r})\right), \text{ if } r = s \in \{2,\dots,p\},\\
	\lime \Cov \left(\inve \Dtri_{\eps}(\tbfX, Y,\theta_{0r}), \ \inve \Dtri_{\eps}(\tbfX, Y,\theta_{0s})\right), \text{ if } r \neq s, r,s \in \{2,\dots,p\}.\\
\end{cases}$
\ethm

The asymptotic normality of $\bhtrue = (\wh{\theta}_1,\htrue)$ follows from Theorem~\ref{thm:normality} with a simple application of the multivariate delta method as  $\wh{\theta}_1 = \sqrt{1 - \ltwoNorm{\htrue}^2},$ implying   $\bhtrue - \btrue = O_P(M^{-1/2}).$
\bcor
\label{cor:full:para:norm}
Under the conditions required for Theorem~\ref{thm:normality},
$$\sqrt{M}(\bhtrue - \btrue ) \toD N_{p}\left(0,J\Lambda(\true)J\t\right),$$
where $J = \left.\left(\left(\frac{\partial \bpara}{\partial \para}\right)\right)\right\vert_{\para = \true}= \left.\begin{pmatrix}
	-\para\t / \sqrt{1-\ltwoNorm{\para}^2}\\
	I_{p-1}
\end{pmatrix}\right\vert_{\para = \true}$ is the Jacobian matrix of size $p \times (p - 1)$.
\ecor

\no Define the intuitive estimator   $\Sighat$ 	for $\Sigma(\true)$ given by $\Sighat = ((\sighat))_{r,s = 2,\dots,p}$,  with \\$\sighat = 
\begin{cases}
	\frac{1}{h M} & \sum_{l=1}^{M} \wh{\Dtri}^2_{h}(\Xl,\Yl,\theta_{0r}) -
	\left(\frac{1}{h M} \sum_{l=1}^{M} \whDtri_{h}(\Xl,\Yl,\theta_{0r})\right)^2, \text{ if } r = s,\\
	\frac{1}{h M} & \sum_{l=1}^{M} \whDtri_{h}(\Xl,\Yl,\theta_{0r}) \whDtri_{h}(\Xl,\Yl,\theta_{0s}) \\
	& \ - \left( \frac{1}{h M} \sum_{l=1}^{M} \whDtri_{h}(\Xl,\Yl,\theta_{0r})\right) \left(  \frac{1}{h M} \sum_{l=1}^{M} \whDtri_{h}(\Xl,\Yl,\theta_{0s})\right), \text{ if } r\neq s.
\end{cases}$\\
The following two propositions imply  consistent estimation of the covariance matrix.
\bprop
\label{lem:normality:tVn:diff:est_var}
Under assumptions~\ref{H1}-\ref{ass:tuning:h},
$\sqrt{M} \left( \mathbf{vec}(\Sighat) - \mathbf{vec}(\Sigma(\true))\right)$  converges to a $(p-1)^2-$  dimensional normal distribution with mean vector $0$ and a finite covariance matrix.
\eprop 
Details about the limiting covariance matrix can be found in Section S$.1.$ of the Supplement. 
A natural   estimate for the asymptotic covariance matrix in Theorem~\ref{thm:normality} is $\wh{\Lambda}(\htrue) :=   \left(\DDVn(\htrue)\right)^{-1}\wh{\Sigma}(\htrue)\left(\DDVn(\htrue)\right)^{-1}.$ 
\bprop
\label{lem:consistent:Vn:diff2}
Under assumptions~\ref{H1}-\ref{ass:tuning:h}, and assumptions (U1)-(U3), and (R1)-(R2) listed in the Supplement Section S.2., 
$$\wh{\Lambda}(\htrue) - \Lambda(\true) \toP 0.$$
\eprop
With  Slutsky's theorem, combining the above propositions with Theorem~\ref{thm:normality}, 
\bcor
\label{cor:normality:est_var}
Under assumptions~\ref{H1}-\ref{ass:tuning:h}, and assumptions (U1)-(U3), and (R1)-(R2) listed in the Supplement Section S.2., 
$$\sqrt{M}(\wh{\Lambda}(\htrue))^{-1/2}(\htrue - \true)\toD N(0,I_{p-1}),$$ 
where $M$ and $a_n$ are as defined in assumption~\ref{ass:tuning:M}.
\ecor
Again it is straightforward to  extend the above result to obtain the limit distribution for the full parameter vector  $\bhtrue = (\wh{\theta}_1,\htrue),$ as due to the constraints  the full parameter vector is a function of the reduced one.
Define the estimate for the  Jacobian matrix of size $p \times (p - 1)$ as $\hat{J} = \left.\left(\left(\frac{\partial \bpara}{\partial \para}\right)\right)\right\vert_{\para = \htrue}= \left.\begin{pmatrix}
	-\para\t / \sqrt{1-\ltwoNorm{\para}^2}\\
	I_{p-1}
\end{pmatrix}\right\vert_{\para = \htrue}$. Then using Corollary~\ref{cor:full:para:norm} and Proposition~\ref{lem:consistent:Vn:diff2}  one has
$
\sqrt{M}(\hat{J}\wh{\Lambda}(\htrue)\hat{J}\t)^{-1/2}(\bhtrue - \btrue) \toD N_{p}\left(0, I_p\right),
$ and furthermore 
\bcor
\label{cor:probConvwithRate:ifr}
Under the conditions required for Corollary~\ref{cor:normality:est_var}, for any $\tbfx \in \mathcal{X} \subset \real^p,$
$$d\left(\hmop{\tbfx \t \bhtrue, \bhtrue}, \mop{\tbfx \t \btrue, \btrue}\right) = O_P(M^{-1/2}).$$
\ecor
In applications of regression models, it is often important to test the statistical significance of added predictors. Based on the above normality results, one can obtain Wald-type statistics to test the significance of certain variables in the linear index. 
Since $\btrue$ is on the surface of the unit sphere, the constraint $\ltwoNorm{\btrue} = 1$ removes one dimension. The actual dimension of the surface of the unit sphere is $p-1$  and the values of $(p-1)$ components of $\btrue$ determine $\btrue$ when  without loss of generality, the value of the first component of $\btrue$ is assumed to be positive. Therefore one can obtain confidence regions for $\btrue$ by  constructing confidence regions for the last $(p-1)$ components of $\btrue$ only. 

A common testing problem concerns the null hypothesis H$_0$: $\theta_k= 0,\ k=r,\dots,p\,$,  for any $2 \le r \le p$.  More  general  tests of a linear null hypothesis H$_0$:  $B\para^{(\mbf{r})}=0$ for a known matrix 
$B$ of full row rank and $\para^{(\mbf{r})}= (\theta_{r},\dots,\theta_p)\t$ are also of interest and are implied  by the  following result, which also
provides (elliptical) asymptotic confidence regions for the components of interest and whereas before $M=M(n)$ is the number of bins in the binning step.
\bcor
\label{cor:chi_sq:hyp:test}
Under the null hypothesis $H_0 : B\para^{(\mbf{r})} = \zeta,$ for some $q\times (p-r+1)$ matrix $B$ with $1 \le q \le p-r+1\,$ of rank $q$, denoting 
the estimated asymptotic covariance matrix for $\htrue^{(\mbf{r})}$ by $\wh{\Lambda}(\htrue^{(\mbf{r})}),$ then under the conditions required for Corollary~\ref{cor:normality:est_var},
$$ T_n= (B\htrue^{(\mbf{r})} - \zeta)\t (B (\wh{\Lambda}(\htrue^{(\mbf{r})})/M)^{-1} B \t)^{-1}  (B\htrue^{(\mbf{r})}  - \zeta) \toD \chi_q^2.$$
Specifying the last $(p-r+1)$ components of the true direction index as \\
$\true^{(\mbf{r})} =(\theta_{0r},\dots,\theta_{0p})\t,$ where $r=2,\dots,p$, 
a $100(1-\gamma)\%$ confidence region for $\true^{(\mbf{r})}$ is 
$$C_{\gamma} = \{ \para \in \real^{p-r+1}  : (\htrue^{(\mbf{r})} - \para)\t (\wh{\Lambda}(\htrue^{(\mbf{r})})/M)^{-1} (\htrue^{(\mbf{r})} - \para) \leq c_\gamma^\ast, \ltwoNorm{\para} <1 \},$$  with $ \prob{(\chi^2_{p-r+1} \leq c_\gamma^\ast)} = 1-\gamma$. Here
$\wh{\Lambda}(\htrue^{(\mbf{r})})$ is the $(p-r+1)$ dimensional sub-matrix of the asymptotic covariance matrix $\wh{\Lambda}(\htrue)$.
\ecor
Observe that for $r=2$, $\true^{(\mbf{r})} =\true.$ Then Corollary~\ref{cor:chi_sq:hyp:test} yields  the confidence region for the parameter $\true$ as $C_{\gamma} = \{ \para \in \real^{p-1}  : (\htrue- \para)\t (\wh{\Lambda}(\htrue)/M)^{-1} (\htrue - \para) \leq c_\gamma^\ast,\,  \ltwoNorm{\para} <1 \},$  with $ \prob{(\chi^2_{p-1} \leq c_\gamma^\ast)} = 1-\gamma$. 
Then the confidence region for $\btrue$ can be obtained immediately through the relationship $\btrue = (\theta_{01},\true)\t$ with $\theta_{01} = \sqrt{1- \ltwoNorm{\true}^2}.$ 

For practical implementation, direct estimation of the asymptotic covariance matrix is tedious  since it involves a tuning parameter to approximate the partial derivative for multiple variables by finite difference quotients. Instead, we use a nonparametric bootstrap approach to provide  a consistent estimator of the asymptotic covariance matrix~\citep{davi:97, shao:12}. 
Consistency of the bootstrap moment estimators for a general M-estimator is a well-studied problem. 
\cite{kato:11} established  uniform integrability of the bootstrap M-estimator, thereby giving sufficient conditions for the consistency of the bootstrap moment estimators. Following similar arguments as Theorem 2.2 in~\cite{kato:11}, we obtain  consistency of the proposed bootstrap covariance matrix estimator.

Let  $(\tbfX^\ast_1,Y^\ast_1), \dots, (\tbfX^\ast_n,Y^\ast_n)$ denote a bootstrap sample, i.e., an independent sample from the empirical distribution of the observed sample $(\tbfX_1,Y_1), \dots, (\tbfX_n,Y_n)$. 
The bootstrap M-estimator 
of $\true$  is $\htrue^\ast = \argmintheta \frac{1}{M} \sum_{l=1}^M d^2\left( \tilde{Y}_l^{\ast}, \hmop{(\tilde{\tbfX}_l^{\ast \intercal} \para)} \right).$ Here $\Yl^\ast$ and $\Xl^\ast$ are the response and predictor values for observations falling in the $l-$th bin, $l=1,\dots,M.$
A bootstrap estimator of the asymptotic covariance matrix is given by~\citep{kato:11,nish:10,buch:95,gonc:05} $$\hat{\Lambda}^\ast := \expect{\left[ M (\hat{\boldsymbol{\theta}}^{\ast} - \htrue) (\hat{\boldsymbol{\theta}}^{\ast} - \htrue)^\intercal | (\tilde{\tbfX}_1,\tilde{Y}_1), \dots, (\tilde{\tbfX}_M,\tilde{Y}_M)\right]}.$$
\bprop
\label{lem:boot:consistency}
Under assumptions~\ref{H1}-\ref{ass:tuning:h}, 
$\hat{\Lambda}^\ast$ is consistent for the true asymptotic covariance matrix $\Lambda(\true).$
\eprop 

Combining the above proposition with Theorem~\ref{thm:normality} using the bootstrap covariance estimator, an analog of Corollary 3 immediately follows,  as $\sqrt{M}(\hat{\Lambda}^\ast)^{-1/2}(\htrue - \true)\toD N(0,I_{p-1}),$ justifying the bootstrap construction of  confidence regions and ensuing inference,  where we 
approximate the bootstrap covariance matrix $\hat{\Lambda}^\ast$ by Monte Carlo estimation. The observed sample $(\tbfX_1,Y_1), \dots, (\tbfX_n,Y_n)$ is resampled with replacement $B$ times and the estimate for the index parameter $\htrue$ computed for each bootstrap sample. 
Based on the $b^{\text{th}}$ bootstrap sample the index parameter is estimated as $\hat{\boldsymbol{\theta}}^{\ast}_{b},$ $b=1,\dots B.$ The bootstrap estimate of the  covariance matrix is then  $\hat{\Lambda}^\ast_B = \frac{1}{B} \sum_{b =1}^B M (\hat{\boldsymbol{\theta}}_b^{\ast} - \htrue) (\hat{\boldsymbol{\theta}}_b^{\ast} - \htrue)^\intercal,$ which is also  consistent for $\Lambda(\true).$

As an example, if  one applies  the  statistic in Corollary~\ref{cor:chi_sq:hyp:test} to test  the null hypothesis
\bal
\label{hyp:null}
H_0 : \theta_{02} = \dots = \theta_{0p} =0, \text{ where  } \theta_{01}=1,
\eal
one  can examine the power of the test for alternatives indexed by a parameter $\delta>0$, 
\bal
\label{hyp:alt}
H_{1\delta} : \theta_{02} = \dots = \theta_{0p} =\delta.
\eal
Under $H_0,$ $T_n = M \htrue\t (\wh{\Lambda}_B^\ast)^{-1} \htrue \sim \chi^2_{(p-1)}$ asymptotically. Noting that $\wh{\Lambda}_B^\ast$ is consistent for  $\Lambda(\true)$ under both $H_0$ and 
$H_{1\delta}$,  
the asymptotic distribution of $T_n$ under $H_{1\delta}$  is the non-central chi-square distribution  $\chi^2_{(p-1)}(\rd)$ with $(p-1)$ degrees of freedom and non-centrality parameter
$\rd =M\td\t  (\Lambda(\td))^{-1}\td$,
where $\td = (\delta,\dots,\delta)$. 
The asymptotic power of the level $\alpha$  test under $H_{1\delta}$ is  $\prob{(T_n > \chi^2_{(p-1)}(1-\alpha))}$, where $T_n \sim \chi^2_{(p-1)}(\rd)$, which demonstrates that for all $\delta>0$ the asymptotic power converges to 1 with the rate $M^{-1}$.


\section{Implementation and simulation studies}
\label{sec:simul}
Implementation of the single index Fr\'echet regression (IFR) model in~\eqref{sim:obj} requires the choice of two tuning parameters,   the bandwidth $b = b(n)$ used  for  the local linear Fr\'echet regression as per \eqref{model:sim} and the number of bins $M = M(n)$  (see assumption~\ref{ass:tuning:M}). 
In applications, the tuning parameters $(b,M)$ can be chosen by leave-one-out cross-validation. The first step is to select the optimal bandwidth parameter $b^\ast$ by minimizing  the mean discrepancy between the local linear Fr\'echet regression estimates and the observed object responses, i.e., 
\[
b^\ast = \underset{b}{\argmin\ } \frac{1}{n}\sum_{i=1}^n d^2(Y_i, \loomop{\tbfX_i \t\bpara, \bpara}), \]
where $\loomop{\tbfX_i \t\bpara, \bpara}$ is the local  linear Fr\'echet regression estimate at $\tbfX_i \t\bpara$ obtained with bandwidth $b$ based on the sample excluding the $i-$th pair $(\tbfX_i,Y_i)$, i.e., 
\[
\loomop{\tbfX_i \t\bpara, \bpara} = \argminomega \frac{1}{(n-1)} \sum_{j\neq i}  \Shat(\tbfX_j \t \bpara,\tbfX_i\t\bpara ,b)d^2(Y_j,\omega).
\]

In practice, we replace leave-one-out cross-validation by $5-$fold cross-validation when $n > 30$. 
Once $b^\ast$ is chosen a second leave-one-out cross-validation step is applied to  select the number of non-overlapping bins $M^\ast,$ where the 
objective function to minimize is the empirical Fr\'echet variance for the binned data,
\[
M^\ast = \underset{M}{\argmin\ } \frac{1}{M} \sum_{l=1}^M d^2(\Yl, m_{\oplus (-l)}^{b^\ast}(\Xl \t\bpara, \bpara)).
\]
Here $m_{\oplus (-l)}^{b^\ast}(\Xl \t\bpara, \bpara)$ is the local linear Fr\'echet regression estimate at $\Xl \t\bpara$ obtained with bandwidth $b$ based on the sample excluding the observation at the $l-$th bin $(\Xl,\Yl)$, i.e., 
\[
m_{\oplus (-l)}^{b^\ast}(\Xl \t\bpara, \bpara) = \argminomega \frac{1}{n} \sum_{i =1}^n  \Shat(\tbfX_i \t \bpara,\Xl \t \bpara ,b^\ast)d^2(Y_i,\omega).
\]

Thus, for each given unit direction $\bpara,$ we first select the optimal tuning parameters $(b^\ast,M^\ast),$ which will generally vary with  $\bpara,$ and then employ them when computing  the loss function $V_n(\bpara).$ Finally,  the index parameter is estimated as  $\bhtrue,$ the unit direction minimizing $V_n(\bpara)$ over $\bpara$ such that $\bpara \t \bpara =1.$ This leads to an iterative scheme,  where for a given unit direction the tuning parameters $(b^\ast, M^\ast)$ are initially selected by cross-validation  and then iteratively updated, in turn with updating  $\bpara$  to minimize  the loss function. We numerically optimize
the empirical loss $V_n(\bpara)$ under the constraint $\bpara\t\bpara =1$ via the following algorithm.
\begin{itemize}
	\item[1.] Take a grid of unit vectors $\bpara$ such that $\bpara\t\bpara =1$. This is achieved by generating $p$ dimensional standard Gaussian random vectors with positive first elements and standardizing them, utilizing the spherical symmetricity of $p$-dimensional standard Gaussian vectors.
	\item[2.] For each $\bpara,$  select  optimal tuning parameters $(b^\ast,M^\ast)$ (for bandwidth and number of non-overlapping bins, respectively) by cross-validation.
	\item[3.] Using $(b^\ast,M^\ast)$ compute the loss function $\Vn(\para) = \frac{1}{M}\sum_{l=1}^M d^2(\Yl,\hmop{\Xl\t\bpara}).$
	\item[4.] Find the minimizer $\htrue$ of $\Vn(\para)$ such that $\bpara\t\bpara  =1$ by searching over all directions $\bpara$ generated in step 1. In our implementation, we considered 500 directions. 
\end{itemize}

The computational challenges to obtain Fr\'echet means vary by metric space. In many cases, the key idea to compute the weighted Fr\'echet means reduces to solving a constrained quasi-quadratic optimization problem and projecting back into the solution space. For random objects such as distributions, positive semi-definite matrices, networks, and Riemannian manifolds among others, obtaining the unique solution is computationally straightforward. 
For our simulations we considered random objects corresponding to samples of univariate distributions equipped with the Wasserstein$-2$ metric and samples of multivariate data with the usual Euclidean metric. 

We generated $500$ Monte Carlo runs for each setting, and for each run obtained a direction estimate  $\bhtrue^{(i)} \ i = 1,\dots, 500.$ The intrinsic Fr\'echet mean of these $500$ estimates on the unit sphere was then  computed as $\wh{\bar{\para}}.$ Given that  both the  $\bhtrue^{(i)}$ and their target    $\btrue$  lie on the unit sphere in $\real^p$,   bias and deviance of the estimator can be  obtained as 
\begin{align}
	\text{bias}(\bhtrue) = \arccos \langle \wh{\bar{\para}},\btrue \rangle, \quad 
	\text{dev}(\bhtrue) = \widehat{\Var}(  \arccos \langle \bhtrue^{(i)}, \bhtrue \rangle).
	\label{simul:bias:var}
\end{align}

To illustrate the performance of  the Wald-type statistic for testing a  linear hypothesis, we again created Monte Carlo runs as described above except that  components of the index were  generated to  follow the null hypothesis in~\eqref{hyp:null}. 
To obtain the power function of the test against the sequence of alternatives given in~\eqref{hyp:alt}, we calculated the test statistic for $500$ simulation runs and determined the fraction of tests that rejected the null hypothesis at the nominal level $\alpha = 0.05.$

\subsection{Distributional responses}
The  space of univariate distributions with the Wasserstein 
metric provides an ideal setting for illustrating the efficacy of the proposed methods. For any two distribution objects $F,G \in (\Omega,d_W)$, the  Wasserstein-2 distance is given by 
\begin{align}
	\label{wass:metric}
	d_W(F,G) = \int_0^1 (F^{-1}(s) - G^{-1}(s))^2 ds,
\end{align}
where $F^{-1}$ and $G^{-1}$ are the quantile functions corresponding to $F$ and $G$ respectively. 
We consider distributions on a bounded domain 
as  responses $Y(\cdot)$ that we represent  by their respective quantile functions $Q(Y)(\cdot)$ 
and that are  paired with a  $p$ dimensional Euclidean predictor vector $\tbfX$.
The true single index projections $\tbfx\t\btrue$ were obtained  by first generating $(Z_1,\dots,Z_p)^\intercal$ from a multivariate Multivariate Gaussian distribution with  $\expect(Z_j) = 0$ and $\Cov(Z_j,Z_{j'}) = \rho = 0.25$. Then the components of $\tbfX= (X_1,\dots,X_p)^\intercal$ were computed as $X_j = 2\Phi(Z_j) -1,$ where $\Phi$ is the standard normal distribution function. Finally, we generated a $p-$dimensional unit vector $\btrue$ such that $\ltwoNorm{\btrue} =1$ and $\bar{\theta}_{01}>0$, and 
computed the projection $\tbfX\t\btrue.$ We selected  $p=4$ 
and random responses were  generated conditional on $\tbfX$, by adding noise to the true regression quantile function
\begin{align}
	Q(m_\oplus(\tbfx))(\cdot) &= \expect{\left(Q(Y)(\cdot)|\tbfX\t\btrue = \tbfx\t\btrue\right)}.
	\label{simul:dens1}
\end{align}

For generating the distributional responses, two simulation settings were examined (see Table~\ref{table:sim}). For both scenarios, three different link functions were considered for the data-generating mechanism, namely $\zeta(z) = z$,  $\zeta(z) = z^2$, and $\zeta(z) = \exp(z).$ In the first setting, the true response was generated as a normal distribution with parameters depending  on $\tbfX\t\btrue.$ For  $\tbfX\t\btrue =\tbfx\t\btrue$, the distribution parameters $\mu(\tbfx) \sim N(\zeta(\tbfx\t\btrue), 0.25)$ and $\sigma(\tbfx) \sim Exp(1/\eta(\tbfx\t\btrue))$ were independently sampled, where $\eta(z) = \frac{\exp(z)}{1 +\exp(z)}$. The corresponding distribution-valued regression function
is given by 
$m_\oplus(\tbfx\t\btrue) = \expect{(Q(Y)(\cdot)|\tbfX\t\btrue = \tbfx\t\btrue)} = \zeta(\tbfx\t\btrue) + \eta(\tbfx\t\btrue) \Phi^{-1}(\cdot),$ where $\Phi(\cdot)$ is the standard normal distribution function.

For the second setting, the distributional parameter $\mu(\tbfx)$ was sampled as before, while the standard deviation parameter was fixed at $\sigma =  0.1.$
The resulting distributions were then subjected to a random transport map $T$  in Wasserstein space  that is uniformly sampled from the collection of transport maps $T_k(a) = a - \sin (ka)/|k|$ for $k \in \{\pm1, \pm 2, \pm 3\}$. The observed distributions are not Gaussian anymore due to the added random transports Nevertheless, the Fr\'echet mean can be shown to equal $ \zeta(\tbfx\t\btrue) + \sigma \Phi^{-1}(\cdot)$. 

In Table~\ref{table:sim}, $T\#p$ is a push-forward measure such that $T\#p(A) = p(\{x : T(x) \in A \})$,  for any measurable function $T : \real \to \real,$
distribution $p \in \mathcal{W}$, and set $A \subset \real.$ 
Here $p$ is a Gaussian distribution with parameters $\mu$ and $\sigma$ as described  above, and $\mathcal{W}$ is the metric space of distributions on a compact support equipped with the 2-Wasserstein metric.  

\begin{table}[!htb]
	\centering
	\caption{Two different simulation settings for distributional objects.}
	\label{table:sim}
	\centering
	\begin{tabular}{l|l}
		\hline
		Setting I &
		Setting II \\ \hline
		\begin{tabular}[c]{@{}l@{}}$Q(Y)(\cdot) = \mu + \sigma \Phi^{-1}(\cdot) $, \\ where \\ $\mu \sim N(\zeta(\tbfx\t\btrue), 0.25),$ \\ $\sigma \sim Exp\left(\frac{1 + \exp(\tbfx\t\btrue)}{\tbfx\t\btrue}\right).$
		\end{tabular} &
		\begin{tabular}[c]{@{}l@{}}$Q(Y)(\cdot) = T \#(\mu +\sigma \Phi^{-1}(\cdot)) $, \\ where\\ $\mu \sim N(\zeta(\tbfx\t\btrue), 0.25), $\ $\sigma = 0.1,$ \\ $T_k(a) = a - \sin(ka)/|a|,  k \in \{\pm1,\pm 2, \pm 3\}.$\end{tabular} \\ \hline
	\end{tabular}
\end{table}
Following these specifications, for each Monte Carlo run we generated  $n$ density objects and multivariate Euclidean predictors from the true model.
The bias and deviance of the estimated direction vectors for varying sample sizes and resulting from 500 Monte Carlo runs are displayed in Table~\ref{tab:dens:set1_2}. The bias due to the local linear Fr\'echet estimation is generally low and the variance of the estimates is seen to diminish with increasing sample size.

\begin{table}[!htb]
	\centering
	\caption{Two different simulation settings for distributional objects. Bias and deviance (within parenthesis) of $\bhtrue$ (measured in radians as per~\eqref{simul:bias:var}) obtained from $500$ Monte Carlo runs,  where the predictor dimension is $p=4$, and the tuning parameters $(b, M)$ were chosen by  $5-$fold cross-validation.}
	\label{tab:dens:set1_2}
	\centering
	\resizebox{\textwidth}{!}{%
		\begin{tabular}{c|c|c|c|c|c|c|c|c|c|c|c|c}
			\hline
			& \multicolumn{6}{c|}{Setting I} & \multicolumn{6}{c}{Setting II} \\ \hline
			& \multicolumn{2}{c|}{\begin{tabular}[c]{@{}c@{}}link1 \\ $(x \mapsto x)$\end{tabular}} & \multicolumn{2}{c|}{\begin{tabular}[c]{@{}c@{}}link2\\ $(x \mapsto x^2)$\end{tabular}} & \multicolumn{2}{c|}{\begin{tabular}[c]{@{}c@{}}link3\\ $(x \mapsto \exp(x))$\end{tabular}} & \multicolumn{2}{c|}{\begin{tabular}[c]{@{}c@{}}link1\\ $(x \mapsto x)$\end{tabular}} & \multicolumn{2}{c|}{\begin{tabular}[c]{@{}c@{}}link2\\ $(x \mapsto x^2)$\end{tabular}} & \multicolumn{2}{c}{\begin{tabular}[c]{@{}c@{}}link3\\ $(x \mapsto \exp(x))$\end{tabular}} \\ \hline
			& bias & dev & bias & dev & bias & dev & bias & dev & bias & dev & bias & dev \\ \hline
			$n$ = 100 & 0.041 & 0.029 & 0.053 & 0.039 & 0.045 & 0.061 & 0.029 & 0.027 & 0.022 & 0.037 & 0.028 & 0.044 \\ \hline
			$n$ = 1000 & 0.023 & 0.013 & 0.027 & 0.012 & 0.029 & 0.012 & 0.010 & 0.012 & 0.011 & 0.014 & 0.017 & 0.021 \\ \hline
		\end{tabular}%
	}
\end{table}

The performance of the proposed method was further evaluated by computing the mean squared deviation (MSD)  between the observed and the fitted distributions. Denoting the simulated true and estimated distribution objects by $\mop{\Xl \t \btrue}$ and $\hmop{\Xl \t \bhtrue}$, respectively,  for $l=1,\dots,M,$ the utility of the estimation was measured quantitatively by 
\begin{align}
	\label{simu:dens:mse:ifr}
	MSD = \frac{1}{M} \sum_{l=1}^M d^2_W(\mop{\Xl \t \btrue, \btrue}, \hmop{\Xl \t \bhtrue, \bhtrue}), 
\end{align}
where $d_W(\cdot,\cdot)$ is the Wasserstein-2 distance between two distributions.

\begin{figure}[!htb]
	\centering
	\begin{subfigure}[b]{0.48\textwidth}
		\centering
		\hspace*{-.15cm}
		\includegraphics[width=.8\textwidth]{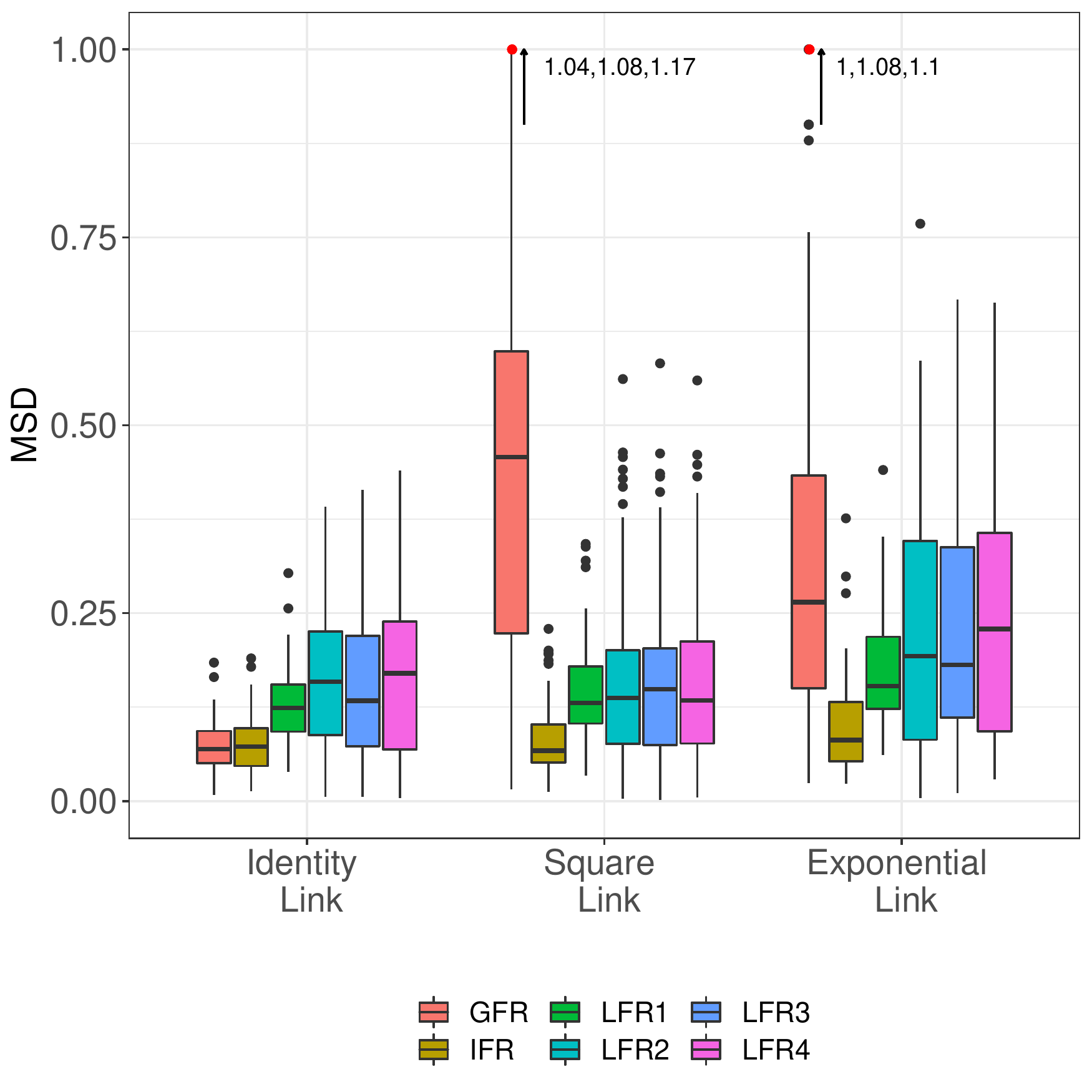}
		\caption{Simulation Setting I.}
		\label{fig:dens_set1:mspe}
	\end{subfigure}
	\hfill
	\begin{subfigure}[b]{0.48\textwidth}
		\centering
		\hspace*{-1.3cm}
		\includegraphics[width=.8\textwidth]{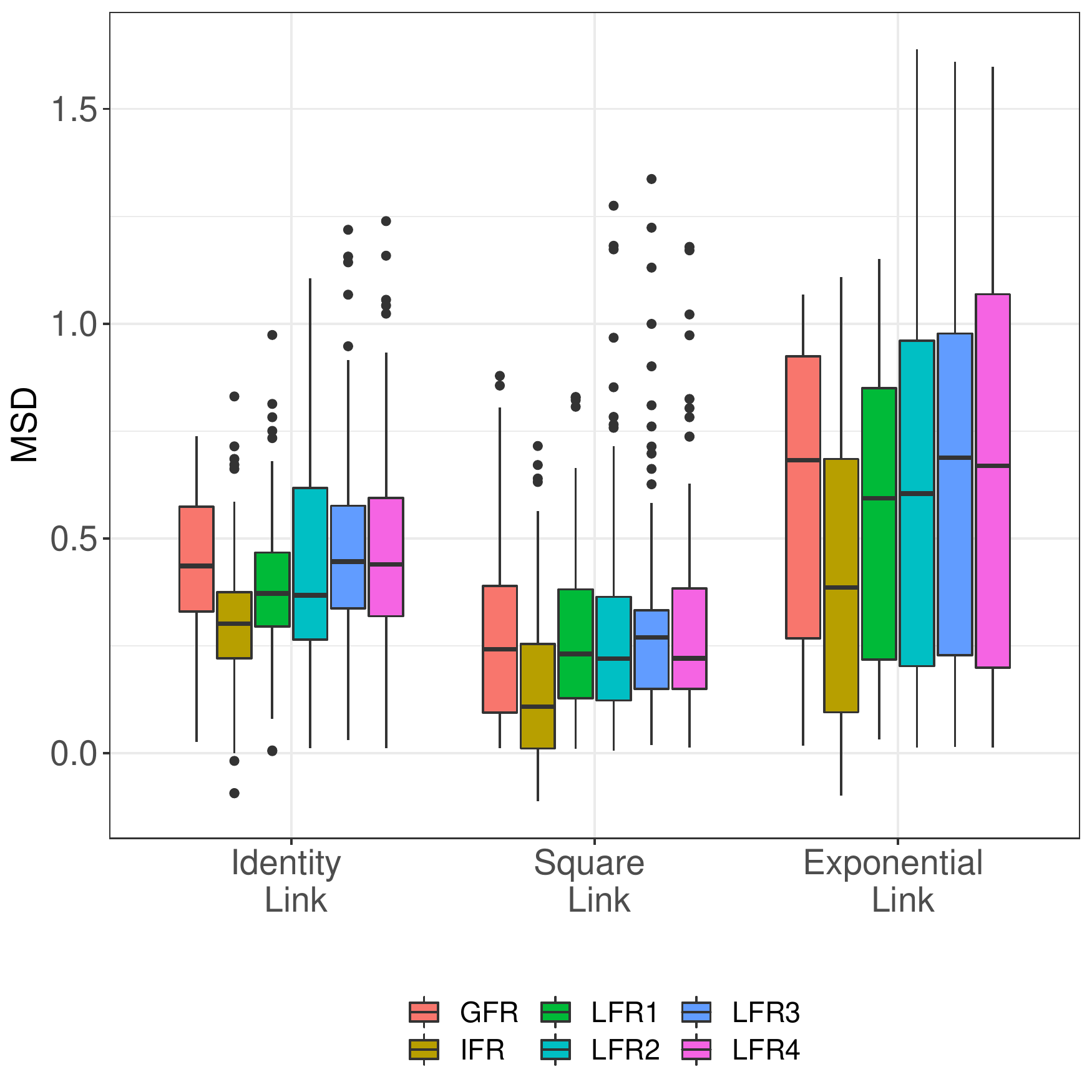}
		\caption{Simulation Setting II.}
		\label{fig:dens_set2:mspe}
	\end{subfigure}
	\centering
	\caption{Boxplot of the mean squared deviation (MSD) of the fits using the single index Fr\'echet regression model (IFR), the Global Fr\'echet regression (GFR) model, and four Local Fr\'echet Regression (LFR) models using the univariate predictor components, for  sample size $n=1000.$ Left and  right panels correspond to  simulation settings I and II, respectively. The left, middle, and right columns in each of the panels correspond to the three different link functions used in the data generation mechanism, namely, identity, square, and exponential link functions, respectively; in all scenarios,  the link functions are estimated from the data.  In the left panel, the outliers having MSD greater than $1$ are marked in red with an upward arrow and the corresponding MSD values are overlaid.}
	\label{fig:dens:mspe}
\end{figure}

We compared the estimation performance of the proposed single index Fr\'echet regression (IFR) method with global Fr\'echet regression (GFR), which directly  handles multivariate predictors as it is a  generalization of global least squares regression \citep{pete:19}. Since  local linear Fr\'echet regression (LFR) is subject to the curse of dimensionality and not suitable for $p=4$ predictors, we fitted four separate LFR models in turn for each of the univariate component predictors and computed the Mean Squared Deviation (MSD) for each of these four fits. No binning is required for either  GFR or LFR model fits. In Figure~\ref{fig:dens:mspe} we denote the MSDs for the four local linear Fr\'echet regression fits as LFR1, LFR2, LFR3, and LFR4, respectively.
Figure~\ref{fig:dens:mspe}, displaying  boxplots of the MSDs over $500$ simulation runs for a sample size of $n=1000.$ The left and right panels correspond to simulation settings I and II, respectively, and in each panel, three cases are considered corresponding to the different link functions used to generate the distributional data. Overall  six Fr\'echet regression methods are compared, for two simulation settings and  three data generation mechanisms.
We observe that the IFR method outperforms the baseline GFR and all four of the LFR methods in all scenarios. The smallest difference  between the IFR and GFR occurs when an identity link function is used in the data generation mechanism. This is as expected since in this case the true model essentially reduces to GFR, the equivalent of a linear model. The individual LFR models have higher MSDs, which  can be attributed to the fact that we are ignoring the effect of the other predictors when fitting  the local model with one predictor at a time.

\begin{figure}[!htb]
	\centering
	\begin{subfigure}[b]{0.32\textwidth}
		\centering
		\includegraphics[width=\textwidth]{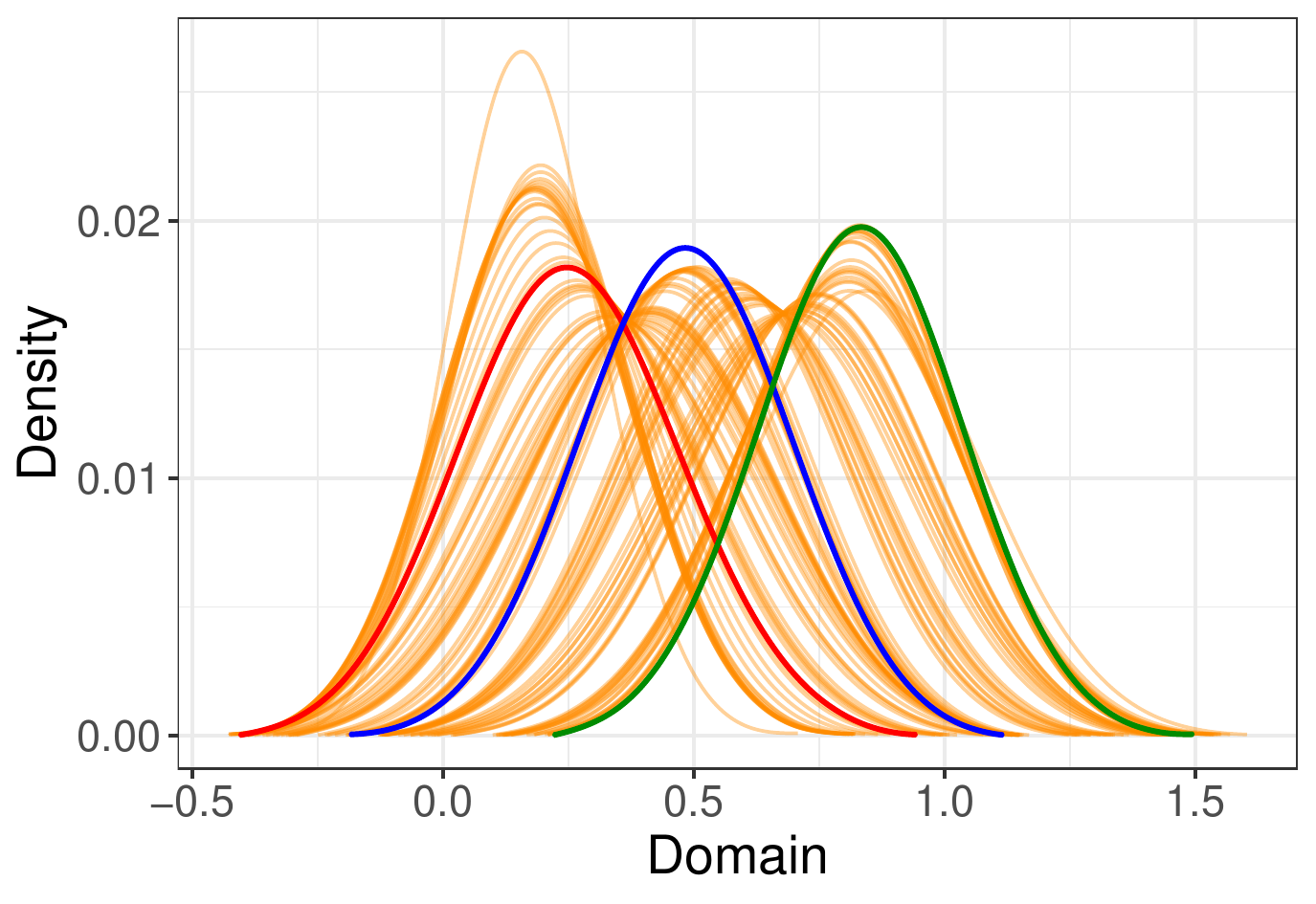}
	\end{subfigure}
	\hfill
	\centering
	\begin{subfigure}[b]{0.32\textwidth}
		\centering
		\includegraphics[width=\textwidth]{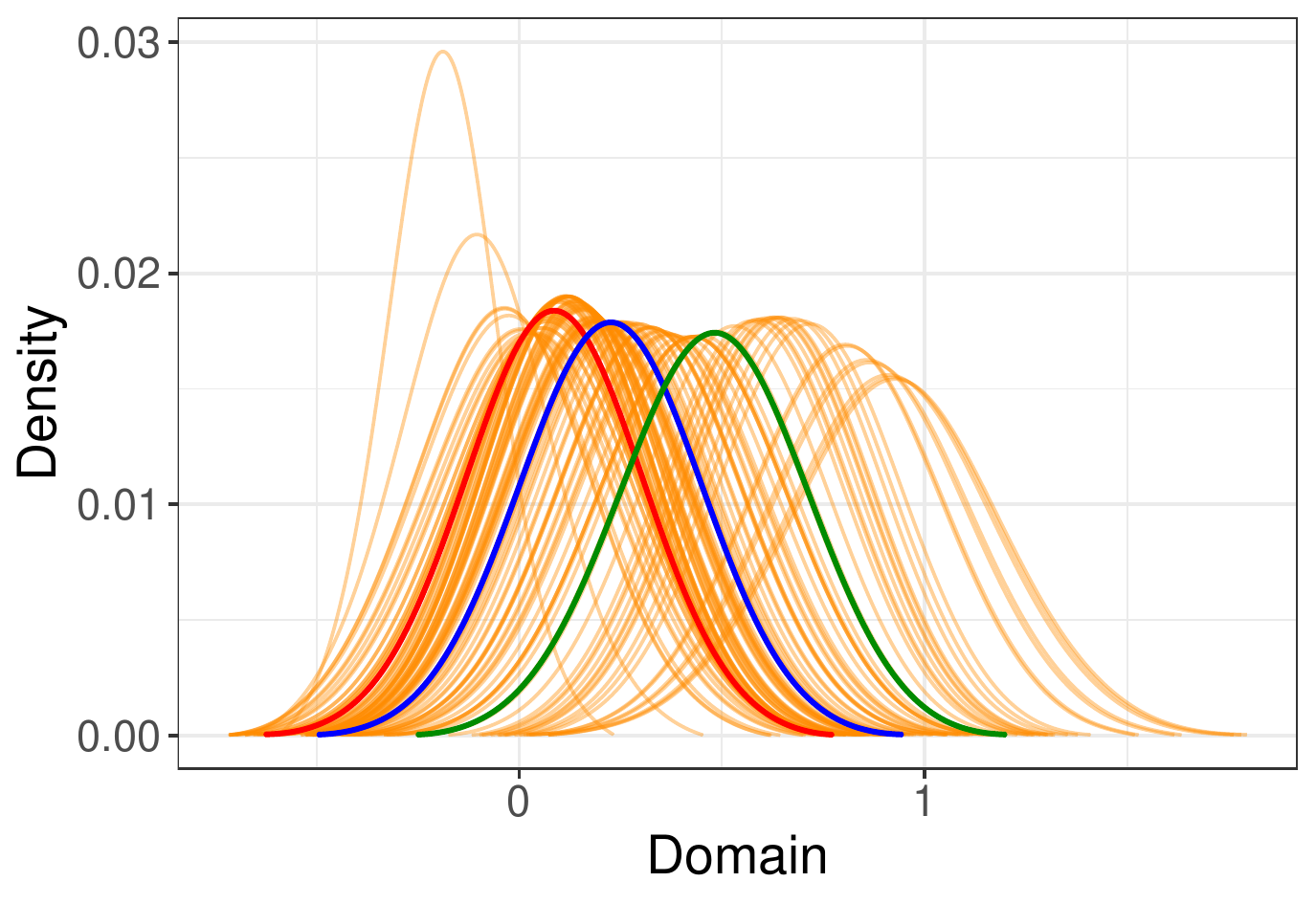}
	\end{subfigure}
	\hfill
	\centering
	\begin{subfigure}[b]{0.32\textwidth}
		\centering
		\includegraphics[width=\textwidth]{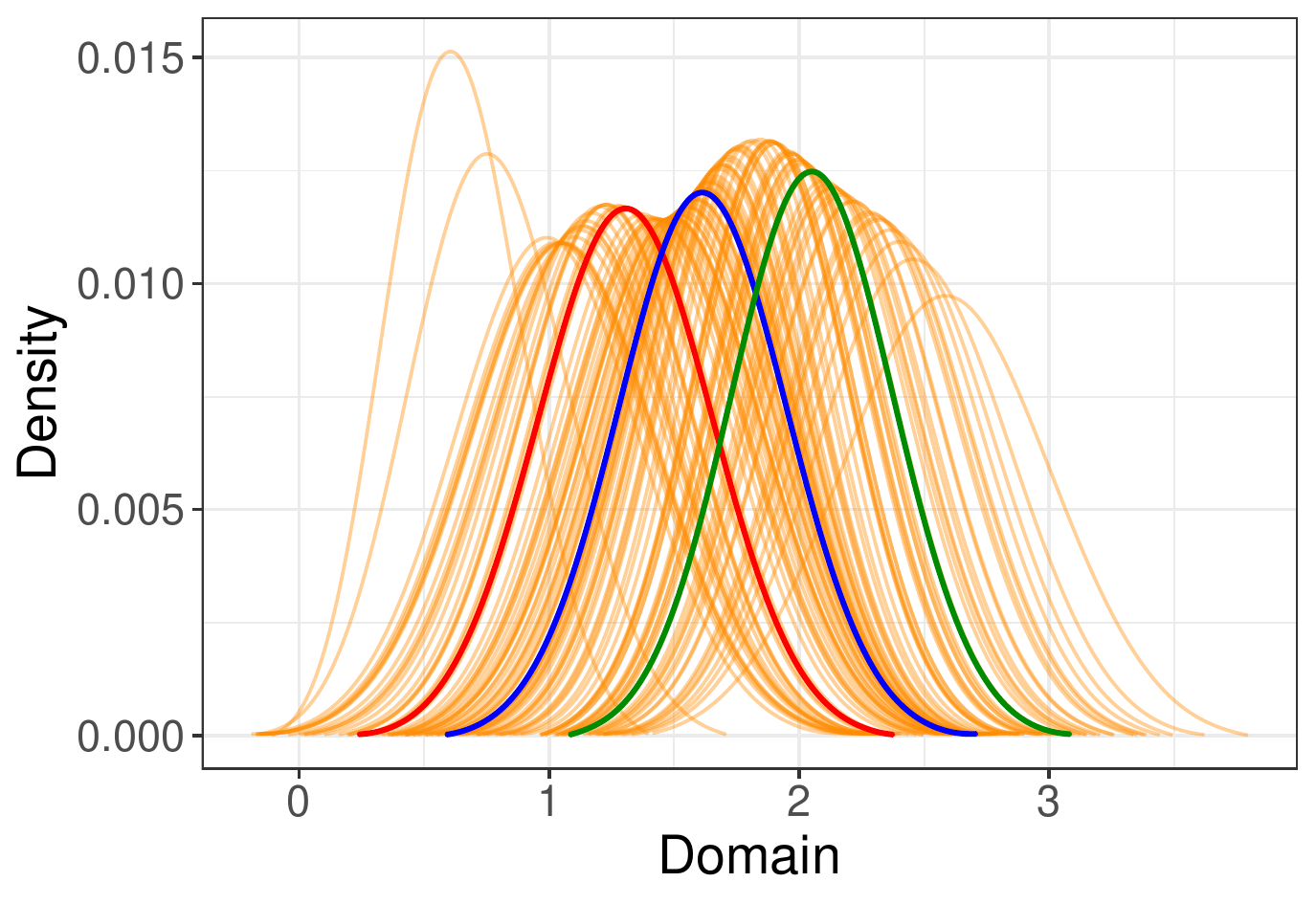}
	\end{subfigure}
	\centering
	\caption{Simulated (orange) and fitted (red, blue, and green) distributional objects represented as densities for simulation setting I for  sample size $n=1000.$ The left, middle, and right panels correspond to three link functions (identity, square, and exponential link)  used in the data generation process. In each case, the IFR model fits are obtained at three different levels of the estimated index values, namely, at $t =$ mean$(\tbfx\t\bhtrue)- 2 \times$ sd($\tbfx\t\bhtrue$) (red), $t =$ mean($\tbfx\t\bhtrue$) (blue) and $t =$ mean$(\tbfx\t\bhtrue)+ 2 \times$ sd($\tbfx\t\bhtrue$) (green).}
	\label{fig:dens:fits}
\end{figure}

Figure~\ref{fig:dens:fits} demonstrates the effect of the index values on the distributional objects  under simulation setting I for the different link functions when responses are represented in the form od densities. 
The three data generation mechanisms are shown in the left, middle, and right panels of Figure~\ref{fig:dens:fits} respectively. For each case, the IFR model was fitted at the mean and mean$\pm2$ sigma levels of the index values, displayed in red, blue, and green lines respectively, while the  observed/simulated densities are overlaid in orange in each panel. In each case, for a higher value of the index level, the fitted densities shift towards the top-right, indicating a positive association of  the single-index values on the mode of the distributions.

To illustrate the out-of-sample prediction performance of the proposed IFR model, the dataset was randomly split into a training set with sample size $n_{\text{train}} = \lfloor 2*n/3 \rfloor$ and a test set with the remaining $n_{\text{test}} = n-\lfloor 2*n/3 \rfloor$ subjects. The IFR method was implemented as follows:  for any given unit direction $\bpara \in \bar{\Theta},$ we partition the domain of the projections into $M$ equal-width non-overlapping bins and consider the representative observations $\Xl$ and $\Yl$ for the data points belonging to the $l-$th bin. The ``true'' index parameter was estimated as $\bhtrue$ as per~\eqref{est:sim:obj}. We then took the fitted index obtained  from the training set and predicted the responses in the test set using the covariates present in the test set. As a measure of the efficacy of the fitted model, we  computed the root mean squared prediction error (RMPE) as
\begin{align}
	\label{rmpe:wass}
	\text{RMPE} &= \left[\frac{1}{M_{n_{\text{test}}}}\sum_{i=1}^{M_{n_{\text{test}}}}  d_W^2\left(\Yl^{\text{test}},  \hmop{\tilde{\mathbf{X}}_l^{\text{test}\intercal}  \bhtrue, \bhtrue} \right)  \right]^{1/2},
\end{align}
where $\Yl^{\text{test}}$ and $\hmop{\tilde{\mathbf{X}}_l^{\text{test}\intercal}  \bhtrue, \bhtrue} $ denote, respectively, the $l^{\text{th}}$ observed and predicted responses in the test set, evaluated at the binned observation $\tilde{\mathbf{X}}_l^{\text{test}}$  and  $d_W$ denotes the Wasserstein-2 metric in~\eqref{wass:metric}. We repeated this process $500$ times and computed RMPE for each split for the subjects separately. The mean and sd of the RMPE  over the repetitions are shown  in Table~\ref{tab:sim:rmpe}. 
The IFR model  is seen to fare best across the different models and scenarios.  

\begin{table}[h!]
	\caption{Mean and sd (in parenthesis) of the RMPE as given in~\eqref{rmpe:wass} comparing the performance of various Fr\'echet regression models: Index Fr\'echet Regression (IFR), Global Fr\'echet Regression (GFR), Local Fr\'echet Regression (LFR). The LFR fits are obtained for four individual predictor components separately.}
	\label{tab:sim:rmpe}
	\begin{tabular}{c|ccc|ccc}
		\hline
		&
		\multicolumn{3}{c|}{Setting I} &
		\multicolumn{3}{c}{Setting II} \\ \hline
		&
		\multicolumn{1}{c|}{\begin{tabular}[c]{@{}c@{}}Identity\\ link\end{tabular}} &
		\multicolumn{1}{c|}{\begin{tabular}[c]{@{}c@{}}Square\\ link\end{tabular}} &
		\begin{tabular}[c]{@{}c@{}}Exponential\\ link\end{tabular} &
		\multicolumn{1}{c|}{\begin{tabular}[c]{@{}c@{}}Identity\\ link\end{tabular}} &
		\multicolumn{1}{c|}{\begin{tabular}[c]{@{}c@{}}Square\\ link\end{tabular}} &
		\begin{tabular}[c]{@{}c@{}}Exponential\\ link\end{tabular} \\ \hline
		IFR &
		\multicolumn{1}{c|}{\begin{tabular}[c]{@{}c@{}}0.0023\\  (0.0012)\end{tabular}} &
		\multicolumn{1}{c|}{\begin{tabular}[c]{@{}c@{}}0.0092 \\ (0.0276)\end{tabular}} &
		\begin{tabular}[c]{@{}c@{}}0.0302\\  (0.0979)\end{tabular} &
		\multicolumn{1}{c|}{\begin{tabular}[c]{@{}c@{}}0.0490 \\ (0.0330)\end{tabular}} &
		\multicolumn{1}{c|}{\begin{tabular}[c]{@{}c@{}}0.1452 \\ (0.0286)\end{tabular}} &
		\begin{tabular}[c]{@{}c@{}}0.1666\\  (0.0988)\end{tabular} \\ \hline
		GFR &
		\multicolumn{1}{c|}{\begin{tabular}[c]{@{}c@{}}0.0136 \\ (0.0002)\end{tabular}} &
		\multicolumn{1}{c|}{\begin{tabular}[c]{@{}c@{}}0.1668\\  (0.0085)\end{tabular}} &
		\begin{tabular}[c]{@{}c@{}}0.1599 \\ (0.0176)\end{tabular} &
		\multicolumn{1}{c|}{\begin{tabular}[c]{@{}c@{}}0.0661 \\ (0.0189)\end{tabular}} &
		\multicolumn{1}{c|}{\begin{tabular}[c]{@{}c@{}}0.2531\\ (0.0095)\end{tabular}} &
		\begin{tabular}[c]{@{}c@{}}0.3413\\  (0.0186)\end{tabular} \\ \hline
		LFR1 &
		\multicolumn{1}{c|}{\begin{tabular}[c]{@{}c@{}}0.0478 \\ (0.0014)\end{tabular}} &
		\multicolumn{1}{c|}{\begin{tabular}[c]{@{}c@{}}0.1671\\  (0.0084)\end{tabular}} &
		\begin{tabular}[c]{@{}c@{}}0.3516 \\ (0.0299)\end{tabular} &
		\multicolumn{1}{c|}{\begin{tabular}[c]{@{}c@{}}0.0679 \\ (0.0191)\end{tabular}} &
		\multicolumn{1}{c|}{\begin{tabular}[c]{@{}c@{}}0.1317 \\ (0.0096)\end{tabular}} &
		\begin{tabular}[c]{@{}c@{}}0.2371\\  (0.0310)\end{tabular} \\ \hline
		LFR2 &
		\multicolumn{1}{c|}{\begin{tabular}[c]{@{}c@{}}0.0479 \\ (0.0015)\end{tabular}} &
		\multicolumn{1}{c|}{\begin{tabular}[c]{@{}c@{}}0.1667\\  (0.0081)\end{tabular}} &
		\begin{tabular}[c]{@{}c@{}}0.3507 \\ (0.0294)\end{tabular} &
		\multicolumn{1}{c|}{\begin{tabular}[c]{@{}c@{}}0.0563 \\ (0.0190)\end{tabular}} &
		\multicolumn{1}{c|}{\begin{tabular}[c]{@{}c@{}}0.1666 \\ (0.0091)\end{tabular}} &
		\begin{tabular}[c]{@{}c@{}}0.1881 \\ (0.0302)\end{tabular} \\ \hline
		LFR3 &
		\multicolumn{1}{c|}{\begin{tabular}[c]{@{}c@{}}0.0476 \\ (0.0020)\end{tabular}} &
		\multicolumn{1}{c|}{\begin{tabular}[c]{@{}c@{}}0.1684\\  (0.0133)\end{tabular}} &
		\begin{tabular}[c]{@{}c@{}}0.3468 \\ (0.0296)\end{tabular} &
		\multicolumn{1}{c|}{\begin{tabular}[c]{@{}c@{}}0.1218 \\ (0.0191)\end{tabular}} &
		\multicolumn{1}{c|}{\begin{tabular}[c]{@{}c@{}}0.1992 \\ (0.0142)\end{tabular}} &
		\begin{tabular}[c]{@{}c@{}}0.1812 \\ (0.0304)\end{tabular} \\ \hline
		LFR4 &
		\multicolumn{1}{c|}{\begin{tabular}[c]{@{}c@{}}0.0454 \\ (0.0062)\end{tabular}} &
		\multicolumn{1}{c|}{\begin{tabular}[c]{@{}c@{}}0.1659\\  (0.0101)\end{tabular}} &
		\begin{tabular}[c]{@{}c@{}}0.3346 \\ (0.0284)\end{tabular} &
		\multicolumn{1}{c|}{\begin{tabular}[c]{@{}c@{}}0.0880 \\ (0.0189)\end{tabular}} &
		\multicolumn{1}{c|}{\begin{tabular}[c]{@{}c@{}}0.2177 \\ (0.0110)\end{tabular}} &
		\begin{tabular}[c]{@{}c@{}}0.2033 \\ (0.0293)\end{tabular} \\ \hline
	\end{tabular}
\end{table}

\begin{figure}[!htb]
	\centering
	\begin{subfigure}[b]{0.48\textwidth}
		\centering
		\hspace*{-.15cm}
		\includegraphics[width=.8\textwidth]{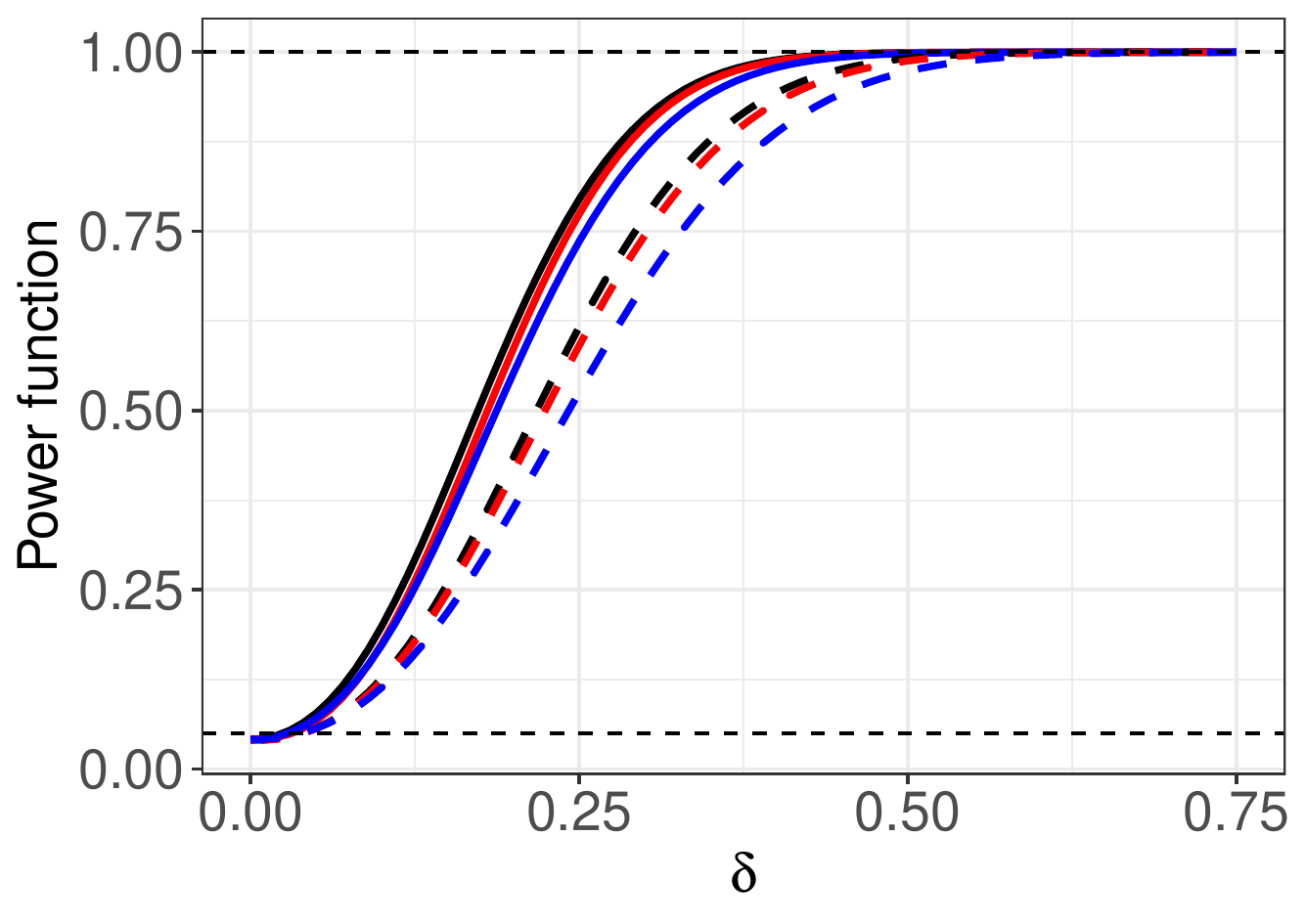}
		\caption{Simulation Setting I.}
		\label{fig:dens_set1:pow}
	\end{subfigure}
	\hfill
	\begin{subfigure}[b]{0.48\textwidth}
		\centering
		\hspace*{-1.3cm}
		\includegraphics[width=.8\textwidth]{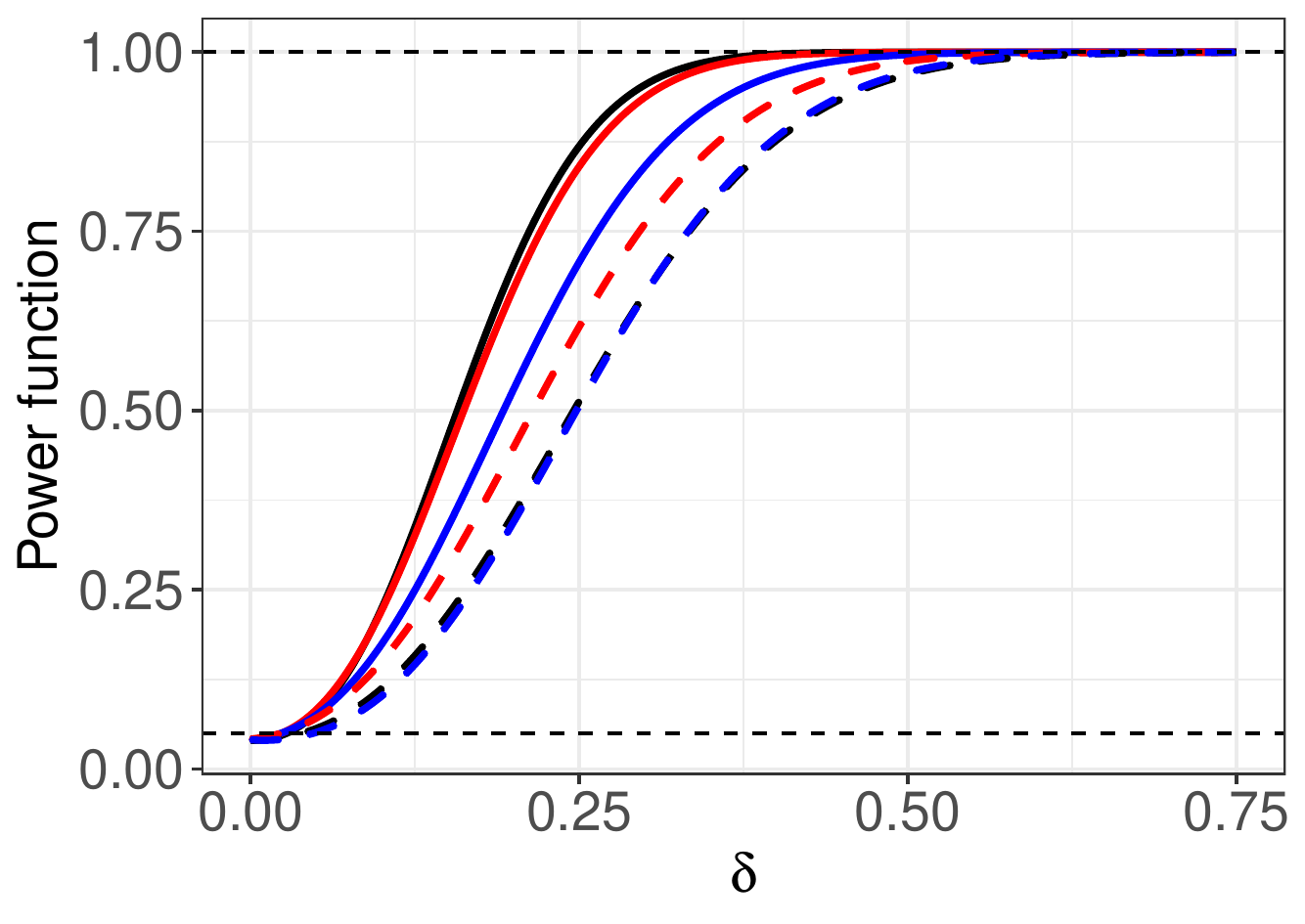}
		\caption{Simulation Setting II.}
		\label{fig:dens_set2:pow}
	\end{subfigure}
	\caption{Empirical power as function of $\delta$ for density object responses. The black, red, and blue curves correspond to the identity, square, and exponential link functions used in the data-generating mechanism, respectively, while the dashed and solid lines correspond to sample sizes $n =100$ and $n = 1000$ respectively. The level of the tests is $\alpha= 0.05$ and is indicated by the dashed line parallel to the x-axis.}
	\label{fig:dens:pow}
\end{figure}

For the case of distributional objects, the linear hypothesis test of $H_0$ in \eqref{hyp:null} against the sequence of alternatives $H_{1\delta}$  in \eqref{hyp:alt} was also carried out. The power functions corresponding to the two simulation settings are shown in Figure~\ref{fig:dens_set1:pow} and ~\ref{fig:dens_set2:pow}, respectively. 
As  $\delta$ increases, the power  is seen to increase rapidly. This shows that the proposed test has non-trivial power (see Figure~\ref{fig:dens:pow}). When $\delta$ is close to $0,$ the test sizes are approximately equal to the nominal significance level of $\alpha=0.05.$  
As expected, power increases with increasing sample size, most notably  under the identity link. In the second simulation setting  when the distributional objects are obtained by transporting a normal distribution, the power function increases at a slower rate, especially under the highly nonlinear (exponential) link function. 

\subsection{Adjacency Matrices as  Responses}
These were generated for weighted graphs as random object responses;  details are in Subsection~\ref{suppl:sec:adj} 
of the Supplement.

\subsection{Euclidean Responses} We applied the new approach targeting general random objects as responses for  the special case of Euclidean responses. It is not specifically designed for this case,  where targeted, well-studied and well-honed 
single index models have a long history. The   numerical results show that the proposed method yields results that are somewhat inferior but overall still comparable to those obtained with specially tailored traditional single index approaches;  see Subsection S.4.4 
of the Supplement.

\section{Data analysis} \label{sec:applications}
\subsection{Resting state functional Magnetic Resonance Imaging: ADNI data}
\label{sec:data:ADNI}
\hfill\\
Resting-state functional Magnetic Resonance Imaging (fMRI) methodology makes it possible to study brain activation and to identify brain regions or cortical hubs that  exhibit similar activity  when subjects are in the resting state \citep{alle:14, ferr:13}.
In resting state fMRI, time series of Blood Oxygen Level Dependent (BOLD) signals are observed in regions of interest (ROI),  where each ROI is represented by the signal of a  seed voxel, which is the voxel in an ROI that has the highest correlation with the signals of nearby voxels. Alzheimer's Disease has been found to be associated  with anomalies in the functional integration of ROIs \citep{damo:12,zhan:10c}.  

Data used in the preparation of this article were obtained from the Alzheimer's Disease Neuroimaging Initiative (ADNI) database (\url{adni.loni.usc.edu}).
BOLD signals for $V= 11$ brain seed voxels for each subject were extracted for the following ROIs: MPFC (Anterior medial prefrontal cortex), PCC (Posterior cingulate cortex), dMFPC (Dorsal medial prefrontal cortex), TPJ (Temporal parietal junction), LTC (Lateral temporal cortex), TempP (Temporal pole), vMFPC (Ventral medial prefrontal cortex), pIPL (Posterior inferior parietal lobule), Rsp (Retrosplenial cortex), PHC (Parahippocampal cortex) and HF$^+$ (Hippocampal formation)  \citep{andr:10}. The pre-processing of the BOLD signals was implemented by adopting standard procedures of slice-timing correction, head motion correction and other standard  steps. The signals for each subject were recorded over the interval $[0, 270]$ (in seconds), with $K=136$ measurements available at two-second intervals. From this the temporal correlations were computed to construct the connectivity correlation matrix, also referred to as  the Pearson correlation matrix in the neuroimaging community. 

The data set in our analysis consists of $n=830$ subjects at  four stages of the disease: $372$ CN (cognitively normal), $113$ EMCI (early mild cognitive impairment), $200$ LMCI (late mild cognitive impairment), and $145$ AD (Alzheimer's)  subjects.  The inter-hub connectivity Pearson correlation matrix for the $i-th$ subject  $Y_{i}$ with elements 
\begin{align}
	(Y_{i})_{qr} = \frac{\sum_{p=1}^{K}(s_{ipq} - \bar{s}_{iq}) (s_{ipr} - \bar{s}_{ir})}{\left[\left(\sum_{p=1}^{K}(s_{ipq} - \bar{s}_{iq})^2\right) \left(\sum_{p=1}^{K}(s_{ipq} - \bar{s}_{iq})^2 \right) \right]^{1/2}}, \,\,q,r = 1,\dots,11	\label{data:ADNI:pearson:corr}
\end{align}
is the response object for each subject, where 
$s_{ipq}$ is the $(p,q)^{\text{th}}$ element of the signal matrix for the $i^{\text{th}}$ subject and $\bar{s}_{iq} := \frac{1}{K}\sum_{p=1}^{K} s_{ipq}$ is the mean signal strength for the $q^{\text{th}}$ voxel. For Alzheimer's disease studies, the ADAS-Cog-13 score (henceforth referred to as C score)  is a widely-used measure of cognitive performance. It quantifies  impairments across cognitive domains that are affected by  Alzheimer's disease \citep{kuep:18}; higher scores indicate more serious cognitive deficiency.  

We considered $p =10$ predictors, namely, $X_{1}=$ stage for the disease (coded as 0-3, indicating Cognitive normal (CN), Early and Late Mild cognitive impairment (EMCI and LMCI), or Alzheimer's Disease (AD), respectively), $X_2 =$ age of the subject (in years),  $X_3 =0$ is the subject is female and $=1$ if male), $X_4 =$C score for the subject at the time of the first scan, and additionally all  pairwise interaction terms between the above predictors, i.e., the products $X_jX_k, \, j \ne k, 1 \le j,k \le 4$ 

In a first step, we test the null hypothesis of no regression effect, i.e.,  with $p=5$, 
$$H_0: \true= \mathbf{0}_{(p-1) \times 1} \ vs. \ H_1: \text{ not all } \theta_{0j} \text{ are }0,  \ j =2,\dots,p, $$ 
where $\btrue = (\theta_{01},\true)\t$ and $\true = (\theta_{02},\dots,\theta_{0p})\t$ with $\theta_{01} = \sqrt{1- \ltwoNorm{\true}^2}.$ The null model has $X_1$ included with $\theta_{01} =1$ since it is known that the stage of cognitive impairment has an effect on brain connectivity/ We obtain 
an estimate of the $(p-1)-$ dimensional vector $\htrue$ as the minimizer of $\Vn(\para)$ as per~\eqref{para:obj} and  $\hat{\theta}_{01} = \sqrt{1 - \ltwoNorm{\htrue}^2}.$
Under the null hypothesis,  $\tilde{T}_n = \htrue \t (\wh{\Lambda}^\ast_B)^{-1} \htrue \overset{approx.}{\sim} \chi^2_{(p-1)}.$ We find that  $\tilde{T}_n = 23.81$,   corresponding to a  $p$ value of $p = 0.0046 <0.005,$ providing evidence that there is indeed a  regression relationship.  
\begin{table}[!htb]
	\caption{Details on step-wise model selection.}
	\label{tab:data:step_reg}
	\begin{center}
		\begin{tabular}{l|c|c|c|c|c|c}
			\hline
			& \multicolumn{2}{c|}{Step 1} & \multicolumn{2}{c|}{Step 2} & \multicolumn{2}{c}{Step 3} \\ \hline
			& Coeff.       & p-value      & Coeff.       & p-value      & Coeff        & p-value      \\ \hline
			Age         & -0.364       & 0.005        & -0.394       & -            & -0.401       & -            \\ \hline
			Gender      & 0.198       & 0.122        & 0.558        & 0.161        &0.173          & 0.113        \\ \hline
			C Score & 0.371         & 0.094        & 0.207        & 0.010        &0.279        & -            \\ \hline
		\end{tabular}
	\end{center}
\end{table}
We also implemented  sequential predictor selection, where we  specified an ``alpha-to-enter''  level  $\alpha = 0.05$ and considered $X_1$ to be in the model  and included each of $X_2,$ $X_3,$ and $X_4$ in the model separately along with $X_1$ then testing the null hypotheses  $\theta_j=0, \ j=2,3,4$ separately. 
Table~\ref{tab:data:step_reg} illustrates the resulting step-wise model  selection.

For example, for testing $\theta_2 = 0$,  we first obtained $\hat{\theta}_2=-0.364$,  $\hat{\theta}_1 = \sqrt{1- (-.364)^2} = 0.931$)   and
$\tilde{T}_n = 7.88$ with a $p$-value of $0.005.$
Thus   $X_2$ (age) was added to the model in step 1, followed by adding $X_4$ (C score) in step 2,  
while  $X_3$ (gender) was not significant.
With $X_1,$ $X_2,$ and $X_4$ in the model, we tested for the significance of the pairwise interaction terms. The null hypothesis for this test  is $H_0 : \theta_5 =\theta_6 = \dots = \theta_{10} =0.$ The p-value was  $0.106,$ and  we did not include interactions in the final model.  
The estimated average Fr\'echet error $\frac{1}{n} \sum_{i=1}^n d^2(Y_i,\hmop{X_{1i}\hat{\theta}_1 + X_{2i}\hat{\theta}_2 + X_{4i}\hat{\theta}_4})$ was
quite small $(0.239).$ 

\begin{figure}[!htb]
	\centering
	\begin{subfigure}[b]{0.32\textwidth}
		\centering
		\includegraphics[width=\textwidth]{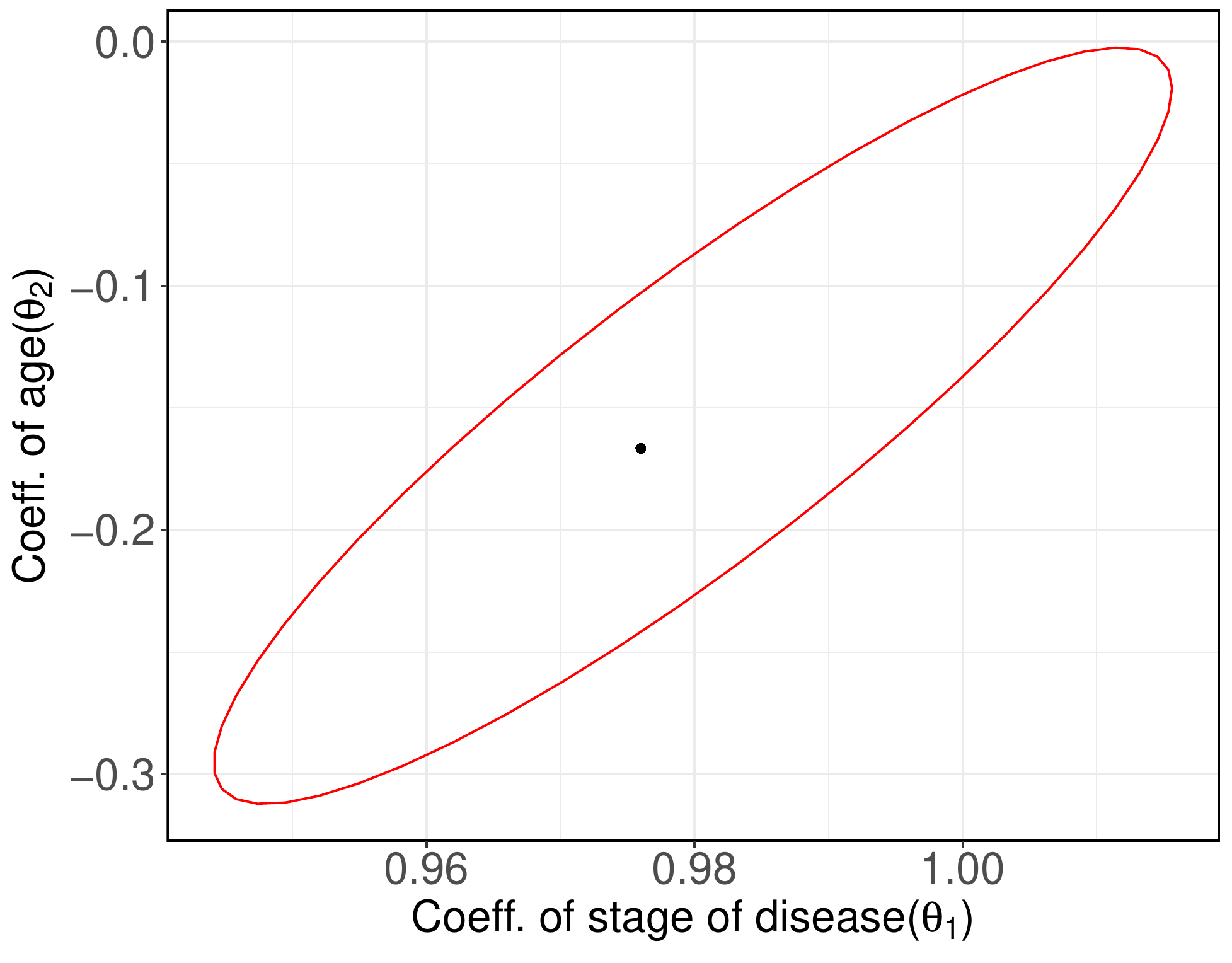}
		\caption{Confidence ellipse for $(\theta_1, \theta_2).$}
		\label{fig:cr:a}
	\end{subfigure}
	\hfill
	\begin{subfigure}[b]{0.32\textwidth}
		\centering
		\includegraphics[width=\textwidth]{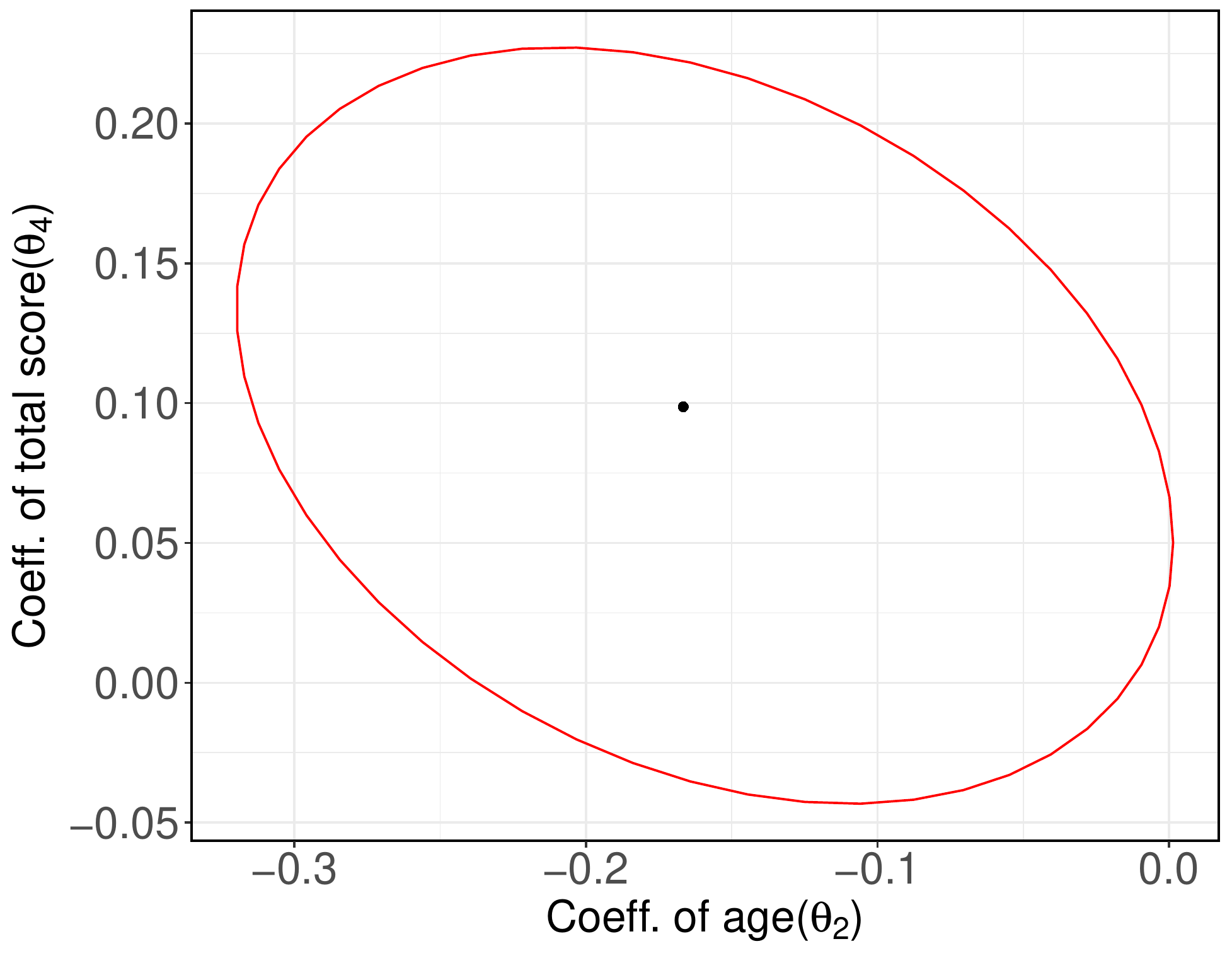}
		\caption{Confidence ellipse for $(\theta_2, \theta_4).$}
		\label{fig:cr:b}
	\end{subfigure}
	\hfill
	\begin{subfigure}[b]{0.32\textwidth}
		\centering
		\includegraphics[width=\textwidth]{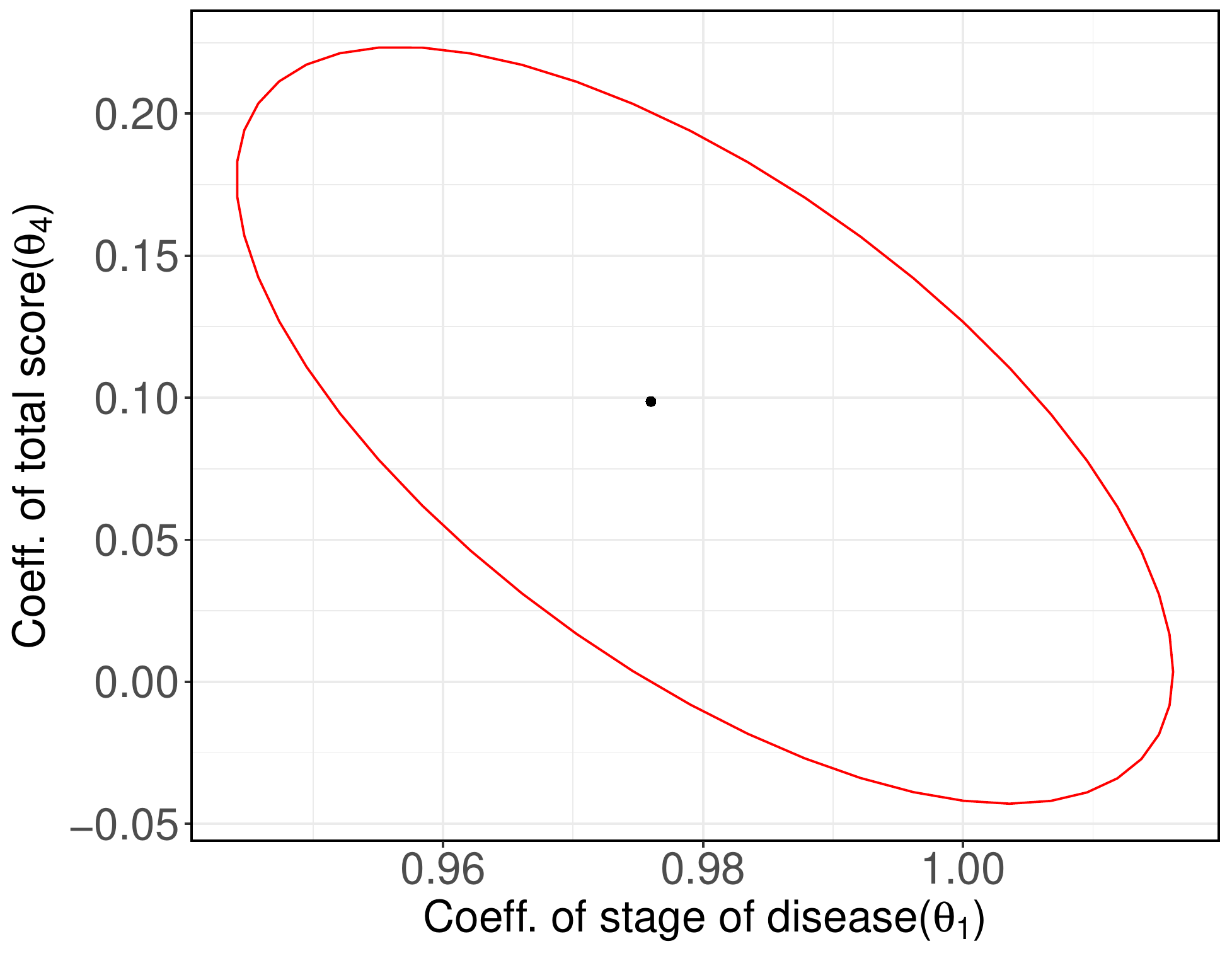}
		\caption{Confidence ellipse for $(\theta_1, \theta_4).$}
		\label{fig:cr:c}
	\end{subfigure}
	\caption{The $95\%$ confidence ellipses for pairs of  coefficients for predictors stage of the disease ($X_1$), age ($X_2$), and C score  ($X_4$).}
	\label{fig:ADNI:CR}
\end{figure}

To construct the confidence regions for the coefficients $(\theta_1,\theta_2,\theta_4)$, we implemented the local linear Fr\'echet regression with the Epanechnikov kernel and used 5-fold cross-validation to select the bandwidth $b$. Using the bootstrap method to obtain the estimated covariance matrix of the limiting distribution we obtained the $95\%$ pairwise confidence ellipses for the coefficients $(\theta_1,\theta_2,\theta_4)$ of the predictors- disease stage, age, and C score, which are displayed in Figure~\ref{fig:ADNI:CR}. 
We observe that none of the pairwise confidence ellipses includes the origin and therefore the p-values are $<0.05,$ implying the significance of the predictors.

\begin{figure}[!htb]
	\centering
	\includegraphics[width=.71\textwidth]{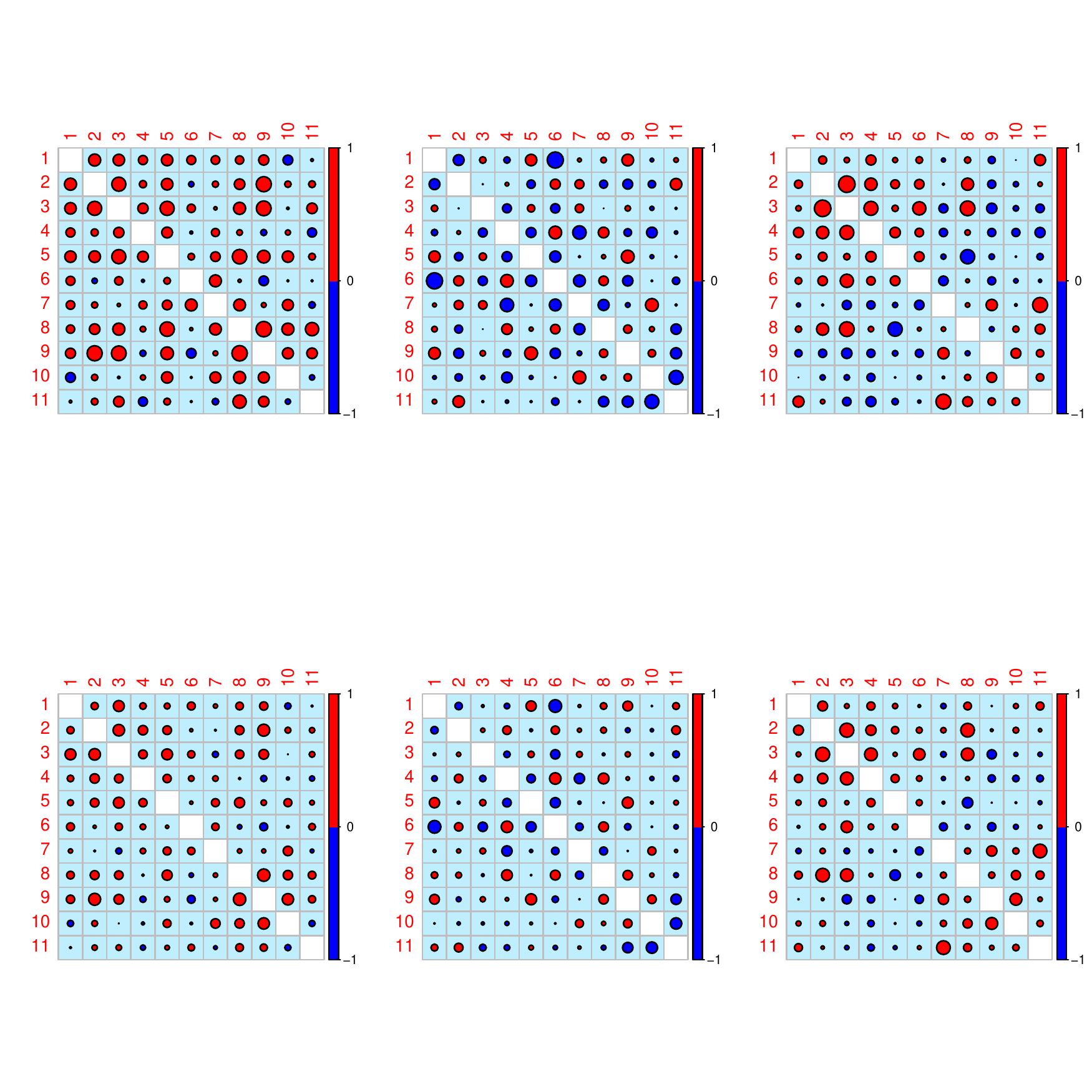}
	\centering
	\caption{Observed and fitted functional connectivity correlation matrices for  different  values of the single index. 
		The panels in the top row, from left to right, depict the observed functional connectivity correlation matrices for those subjects for whom the estimated index values are  closest to the $25\%, 50\%,$ and $75\%$ quantile of all indices across subjects,  respectively. The bottom row shows the fitted functional connectivity correlation matrices for the same subjects, 
		(from left to right). Positive (negative) values for correlations are drawn in red (blue), where larger circles correspond to larger absolute values.}
	\label{fig:ADNI:obs_vs_fits}
\end{figure}

To illustrate the effect of the single index on the response, we computed the estimated index  of  the fitted  model for each subject 
and then obtained the $25\%, 50\%,$ and $75\%$ quantiles across all subjects, with values   $q_1 = 15.048,$ $q_2 = 16.430$ and $q_3 = 18.250,$ respectively. The values of  the four covariates for the subjects with estimated  index values closest to $q_1,$ $q_2,$ and $q_3$  are  in Table~\ref{tab:ADNI:fits}, and their  observed and fitted functional connectivity correlation matrices are illustrated in 
Figure~\ref{fig:ADNI:obs_vs_fits}. The fitted correlation matrices correspond to the values of the estimated object link function at the three index values and are contrasted with the observed correlation matrices for the three subjects.   This gives an  idea of how the fitted correlation matrix changes as the index move through the three quantile levels.  
\begin{table}[!htb]
	\centering
	\caption{Covariate values  for the subjects with estimated index values closest to the first three quantiles of the estimated index when considered across all subjects,   $q_1 (15.048),$ $q_2 (16.430),$ and $q_3 (18.250),$ respectively. Subject $726$ has an estimated index value that is closest to $q_1$, subject 695 closest to $q_2$, and subject 556 closest to $q_3.$}
	\label{tab:ADNI:fits}
	\centering
	\begin{tabular}{c|c|c|c|c|c}
		\hline
		\begin{tabular}[c]{@{}c@{}}Subject\\ number\end{tabular} &
		\begin{tabular}[c]{@{}c@{}}Estd.\\ index value\end{tabular} &
		\begin{tabular}[c]{@{}c@{}}Stage of the\\ disease\end{tabular} &
		Age &
		Gender &
		C score \\ \hline
		726 & 15.045 & 2 & 66.10 y & M & 20.33 \\ \hline
		695 & 16.430 & 2 & 78.12 y & M & 14    \\ \hline
		556 & 18.252 & 1 & 72.55 y & M & 51.67 \\ \hline
	\end{tabular}
\end{table}

We observe that the fits match the general pattern of the observed matrices quite well. The Frobenius distances between the observed and the estimated matrices at $q_1,$ $q_2,$ and $q_3$ are  $1.68,$ $1.10,$ and $0.79,$ respectively. The fitted model reflects the trends seen in the observed correlation matrices and  illustrates the nonlinear  dependence of functional connectivity on the index value.  

We also studied  the out-of-sample prediction performance of the proposed IFR model,  
for which we used the root mean squared prediction error 
\begin{align}
	\text{RMPE} = \left[\frac{1}{M_{n_{\text{test}}}}\sum_{i=1}^{M_{n_{\text{test}}}}  d_F^2\left(\Yl^{\text{test}},  \hmop{\Xl \t \bhtrue, \bhtrue} \right)  \right]^{1/2},
\end{align}
where $\Yl^{\text{test}}$ and $\hmop{\Xl \t \bhtrue}$ denote, respectively, the $l^{\text{th}}$ observed and predicted responses in the test set, evaluated at the binned observation $\Xl.$  Here, $n_{\text{train}}$ and $n_{\text{test}}$ denote the sample sizes of the training and testing sets formed by randomly splitting the data. We repeated this process $200$ times, and computed RMPE for each split for the subjects separately. 
The tuning parameters $(b,M)$ were chosen by a $5-$fold cross-validation method for each replication of the process.
The prediction performance of the IFR model was compared with other applicable Fr\'echet regression models, namely, the global Fr\'echet regression (GFR) model with the three-dimensional predictor $(X1, X_2, X_4)$ and two separate local linear Fr\'echet regression (LFR) models, one with the single predictor $X_2$ (age) and the other with the single predictor $X_4$ (C score). 
When comparing the performance of these models 
(Table~\ref{tab:ADNI:rmpe}), we find 
\begin{table}[!htb]
	\centering
	\caption{Mean and sd (in parenthesis) of the root mean prediction error (RMPE) over $200$ Monte Carlo simulation runs for various object regression methods.
		The methods compared are index Fr\'echet regression (IFR);  global Fr\'echet Regression (GFR) with the three  predictors stage of the disease, age, and ADSA score; and two local linear Fr\'echet regression (LFR) models with separate  one-dimensional predictors.}
\label{tab:ADNI:rmpe}
\begin{tabular}{c|c|c|c}
	\hline
	IFR            & GFR            & \begin{tabular}[c]{@{}c@{}}LFR1\\ (Predictor Age)\end{tabular} & \begin{tabular}[c]{@{}c@{}}LFR2\\ (Predictor C  Score)\end{tabular} \\ \hline
	0.3066 (0.012) & 0.5083 (0.011) & 0.5076 (0.012)                                          & 0.5326 (0.013)                                                  \\ \hline
\end{tabular}
\end{table}
that the out-of-sample prediction error is low for the IFR model, as compared to the global and local Fr\'echet regression approaches.  In fact, it is not far from  the in-sample prediction error $(0.251)$, calculated as the average distance between the observed training sample and the predicted objects based on the covariates in the training sets. This motivates  the proposed IFR models.

\subsection{Human mortality data: Age-at-death distributions as responses}
Lifetables reflecting human mortality across 40 countries correspond to  distributional responses,  coupled with various country-specific covariates. We implement an  overall test for the regression effect for these data. Details about this analysis  are in the Supplement, subsection S.4.1.  

\subsection{Emotional well-being of unemployed workers: Compositional data as responses.}
We further demonstrate the proposed IFR method for the analysis of mood compositional data. Here the object-valued responses lie on a manifold (sphere) with positive curvature. Thus the sufficient (but not necessary) condition for assumption~\ref{ass:convexity:d_m} that the underlying metric space behaves like a CAT(0) space is not satisfied, however, the numerical performance of the IFR method remains quite  good; see Supplement, subsection S.4.2. 
This  suggests a certain degree of model   robustness.

\section{Discussion}
\label{sec:disc}
Binning the data to reduce the effective sample size is not necessary for the basic consistency results without rates. As discussed at the end of Section~\ref{sec:model:methods}, the binning method is introduced in order to invoke the uniform consistency rate for the local Fr\'echet regression and the effective sample size $M = M(n)$ is tied to this rate by virtue of assumption~\ref{ass:tuning:M}. To avoid confusion, we discuss the binning approach throughout.
The rate of convergence for $\bhtrue -\btrue$ is  $M^{-1/2}$. Since our rate results and proofs rely on the uniform convergence rate of local Fr\'echet regression, this rate cannot be improved within the current  framework  and overcoming these limits would require a fundamentally different approach.  

The assumptions required to obtain the technical results are 
essentially the same as those used before in the Fr\'echet regression literature, specifically in~\cite{chen:22}. We require curvature and entropy conditions to hold uniformly across all index values and direction parameters. The curvature and entropy conditions can be verified for commonly observed objects such as univariate probability distributions, positive definite matrices, or data on the surface of a sphere, as well as other random objects under suitable metrics.  The Lipschitz condition~\ref{ass:reg:cont} on the link function is standard in single-index models, while assumption~\ref{ass:convexity:d_m} reflects the interplay between the properties of the metric and the link function. 
Assumption~\ref{ass:convexity:d_m} is implied by the easier-to-interpret assumption~\ref{ass:CAT0}-\ref{ass:reg:cont2} (see Appendix~\ref{append:curvature:assump}).

The classical single index model for Euclidean responses has been recently extended to a single index coefficient model for  quantile regression~\citep{zhao:17}. This is a desirable extension for the object case of index Fr\'echet regression as well. One problem to resolve in this case is to define quantiles in the metric space where the object responses lie since there is no order. The problem of defining quantiles is already difficult and ambiguous for multivariate Euclidean objects. This is  a potentially interesting topic for future research.

Finally,  inference results for object regression are scarce. For example, the  Wasserstein $F$-tests proposed by~\cite{pete:21} are exclusively aimed at  univariate distribution quantiles within the specific setting of global Fr\'echet regression. We provide here a general framework to obtain inference for the case of vector predictors coupled with object responses, which  includes generalized versions of inference for model comparisons and  for assessing the  significance of individual predictors.  

\Appendix

\section{Geodesics and curvature} 
\label{append:curvature}
The length of a curve $\phi : [0,1] \to \Omega$ connecting two distinct points $x,y \in \Omega$ can be measured by taking partitions $P =\{t_{0}\leq t_{1}\leq\cdots\leq t_{k}\} \subset [0,1]$ and finding the supremum polygonal length 
\[
|\phi| := \sup_{P\in \mathcal{P}} \sum_{j=1}^{k}d(\phi(t_{j}),\phi(t_{j-1})),
\]
where $\mathcal{P}$ is any collection of subsets of $[0,1]$ with finite cardinality.
The metric space $(\Omega,d)$ is a length space
if $d(x,y)=\inf_{\phi}|\phi|$, where the infimum ranges over
all curves $\phi:[0,1] \to \Omega$ connecting two distinct points $x$ and $y$, that is, 
i.e., $\phi(0)=x$ and $\phi(1)=y$. 
A geodesic on $\Omega$ connecting two distinct points $x$ and $y$ is the shortest path connecting the two points. Geodesics in a metric space are analogous to straight lines in a Euclidean space.

Unlike Euclidean spaces, a general metric space may  not be flat, and curvature is used to measure the amount of deviation from being flat. The curvature of a given geodesic space is classified by comparing the geodesic triangles
on the metric space to those on the corresponding reference spaces $M_{\kappa}^{2}.$ When $\kappa=0$, $M_{\kappa}^{2}=\real^{2}$ with the standard Euclidean distance $d_E(x,y) = ||x-y||_E,$ for any $x,y \in \real^2.$
A geodesic triangle with vertices $p,q,r$ in a geodesic space  $\Omega$, denoted by $\triangle(p,q,r)$, consists of three geodesic segments that connect
$p$ to $q$, $p$ to $r$ and $q$ to $r$, respectively. A comparison triangle $\triangle(\bar p,\bar q, \bar r)$ in the reference space $M_k^2 =\real^2$ is a geodesic triangle in $\real^2$ formed by the vertices $\bar{p},$ $\bar{q},$ and $\bar{r}$ such that,
\begin{align}
\label{geodesic:conseq}
d(p,q) = ||\bar{p} - \bar{q}||_E,\ 
d(q,r) = ||\bar{q} - \bar{r}||_E,\
d(p,r) = ||\bar{p} - \bar{r}||_E.
\end{align}
$\Omega$ is said to have a non-positive curvature if there exists a comparison triangle $\triangle(\bar p,\bar q,\bar r)$ in the reference space $\real^2$ such that $d(x,y)\leq ||\bar x -\bar y||_E$ for all $x\in \overline{pq}$ and $y\in \overline{pr}$ and their comparison points $\bar x$ and $\bar y$ on $\triangle(\bar p,\bar q,\bar r)$.
A geodesic space with curvature upper bounded by $0,$ in which every geodesic triangle $\triangle(p,q,r)$  satisfies the following $\mathrm{CAT}(0)$ inequality is  a CAT(0) space, 
\begin{align}
\label{CAT0:ineq}
d(x,y)\leq ||\bar{x}-\bar{y}||_E \text{ for all } x\in\overline{pq} \text{ and } y\in\overline{pr} \text{ and their comparison points } \bar x,\bar y\in\real^2.
\end{align}
Every CAT(0) space is uniquely geodesic. Examples of CAT(0) spaces include Euclidean space, the space of symmetric positive definite matrices, Wasserstein-2 spaces, or phylogenetic tree spaces. For a  detailed introduction  to  metric geometry, we refer to  \cite{bura:01}. A compilation of the most relevant facts can be found  in \cite{lin:19b}.
\begin{figure}
\centering
\includegraphics[width =.8\textwidth]{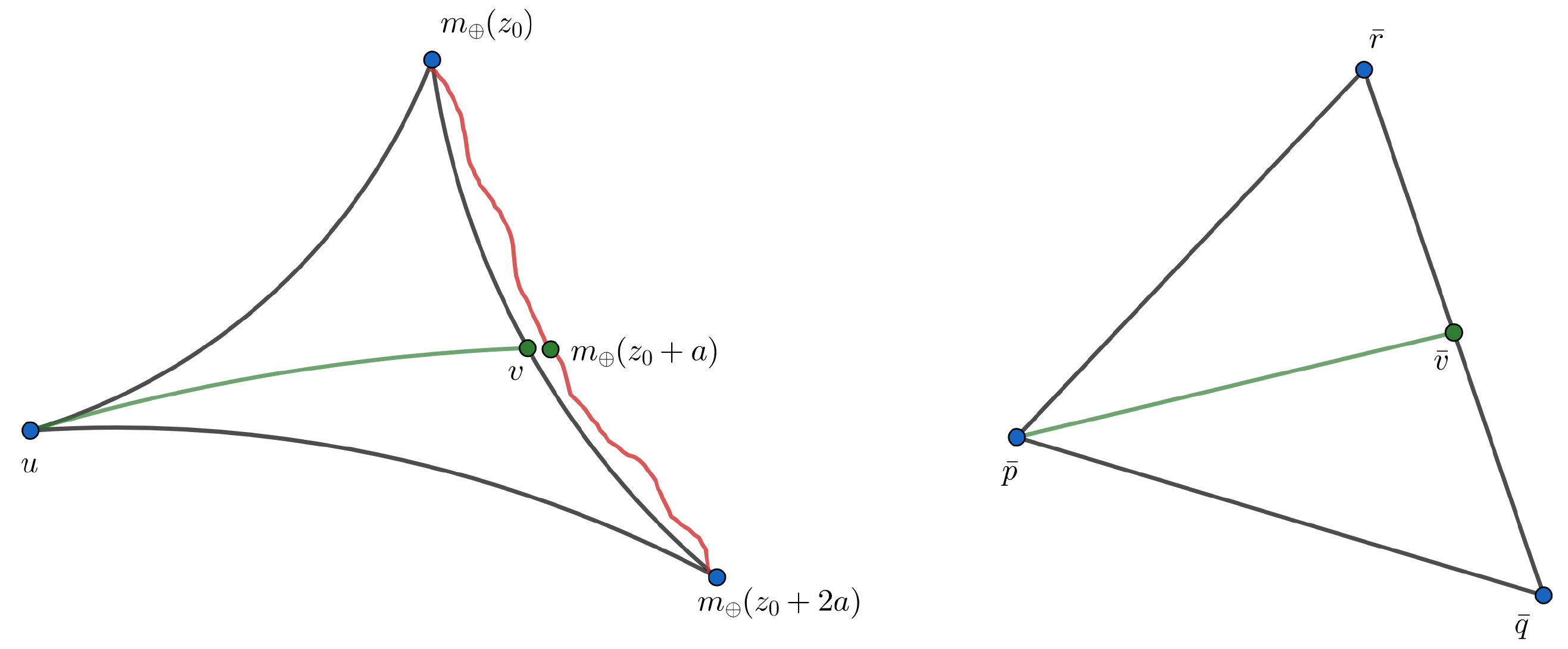}
\caption{Left figure: Geodesic triangle formed by the three points $u,$ $\mop{z_0},$ $\mop{z_0+2a}$, where  $v$ is the midpoint of the geodesic connecting the points $\mop{z_0}$ and $\mop{z_0+2a}.$ The red line depicts the true regression function $m_\oplus.$ $\mop{z_0 +a}$ is closely approximated by $v$ lying on a geodesic that connects   $\mop{z_0}$ with $\mop{z_0+2a}.$ Right figure:  Reference triangle in $\real^2$ as an illustration of the CAT(0) inequality. }
\label{fig:geodesic}
\end{figure}

\section{Sufficient conditions for assumption (A5)} 
\label{append:curvature:assump}
We discuss here  sufficient conditions under which  assumption~\ref{ass:convexity:d_m} holds. For this we consider the  following assumptions:
\ben[label = (K\arabic*), series = fregStrg, start = 1]
\item \label{ass:CAT0} $(\Omega,d)$ is a CAT(0) space, that is every geodesic triangle satisfies the CAT(0) inequality in~\eqref{CAT0:ineq}. 
\een
For any $z_0 \in \real,$  and $u\in \Omega,$ there exists some $a_0>0,$ such that for small enough $a \in (0,a_0],$ we may consider the geodesic triangle formed by $u,$ $\mop{z_0},$ $\mop{z_0+2a}$ for $z_0,z_0+2a \in \mc{T},$ for which we assume the following. 
\ben[label = (K\arabic*), series = fregStrg, start = 2]
\item \label{ass:geodesic:approx}  Defining the midpoint $v$ of the geodesic path connecting $\mop{z_0}$ and $\mop{z_0+2a}$ such that
\begin{align}
\label{ass:geodesic:midpt}
d(\mop{z_0},v) = d(\mop{z_0+2a},v) = \frac{1}{2} d(\mop{z_0+2a},\mop{z_0}),
\end{align}
we require 
\begin{align}
\label{ass:geodesic:approx:midpt}
&d(\mop{z_0+a},v) \leq C_\ast a^2,
\end{align}
where $C_\ast>0$ does not depend on $z_0,$ and is such that, $L_\ast^2>2DC_\ast,$ $L_\ast$ and $D$ being the lower Lipschitz constant for $m_\oplus$ from assumption~\ref{ass:reg:cont}, and the diameter of the metric space $\Omega,$ respectively.
\item \label{ass:reg:cont2} There exist real constants $L_\ast >0$ such that, for all  $\tbfx$ with norm bounded both above and below, and for all $\bpara_1,  \bpara_2 \in \bar{\Theta},$
\[
d\left(\mop{\tbfx \t\bpara_1, \bpara_1}, \mop{\tbfx \t\bpara_2, \bpara_2}\right) \geq L_\ast\ltwoNorm{ \bpara_1-\bpara_2}.
\]
\een
Figure~\ref{fig:geodesic} illustrates the geometry of the geodesic triangles in $\Omega$ and its reference space $\real^2.$ 
Assumption~\ref{ass:geodesic:approx} can be verified when  the link function $m_\oplus$ is smooth enough for the case of conventional Euclidean single index models. It thus provides an extension of the  usual smoothness assumption in the case of random object responses.
In section C of the Supplement 
We discuss this further in the context of Euclidean responses and in the case where the responses lie in the space of distributions equipped with Wasserstein-2 metric, and derive assumption~\ref{ass:convexity:d_m} under the sufficient conditions~\ref{ass:CAT0},~\ref{ass:geodesic:approx}, and~\ref{ass:reg:cont2}.

Assumption~\ref{ass:reg:cont2} in conjunction with assumption~\ref{ass:reg:cont} implies that the link function $m_\oplus$ is bi-Lipschitz. This limits the rate at which the object $m_\oplus$ can change, essentially it cannot change too fast or too slowly. A bi-Lipschitz function is an injective Lipschitz function whose inverse function is also Lipschitz. The bi-Lipschitz condition is stronger than the common assumption of a  monotone link function in classical single index modeling with Euclidean responses. In the special case of $\Omega = \real$ this reduces to requiring a monotone differentiable function with strictly positive derivative almost everywhere and restricts the monotonicity to a smaller subclass of strictly monotone functions. In the special case of Euclidean responses, this simplifies to the assumption that  the link function $m_\oplus = m$ is monotone and differentiable such that $m'(x)$ is strictly monotone with continuous derivative bounded away from zero. Such technical assumptions are commonly used for deriving distributional results in the existing single index literature, by virtue of a Taylor expansion of the link function $m$ in the Euclidean case.
\unappendix
\bibliographystyle{apalike}
\bibliography{./sim_ref}

\clearpage

\section*{Supplementary Material}
\section*{S.1. Technical assumptions~\ref{ass:minUnif}-~\ref{ass:curvatureUnif},~\ref{ass:ker}-~\ref{ass:jointdtn} } 
\label{appen:assump}
\hspace*{\fill} \\
In this section, we describe the technical assumptions needed to establish the uniform rate of convergence for the local linear Fr\'echet regression estimator in Lemma 1 
in Section 3 
of the main manuscript. We also provide motivation and discuss suitable examples regarding the assumptions.

The assumptions required to obtain the technical results are 
essentially the same as those used before in the Fr\'echet regression literature, specifically in~\cite{chen:22}. To adapt these assumptions to  the present situation, we require the curvature and entropy conditions to hold uniformly across all index values and direction parameters. The curvature and entropy conditions can be verified for commonly observed objects such as univariate probability distributions, positive definite matrices, or data on the surface of a sphere, as well as other random objects under suitable metrics.

Denote by $\mcTbpara$ the support of the random variable $T = \tbfX \t \bpara$ for any given unit direction $\bpara \in \bar{\Theta},$ where $\bar{\Theta}$ is defined in equation (2.5) of the main manuscript. 
Under assumption (A3), 
for bounded random variables $\tbfX,$ we can write $\mcTbpara \subset \mcT$ for some bounded subset $\mcT$ of $\real.$ For a  given direction $\bpara \in \bTheta$ such that $\tbfX \t \bpara = t,$ where $\bTheta$ is as given in equation (2.5), 
the conditional Fr\'echet mean is given by 
\bal
\label{loc:fr:target}
\mop{t,\bpara} = \argminomega \ M(\omega,t,\bpara); \quad M(\omega,t,\bpara) := \expect{(d^2(Y,\omega)|\tbfX\t \bpara = t)},
\eal
and the local linear Fr\'echet regression estimate by 
\bal
\label{loc:fr::inter:target}
\hmop{t,\bpara} = \argminomega \ \hL(\omega,t,\bpara);\quad  \hL(\omega,t,\bpara) :=\frac{1}{n}\sum_{i=1}^n\Shat(\tbfX_i \t \bpara, t, b)d^2(Y_i,\omega)),
\eal
where $\Shat$ is the empirical estimate (from equation (2.10))  
of the nonparametric weight function (described in equation (2.8)) 
in Section 2 
of the main manuscript and $b$ is the bandwidth parameter for the kernel involved in the localized Fr\'echet mean.
We also define the intermediate localized weighted Fr\'echet means as
\bal
\label{loc:fr:estd}
\tmop{t,\bpara} = \argminomega \ \tL(\omega,t,\bpara); \quad \tL(\omega,t,\bpara):=\expect{(S(\tbfX\t\bpara, t, b)d^2(Y,\omega))},
\eal
where the nonparametric weight function is described in equation (2.8) 
in the main manuscript.
The following additional assumptions are required, which are analogous  versions of the assumptions  in \cite{chen:22}. 
\ben[label=(U\arabic*)]
\item \label{ass:minUnif} 
For all $t \in \mcT$  and $\bpara \in \bTheta,$
the minimizers $\mop{t,\bpara}$, $\hmop{t,\bpara}$, and $\tmop{t,\bpara}$ exist and are unique, the latter two almost surely. In addition, for any $\eps>0$, 
\bgt
\inf_{t \in \mcT}
\inf_{d(\mop{t,\bpara},\omega)>\eps} [M(\omega,t,\bpara)-M(\mop{t,\bpara},t,\bpara)]>0,\\
\liminf_{b \to 0} \inf_{t \in \mcT} 
\inf_{d(\omega, \tmop{t,\bpara}) > \eps} [\tL(\omega,t,\bpara) - \tL(\tmop{t,\bpara},t,\bpara)]>0,
\egt
and there exists $c = c(\eps)>0$ such that
\bgt
\prob\left(\inf_{t \in \mcT} \inf_{d(\hmop{t,\bpara},\omega) >\eps} [\hL(\omega,t,\bpara) - \hL(\hmop{t,\bpara},t,\bpara)]\geq c \right) \to 1.
\egt
\item \label{ass:entropyUnif} 
Let $\mc{B}_r(\mop{t,\bpara}) \subset \Omega$ be a ball of radius $r$ centered at $\mop{t,\bpara}$ and\\ $\mc{N}(\eps,\mc{B}_r(\mop{t,\bpara}),d)$ be its covering number using balls of radius $\epsilon.$ Then
\bgt
\underset{r\to 0+}{\lim} \int_0^1  \sup_{t \in \mcT}  \sqrt{1 + \log \mc{N}(r\eps,\mc{B}_r(\mop{t,\bpara}),d)} d \epsilon = O(1).
\egt
\item \label{ass:curvatureUnif} 
There exists $r_1,r_2>0,$ $c_1,c_2>0,$ and $\beta_1,\beta_2>1$ such that
\begin{align}
&\inf_{t \in \mcT} \ \inf_{d(\mop{t,\bpara},\omega)<r_1}  [M(\omega, t,\bpara) - M(\mop{t,\bpara},t,\bpara) - c_1d^2(\omega,\mop{t,\bpara})^{\beta_{1}}] \geq 0, \nonumber \\
&\underset{b \to 0}{\liminf} \ \inf_{t \in \mcT} 	\inf_{d(\omega, \tmop{t,\bpara}) <r_2}\ [\tL(\omega, t,\bpara) - \tL(d(\tmop{t,\bpara}, t,\bpara) - c_2d^2(\omega, \tmop{t,\bpara})^{\beta_2}] \geq 0.
\end{align}
\een
Furthermore, we require the following assumptions for kernels and distributions. 
\ben[label = (R\arabic*)]
\item \label{ass:ker} 
The kernel $K$ is a probability density function, symmetric around zero, uniformly continuous on $\real$ such that $\int_\real K(x)^j x^k < \infty,$ for $j,k=1,\dots 6.$  The derivative $K'$ exists and is bounded on the support of $K$, i.e., $\sup_{x : K(x)>0}|K'(x)| < \infty.$ Additionally, $\int_\real x^2 |K'(x)|\sqrt{|x\log|x||} dx <\infty.$
\item \label{ass:jointdtn} 
For any given unit direction $\bpara \in \bar{\Theta},$ the marginal density $f_{T,\bpara}$ of $T= \tbfX \t \bpara$ and the conditional densities $f_{T,\bpara|Y}(\cdot,y)$ of $T$ given $Y= y$ exist and are twice continuously differentiable in the interior of $\mcT$ for all $\bpara \in \bar{\Theta}$, the latter for all $y \in\Omega$. The marginal density $f_{T,\para}$ is bounded away from zero on its support $\mcT$ for all $\bpara \in \bar{\Theta}$ i.e., 
$ \inf_{t\in\mcT}\ f_{\tbfX \t \bpara}(t) >0.$

The second-order derivative $f_{T,\bpara}''$ is uniformly bounded for all $t\in \mcT,$ $\bpara \in \bar{\Theta},$ that is,\\ 
$\sup_{t\in\mcT} |f_{T,\bpara}''(t)|<\infty$. 
The second-order partial derivatives
$(\partial^2 f_{T, \bpara|Y}/\partial t^2)(\cdot,y)$ are uniformly bounded, uniform over all $\bpara \in \bTheta$, i.e., \\
$ \sup_{t \in \mcT} \ \sup_{y \in \Omega} |(\partial^2 f_{T,\bpara|Y}/\partial t^2)(\cdot,y) | < \infty.$

Additionally, for any open set $B\subset \Omega$, $\prob(Y\in B | \tbfX \t \para = t)$ is continuous as a function of $t$ and $\bpara.$ For any $t\in \mcT$ and $\bpara \in \bTheta,$ $M(\omega,t,\bpara)$ is equicontinuous, that is,
\[
\limsup_{\bpara_1 \to\bpara_2} \ \sup_{t\in \mcT}\ \sup_{\omega \in \Omega} \left| M(\omega, t,\bpara_1) - M(\omega, t,\bpara_2)\right| = 0.
\]
\een
Similar yet weaker assumptions have been made in~\cite{pete:19} for pointwise
rates of convergence for local linear Fr\'echet regression estimators. ~\cite{chen:22} made stronger assumptions in this regard to establish uniform convergence results over univariate predictor values. In the above assumptions~\ref{ass:minUnif}-~\ref{ass:curvatureUnif} we adapt those in~\cite{chen:22}, incorporating  uniform bounds over the index parameter as well as over the values of the single index. Since the objective function for the local  Fr\'echet regression involves both the index value $\tbfx\t\bpara = t$ and the index parameter $\bpara$, conditions on the well-separatedness, entropy, and curvature needs to be extended for all values of $t$ and $\bpara.$  These assumptions are adapted from empirical process theory, guarantee the asymptotic uniform
equicontinuity of $\tilde{L}_b$, and control the behavior of  $\tilde{L}_b -M$  and $\hat{L}_n - \tilde{L}_b$ near the minimizers $m_\oplus(t,\bpara)$ and $\tilde{m}_\oplus(t,\bpara)$, respectively, uniformly over $t$ and $\bpara.$ assumption~\ref{ass:minUnif} is commonly
used to establish the uniform consistency of M-estimators~\citep{vand:00} by showing the weak convergence of the respective empirical processes.  In
conjunction with the assumption that the metric space $\Omega$ is
totally bounded, this implies the pointwise convergence of the minimizers for any given $t$ and $\bpara$; it also ensures
that the asymptotic uniform equicontinuity of $\tilde{L}_b$ and $\hat{L}_n$, and implies the (asymptotic) uniform
equicontinuity of $\tilde{m}_\oplus$ and $\hat{m}_\oplus$, whence the uniform convergence of the minimizers follows as the support of $\tbfx\t\bpara$ is compact for any $\bpara$.

Assumptions~\ref{ass:minUnif}-~\ref{ass:curvatureUnif} are easily  verified for specific metric space-valued objects. 
\begin{itemize}
\item[\emph{Example 1}] Let $\Omega$ be the set of probability distributions on a closed interval of $\real$ with finite second moments, endowed with the Wasserstein-2 distance $d_W$, i.e.,  for any two distributional objects $Y_1$ and $Y_2$ with cdfs $F_{Y_1}$ and $F_{Y_2}$ respectively,
\begin{align*}
	d_W(Y_1,Y_2) = \int_0^1 (F_{Y_1}^{-1}(z)-F_{Y_2}^{-1}(z))^2 dz,
\end{align*}
where $F_{Y_j}^{-1}(z)$ is the quantile function for $Y_j$, $j=1,2.$ The Wasserstein space $(\Omega,d_W)$
satisfies assumptions~\ref{ass:minUnif}-~\ref{ass:curvatureUnif} with $\beta_1 = \beta_2 =2.$
\item[\emph{Example 2}] Let $\Omega$ be the space of $r$-dimensional correlation matrices, i.e., symmetric, positive
semidefinite matrices in $\real^{r\times r}$ with diagonal elements equal to $1$, endowed with the Frobenius
metric $d_F$ . Specifically for any two elements $Y_1, Y_2\in\Omega$, 
\begin{align*}
	d_F(Y_1,Y_2) = \sqrt{\trace{((Y_1-Y_2)^\intercal(Y_1-Y_2))}}.
\end{align*}
The space $(\Omega,d_F)$ satisfies assumptions~\ref{ass:minUnif}-~\ref{ass:curvatureUnif} with $\beta_1 = \beta_2 =2.$
\end{itemize}

For Examples 1-2, we note that since the Wasserstein space for one-dimensional distributions and the space of correlation matrices are Hadamard spaces,  there exists a unique minimizer of $M(\cdot,t,\bpara)$ for any $t\in \mathcal{T}$ and $\bpara \in \bTheta$~\citep{stur:03}. Examples 1-2 follow from similar arguments as those in the proofs of
Propositions 1-2 of~\cite{pete:19} by observing that the arguments hold uniformly across  $t$ and $\bpara.$
Assumptions~\ref{ass:ker} and~\ref{ass:jointdtn} are  standard distributional
assumptions for local nonparametric regression and are needed to show the convergence of the bias and stochastic parts for the local linear Fr\'echet estimator uniformly over all  $t$ and $\bpara.$
In particular, Assumption~\ref{ass:ker} can be verified for a general class of kernel functions given by
\begin{align*}
c_\kappa(1-x^2)^\kappa\indicator{[-1,1]},\ \kappa\in \mathcal{Z},
\end{align*}
where $c_\kappa = \frac{\Gamma(k+\frac{3}{2})}{\sqrt{\pi}\Gamma(k+1)}$ is such that $\int_{-1}^1 	c_\kappa(1-x^2)^\kappa dx = 1$ and the indicator function is defined as $\indicator{A} = 1 \text{ if } \tbfX \in A, \text{ and } 0 \text{ otherwise.}$
The Epanechnikov kernel  $K(x)= \frac{3}{4}(1-x^2)\indicator{[-1,1]}$ belongs to this class of kernel functions for $\kappa = 1$ with $c_\kappa = 3/4.$ 
\section*{S.2. Further discussion of assumption (A5)} 
\label{suppl:assump}
\hspace*{\fill} \\
Assumption (A5) 
in Section 3 
of the main manuscript intuitively means that $m_{\oplus}$ can be locally approximated by straight lines in Euclidean space and geodesics in geodesic spaces.  In the Euclidean case, it  is satisfied for twice differentiable functions $m_\oplus$, a common assumption for classical single index modeling. Beyond the Euclidean special case, assumption (A5) 

Consider first the Euclidean case,  where  $\Omega$ is a compact subset $\mc{M} \subset\real$ and denote the link function $m_\oplus$ by $m.$ Noting that the map $h:\para\mapsto \bpara$ is continuous, and $\mop{\bs{\mathbf{z}}\t\bpara,\bpara} := \phi(\bpara) = \phi(h(\para))$, for some function $\phi$ of $\bpara \in \bTheta$ and for any given $\bs{\mathbf{z}} \in \mathcal{X} \subset \real^p$, with a slight abuse of notation, we write $\mop{\bs{\mathbf{z}}\t\para,\para}$ instead of $\mop{\bs{\mathbf{z}}\t\bpara,\bpara}$. For any given $z \in \mathcal{X} \subset \real^p$ and $\para \in \Theta$ such that $\para \t \para < 1,$ denote
$m\left( z\t\para,\para\right) = m(z_0,\para)$ by $m(z_0),$ where $z_0 = z\t\para \in \real$ and for a small enough $a \in (0,a_0),$ such that $z_0,z_0+2a \in \mcT,$ we have $m(z_0),m(z_0+a),m(z_0+2a) \in \mc{M}.$ If $m(\cdot)$ is twice continuously differentiable in any open subset containing $z_0$ such that the derivatives are uniformly bounded, the midpoint on the straight line (geodesic path) connecting $m(z_0)$ and $m(z_0+2a)$ is given by $v = \frac{1}{2}[m(z_0)+m(z_0+2a)].$ Using a second-order Taylor expansion for the function $m$ around $z_0$, we have
\begin{align*}
&\lVert v-\mop{z_0+a} \rVert_E\\ =& \lVert \frac{1}{2}[m(z_0)+m(z_0+2a)] -\mop{z_0+a} \rVert_E\\
=&  \lVert [\frac{1}{2}m(z_0) + \frac{1}{2}m(z_0) + a m'(z_0) + \frac{1}{2} \frac{(2a)^2}{2}m''(\zeta_1) ]
- [m(z_0) + am'(z_0) + \frac{a^2}{2!}m''(\zeta_2)] \rVert_E\\
=& \lVert a^2 [m''(\zeta_1) - \frac{1}{2}m''(\zeta_2) ]\rVert_E,
\end{align*}
where $z_0<\zeta_1<z_0+2a,$ and $z_0<\zeta_2<z_0+a.$ Assuming a uniform bound on the second derivative of $m,$ such that $|m''(z)|\leq C$ for some $C>0$ and for all $z \in \mcT,$ we have that $\lVert v-\mop{z_0+a} \rVert_E \leq \frac{3C}{2}a^2.$ Thus, assumption (K2) 
holds for $C_\ast = 3C/2,$ as long as the bound $C$ on the second derivative of $m$ is sufficiently small.
\begin{figure}
\centering
\includegraphics[width =.8\textwidth]{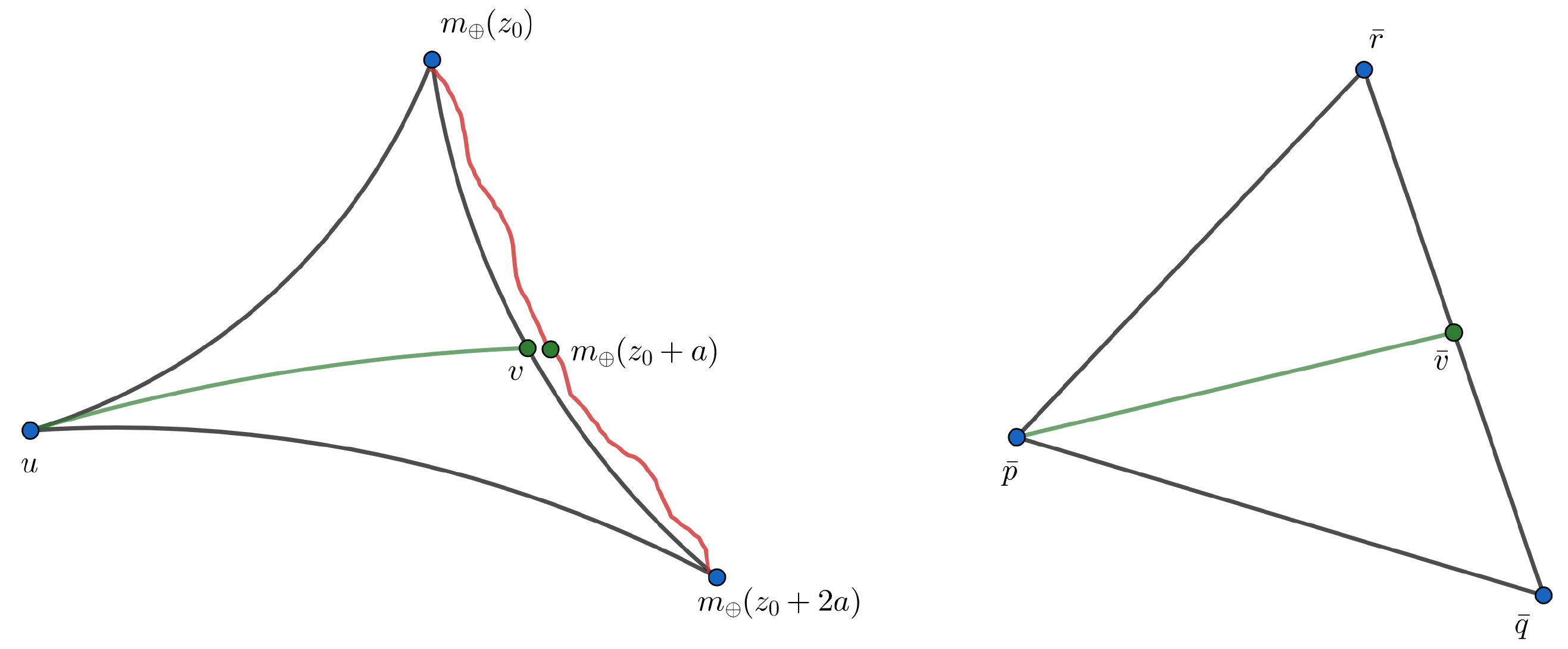}
\caption{The left figure shows the geodesic triangle formed by the three points $u,$ $\mop{z_0},$ $\mop{z_0+2a}$, where  $v$ is the midpoint of the geodesic connecting the points $\mop{z_0}$ and $\mop{z_0+2a}.$ The red line depicts the true regression function $m_\oplus.$ $\mop{z_0 +a}$ is closely approximated by $v$ lying on a geodesic that connects   $\mop{z_0}$ with $\mop{z_0+2a}.$ The right hand side shows the reference triangle in $\real^2$ as an illustration of the CAT(0) inequality. }
\label{fig:geodesic}
\end{figure}

Next, we consider $\Omega$ to be the space of univariate distributions, $\mc{F},$ endowed with the Wasserstein-2 metric $d_W.$ The quantile functions for the distributional objects $\mop{z_0},$ $\mop{z_0+a},$ and $\mop{z_0+2a}$ are denoted by $Q(\mop{z_0})(\cdot),$ $Q(\mop{z_0+a})(\cdot),$ and $Q(\mop{z_0+2a})(\cdot),$ respectively. Similarly, the quantile function of the midpoint $v$ of the geodesic path connecting $\mop{z_0}$ and $\mop{z_0+2a}$ is given by
\[
Q(v)(\cdot) =  \frac{1}{2}[Q(\mop{z_0})(\cdot) + Q(\mop{z_0+2a})(\cdot) ].
\]
We write  $q(z_0)(\cdot)=Q(\mop{z_0})(\cdot) = q(z_0)(\cdot)$, analogously for related quantities.  The Wasserstein distance between $v$ and $\mop{z_0+a}$ is then given by
\begin{align*}
d_W^2(v,\mop{z_0+a}) &= \int_{0}^{1} \left( Q(v)(t) -  Q(\mop{z_0+2a})(t)\right)^2 dt\\
& = \int_{0}^{1} \left(  \frac{q(z_0)(t) + q(z_0+2a)(t)}{2}-  q(z_0+2a)(t)\right)^2 dt
\end{align*}
We assume that for every $t \in [0,1],$ $q(z)(t)$ is twice continuously differentiable as a function of $z,$ for any $z$ in an open subset containing $z_0$ such that derivatives of $q(z)(t)$ are uniformly bounded for each $t \in [0,1]$. Using a second-order Taylor expansion of $q(\cdot)(t)$ pointwise $t \in [0,1],$ and following a similar argument as in the Euclidean case, we have
\begin{align*}
d_W^2(v,\mop{z_0+a})
& = \int_{0}^{1} \left( a^2 [q''(\zeta_1)(t) - \frac{1}{2}q''(\zeta_2)(t)]\right)^2 dt,
\end{align*}
Lastly, under the assumption that the $|q''(z)(t)| \leq r(t),$ such that $\int_0^1 r^2(t) <C,$ assumption (K2) 
holds for $C_\ast = 3/2C,$ as long as the bound $C$  is sufficiently small.

We further illustrate the argument for assumption (K2) 
for distributional objects in the specific context of 
a location-scale family of univariate distributions, $\mc{F},$ endowed with the Wasserstein-2 metric $d_W.$ Denoting the location and scale parameters as $\mu(\cdot)$ and $\sigma(\cdot)$ respectively, the quantile function corresponding to the distribution object $\mop{z_0}\in \mc{F}$ will be given by
\[
Q(\mop{z_0})(\cdot) = \mu(z_0) +\sigma(z_0)F^{-1}(\cdot),
\]
where $F^{-1}(\cdot)$ is the quantile function for the distribution object $\mop{z_0}.$ The quantile functions for  $\mop{z_0+a}$ and  $\mop{z_0+2a}$ can be similarly defined. Also, the quantile function of the midpoint of the geodesic path connecting $\mop{z_0}$ and $\mop{z_0+2a}$ is given by
\[
Q(v)(\cdot) =  \frac{1}{2}[\mu(z_0) +\mu(z_0+2a)]  + \frac{1}{2}[\sigma(z_0) +\sigma(z_0+2a)] F^{-1}(\cdot).
\]
The Wasserstein distance between $v$ and $\mop{z_0+a}$ is given by
\begin{align*}
d_W^2(v,\mop{z_0+a}) &= \left|\frac{\mu(z_0) +\mu(z_0+2a)}{2} - \mu(z_0+a) \right|^2 \\  
&\quad \hspace{-2cm} + \left|\frac{\sigma(z_0) +\sigma(z_0+2a)}{2} + \sigma(z_0+a) - 2\left( \frac{\sigma(z_0) +\sigma(z_0+2a)}{2}\sigma(z_0+a) \right)^{1/2}\right|^2\\
& \hspace{-2cm} \leq \left|\frac{\mu(z_0) +\mu(z_0+2a)}{2} - \mu(z_0+a) \right|^2  + \left|\frac{\sigma(z_0) +\sigma(z_0+2a)}{2} - \sigma(z_0+a) \right|^2,
\end{align*}
where the last inequality holds because  $\frac{1}{2}\sigma(z_0) +\sigma(z_0+2a)$ and $\sigma(z_0+a)$ are both positive.
Assuming $\mu(\cdot)$ and $\sigma(\cdot)$ are twice continuously differentiable in any open subset containing $z_0$ such that their derivatives are uniformly bounded, the result follows in a similar manner to the Euclidean case.

We next show that assumption (A5) 
holds under the sufficient conditions (K1), (K2),and (K3), 
that is, for any $u \in \Omega,$ and $z_0 \in \mcT,$ there exists some $\kappa>0,$ such that, for any small $a>0,$
\begin{align}
\label{eq:cat0:res}
\frac{1}{a^2}[d^2(u, \mop{z_0+2a}) -2d^2(u,\mop{z_0+a}) +d^2(u,\mop{z_0})] \geq \kappa
\end{align}

Observe that
\begin{align}
\label{eq:cat0:prf}
&\frac{1}{a^2}[d^2(u,\mop{z_0+2a}) -2d^2(u,\mop{z_0+a}) +d^2(u,\mop{z_0})] \\  \nonumber	=& \frac{1}{a^2}[d^2(u,\mop{z_0+2a}) -2d^2(u,v) +d^2(u,\mop{z_0})] \\ \nonumber +& \frac{1}{a^2}[2d^2(u,v)  - 2d^2(u,\mop{z_0+a})].
\end{align}
Assumption (K3) 
in conjunction with assumption (A2) 
implies that $m_\oplus$ is bi-Lipschitz with constants $0\leq L_\ast \leq L$. We have 
\begin{align}
\label{eq:cat0:prf2}
2aL_\ast\leq d(\mop{z_0+2a},\mop{z_0}) \leq 2La.
\end{align}
Thus the first term of~\eqref{eq:cat0:prf} becomes
\begin{align}
\label{eq:cat0:prf3}
&\frac{1}{a^2} \left[ d^2(u,\mop{z_0+2a}) - 2d^2(u,v) + d^2(u,\mop{z_0})\right] \\ \nonumber	& \geq \frac{4L_\ast^2}{d^2(\mop{z_0+2a},\mop{z_0})} \left[ d^2(u,\mop{z_0+2a}) - 2d^2(u,v) + d^2(u,\mop{z_0})\right],
\end{align}
where this inequality follows from  assumptions (A2), 
using~\eqref{eq:cat0:prf2}. 
Assuming $\Omega$ is  a geodesic CAT(0) space, the geodesic triangle $\triangle(u,\mop{z_0},\mop{z_0+2a}),$ formed by the vertices $u,$ $\mop{z_0},$ and $\mop{z_0+2a},$ will have a comparison triangle $\triangle(\bar{p},\bar{q},\bar{r})$ in the reference space $\real^2$ for some points $\bar{p},\bar{q},\bar{r} \in \real^2.$ This implies 
\begin{align}
\label{eq:cat0:prf3a}
d(u,\mop{z_0}) &= ||\bar{p} - \bar{q}||_E, \quad
d(u,\mop{z_0+2a}) = ||\bar{p} - \bar{r}||_E,  \\  \nonumber	d(\mop{z_0},v) &= ||\bar{q} - \bar{v}||_E, \quad
d(\mop{z_0+2a},v) = ||\bar{r} - \bar{v}||_E.
\end{align}
By virtue of assumption (K1), 
\begin{align}
\label{eq:cat0:prf3b}
d(u,v) \leq ||\bar{p} - \bar{v}||_E.
\end{align}
Thus combining \eqref{eq:cat0:prf3}-- \eqref{eq:cat0:prf3b} one obtains 
\begin{align}
\label{eq:cat0:prf3c}
&\frac{1}{a^2} \left[ d^2(u,\mop{z_0+2a}) - 2d^2(u,v) + d^2(u,\mop{z_0})\right] \\ \nonumber	\geq &  2L_\ast^2 \  \frac{\frac{\ltwoNorm{\bar{p}-\bar{r}}_E^2 - \ltwoNorm{\bar{p}- \bar{v}}_E^2}{||\bar{r} - \bar{v}||_E} - \frac{\ltwoNorm{\bar{p}-\bar{v}}_E^2 - \ltwoNorm{\bar{p}-\bar{q}}_E^2}{||\bar{q} - \bar{v}||_E}}{||\bar{r} - \bar{q}||_E}
= 2L_\ast^2 >0.
\end{align}
This uses  the fact that $\bar{r},$ $\bar{v},$ $\bar{q}$ are co-linear in the Euclidean space with $\bar{v}$ being the midpoint between $\bar{r}$ and $\bar{q},$ and hence the second order difference is just $1.$ 
Thus the first term of~\eqref{eq:cat0:prf} is seen to be greater than or equal to $2L_\ast^2.$

As for the second term of~\eqref{eq:cat0:prf}, by simple algebra and the triangle inequality, 
\begin{align}
\label{eq:cat0:prf4}
& \left|\frac{2}{a^2}[d^2(u,v)  - d^2(u,\mop{z_0+a})]\right| \\   \nonumber	=& \frac{2}{a^2} \left|(d(u,v)  + d(u,\mop{z_0+a})) \right|\ \left|(d(u,v)  - d(u,\mop{z_0+a})) \right|\\  \nonumber	\leq & \frac{4D}{a^2}d(v,\mop{z_0+a}) \leq 4DC_\ast.
\end{align}
The last inequality follows from equation (B.2) 
in assumption (K2). 
In assumption (K2), 
given $L$ and $D,$ $C_\ast$ can be chosen sufficiently small such that $2L_\ast^2 > 4DC_\ast.$ Thus, combining~\eqref{eq:cat0:prf3c} and~\eqref{eq:cat0:prf4}  with~\eqref{eq:cat0:prf}, the result follows for $\kappa = 2L_\ast^2 - 4DC_\ast>0.$\\
\section*{S.3. Additional data illustrations and simulations}
\label{suppl:data}
\hspace*{\fill} \\
This section provides further illustrations of data applications and simulations. 
Random objects considered in the additional data demonstrations discussed in this section are univariate probability distributions with compact support endowed with the Wasserstein-2 metric (applied to human mortality data) and  compositional data that are mapped to the positive segment of a sphere,    endowed with the geodesic distance  and 
applied to the mood compositional data. Further illustrations of the proposed method include
an additional plot for the ADNI study and a simulation study with Euclidean responses. 

\subsection*{S.3.1. Human mortality and age-at-death distributional object responses}
\label{suppl:mort}
\hspace*{\fill} \\
The performance of the proposed model is demonstrated with an application to  human mortality data across countries. We view the age-at-death distributions as random object responses of interest and aim to find their association with Euclidean predictors such as economic, social, and healthcare indices among other relevant factors, aiming at a comprehensive understanding of human longevity and health conditions.

\begin{figure}[!htb]
\centering
\begin{subfigure}[b]{0.45\textwidth}
\centering
\includegraphics[width=1.1\textwidth]{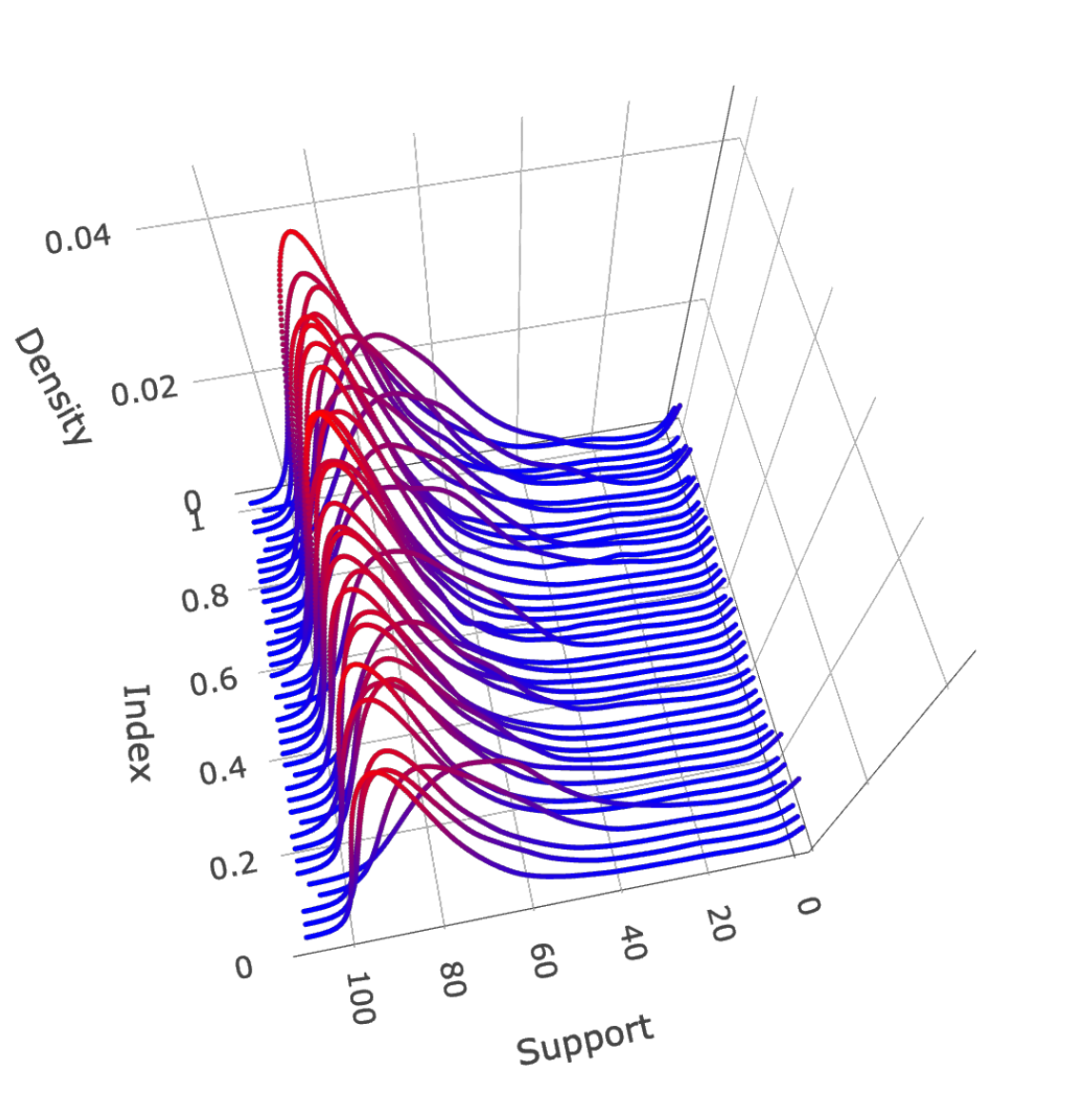}
\end{subfigure}
\hfill
\begin{subfigure}[b]{0.45\textwidth}
\centering
\includegraphics[width=1.1\textwidth]{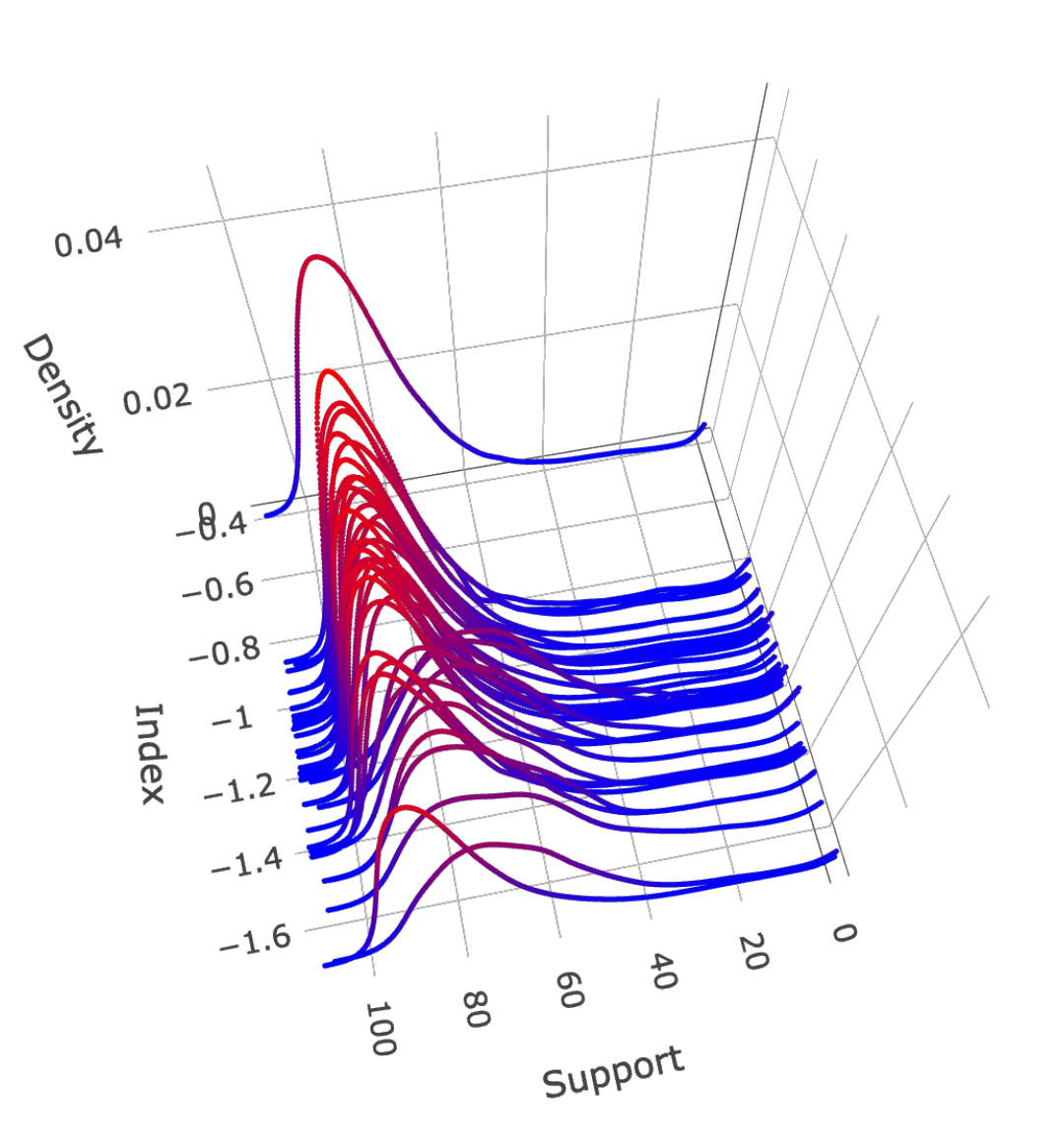}
\end{subfigure}
\caption{Data visualization for age-at-death densities for $40$ countries at the calendar year 2010. The left panel shows the observed densities at random order while the right panel plots the observed densities against the estimated index values from the proposed Index Fr\'echet Regression (IFR) model.}
\label{fig:obs_vs_index}
\end{figure}

For this analysis, we used the lifetables for males aggregated yearly in age groups varying from age $0$ to $110$ for $40$ countries in  the calendar year 2010. The data consist of period lifetables for each country and each calendar year and were obtained from the Human Mortality Database (\url{https://www.mortality.org/}). We  computed histograms of age-at-death from the lifetables for each country and calendar year, which were then smoothed with local least squares to obtain smooth estimated probability density functions for age-at-death using the R package frechet~\citep{R:chen:20:fr}. After this preprocessing step, the data are a  sample of univariate probability distributions for $n= 40$ countries was obtained, shown in the left  
panel of Figure~\ref{fig:obs_vs_index}. We equipped the  sample of age-at-death distributions  with the Wasserstein-2 metric $(\Omega,d_W)$ and 
selected the following six socio-economic predictors measured at the calendar year 2010:  
$X_1 =$ Population density (people per sq. km of land area), $X_2 =$ Fertility rate, total (births per woman), $X_3 =$ GDP per capita, at Purchasing Power Parity (PPP), $X_4 =$ Access to electricity (\% of the population), $X_5 = $ Current health expenditure (\% of GDP), and $X_6 =$  Unemployment, total (\% of the total labor force) (national estimate). The data were obtained  from the World Bank Database at \url{https://data.worldbank.org}.

We first standardized all predictors separately, then applied the proposed Index Fr\'echet Regression (IFR) method to obtain the estimated unit direction parameter (rounded to 4 decimal places)
$$ \bhtrue =  (0.0173,  0.7875,  0.5879,  0.0167,  0.1646, -0.0807)^\intercal.$$ 
The estimated coefficient for the predictor Fertility Rate ($X_2$) has the highest absolute value,
indicating its heavy influence relative to the other five predictors on the index $\tbfX\t\bhtrue,$ and hence on the fitted value for the IFR model. 
The estimated index $\tbfX\t\bhtrue$ can be also perceived as the first sufficient predictor, which  reduces the dimension of the predictor space without losing the information about the response. This 
aligns with the sufficient dimension reduction methods for Fr\'echet regression~\citep{zhan:21} and provides an insight into the overall dependence of the predictors on the object response.

In the right panel of Figure~\ref{fig:obs_vs_index},  the age-at-death densities are plotted against the estimated index values, aka the first sufficient predictors, $\tbfX\t\bhtrue$. It is evident that countries with low index values have modes of the distribution at lower ages, while for countries with high values of the index, the modes of mortality distributions are significantly higher. Further, the countries with higher index values indicate very low infant mortality rates.

\begin{figure}[!htb]
\centering
\begin{subfigure}[b]{0.45\textwidth}
\centering
\includegraphics[width=1.1\textwidth]{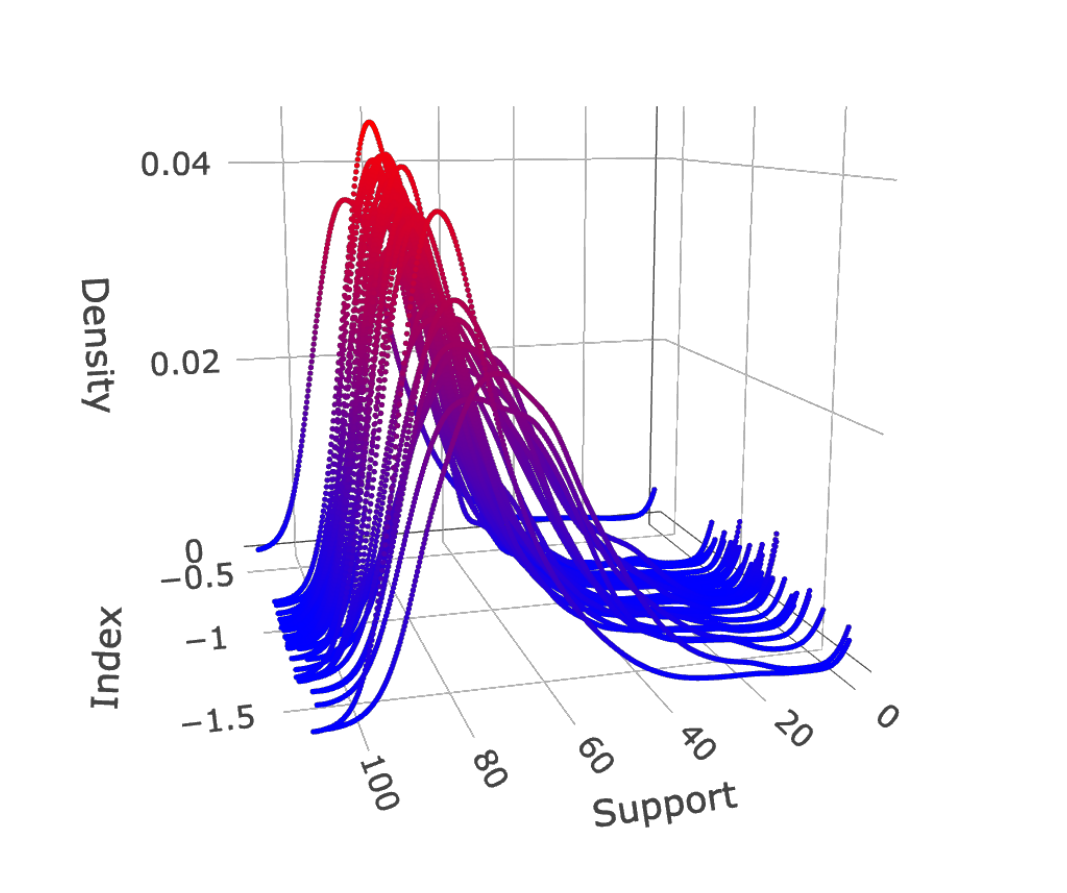}
\end{subfigure}
\hfill
\begin{subfigure}[b]{0.45\textwidth}
\centering
\includegraphics[width=1.1\textwidth]{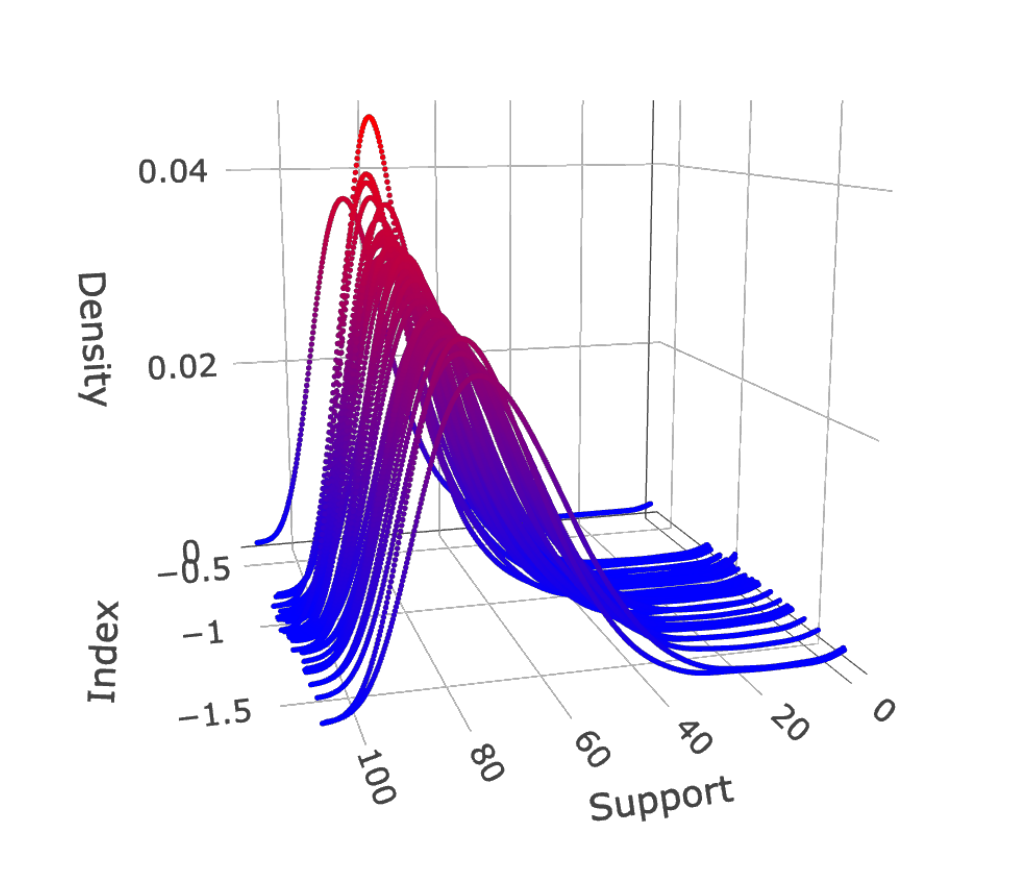}
\end{subfigure}
\caption{The observed and estimated age-at-death distributions for $40$ countries at the calendar year 2010 are displayed in the left and right panel of figure, respectively. The distributions are plotted over the support of the age interval $[0,110]$ against the index values estimated by the IFR model.}
\label{fig:obs_vs_estd_index}
\end{figure}

The plots of the observed and estimated age-at-death densities over the support of age $[0,110]$ and against the estimated index values, aka the first estimated sufficient predictor, 
are shown in Figure~\ref{fig:obs_vs_estd_index}. It is interesting to observe that the estimated index values are associated with the location and variation features of the age-at-death distributions. Specifically, with the increase in the values of the index, the mean of the mortality distribution increases non-linearly while the standard deviation diminishes, indicating the death age more concentrates between 70 and 80. This finding is in line with the observations of~\cite{zhan:21}, who employed several sufficient dimension reduction (SDR) techniques to the mortality distributions.

Further, the importance of various predictors can be inferred from the estimated coefficients  $\bhtrue.$ As before we keep the first predictor ($X_1 =$ Population density) with the corresponding coefficient $\hat{\theta}_1 = 0.0173 >0$ in the model and test for the following hypothesis: $H_0: \theta_{02} =\dots = \theta_{0p} = 0$ vs. $H_1$, the complement of $H_0$, which is the test for overall regression effect for object responses.  Writing  $\htrue = (\hat{\theta}_2,\dots,\hat{\theta}_6),$ the test statistic is constructed as 
$\tilde{T}_n = \htrue \t (\wh{\Lambda}^\ast_B)^{-1} \htrue \overset{approx.}{\sim} \chi^2_{5}$ under $H_0$ (see Section 5.1), 
where $\wh{\Lambda}^\ast_B$ is the bootstrap estimator for asymptotic covariance matrix as described in Proposition 5. 
The null hypothesis is rejected at level $\alpha$ if $\tilde{T}_n > \chi^2_{5} (1-\alpha).$ From our analysis, $\tilde{T}_n = 18.883>  11.0705 = \chi^2_{5} (1-\alpha)$ for the level $\alpha = 0.05.$ The p-value  is actually $0.002$ and the null hypothesis is thus clearly rejected, demonstrating there is a regression effect.  Upon further analysis it is found that the most significant predictors, in order, are 
$X_2 =$ Fertility rate, total (births per woman), $X_3 =$ GDP per capita, at Purchasing Power Parity (PPP),  and $X_5 =$ Current health expenditure (\% of GDP). 

\begin{figure}
\centering
\includegraphics[width =.8\textwidth]{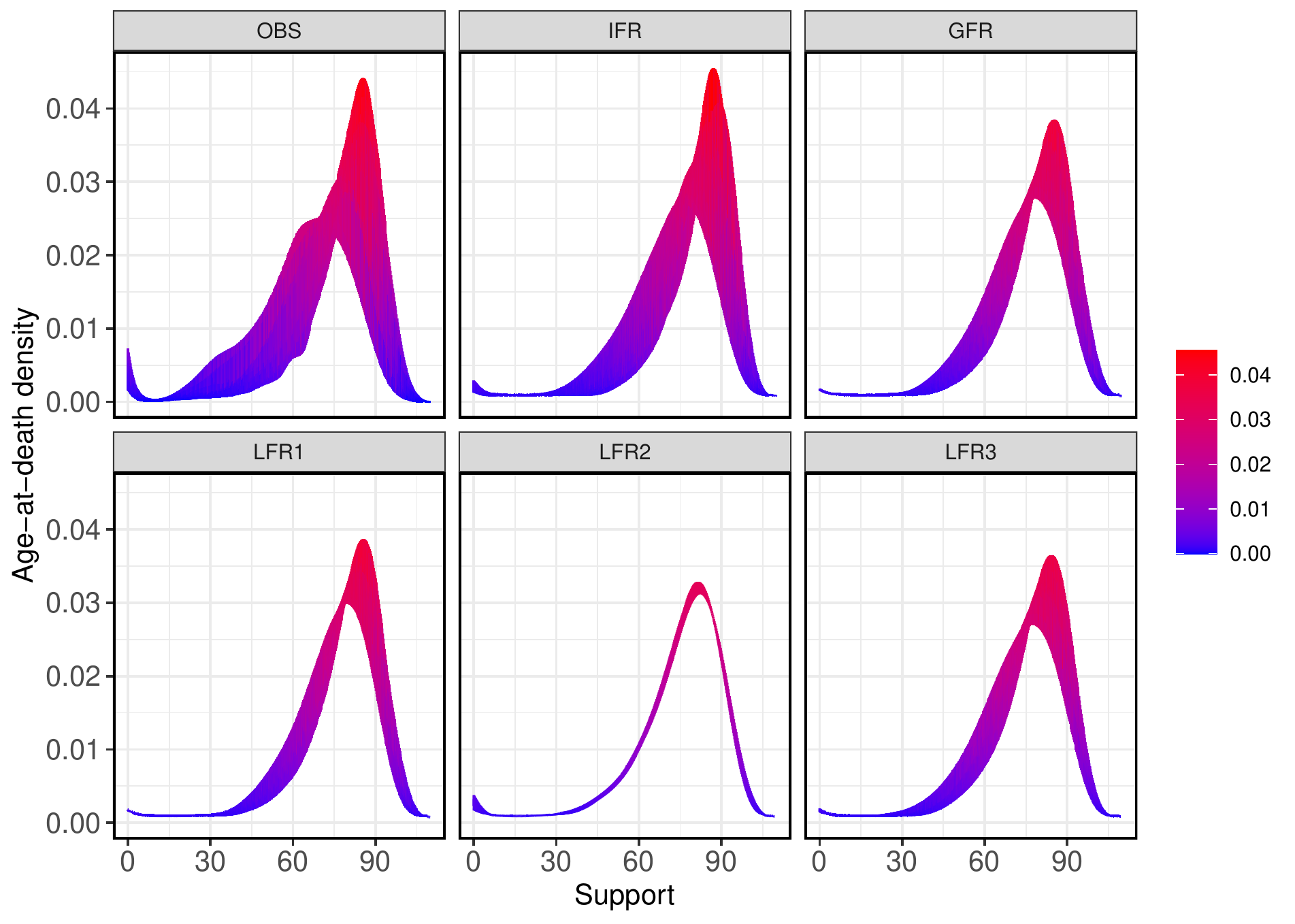}
\caption{Figure displaying the observed and predicted smooth densities. Clockwise, from top-left the observed densities (OBS), the fitted densities using Index Frechet Regression (IFR), Global Fr\'echet Regression (GFR), and Local Fr\'echet Regression (LFR).  The predictors used for the LFR fits are Fertility Rate (LFR1), GDP per capita (LFR2) and 
Health Expenditures (LFR3), respectively. Densities are color-coded (blue to red indicating low to high) by the mode of the age-at-death distribution.}
\label{fig:2dfits}
\end{figure}

We proceed to compare  fits for the year 2010 from the IFR model  with the Global Fr\'echet Regression (GFR) model with the $6-$dimensional predictors, as well as with three separate Local Fr\'echet Regression (LFR) models, where the three important predictors Fertility Rate, GDP per capita and Health Expenditure are considered  in each LFR model separably as univariate predictors. The global Fr\'echet model suffers from model-induced bias, while the local linear Fr\'echet Regression models with individual univariate predictors
lack relevant information from other variables. The IFR model is  a semiparametric approach that  combines the strengths of both of these models. Figure~\ref{fig:2dfits} displays the observed as well as the fitted distributions (as densities) for these five models. The superiority of the IFR model compared to the local linear Fr\'echet fits,
using only the relatively important predictor variables individually indicates that 
all predictors simultaneously  play an important role in the overall prediction through the estimated index $\tbfx\t\bhtrue.$
To study the effect of the most important predictors,  GDP per capita, fertility rate, and Health expenditure percentage on the age-of-death densities, we fitted the IFR model when varying the value of one predictor, while keeping the other two fixed at their mean levels.
\begin{figure}[!htb]
\centering
\begin{subfigure}[b]{0.32\textwidth}
\centering
\includegraphics[width = \textwidth]{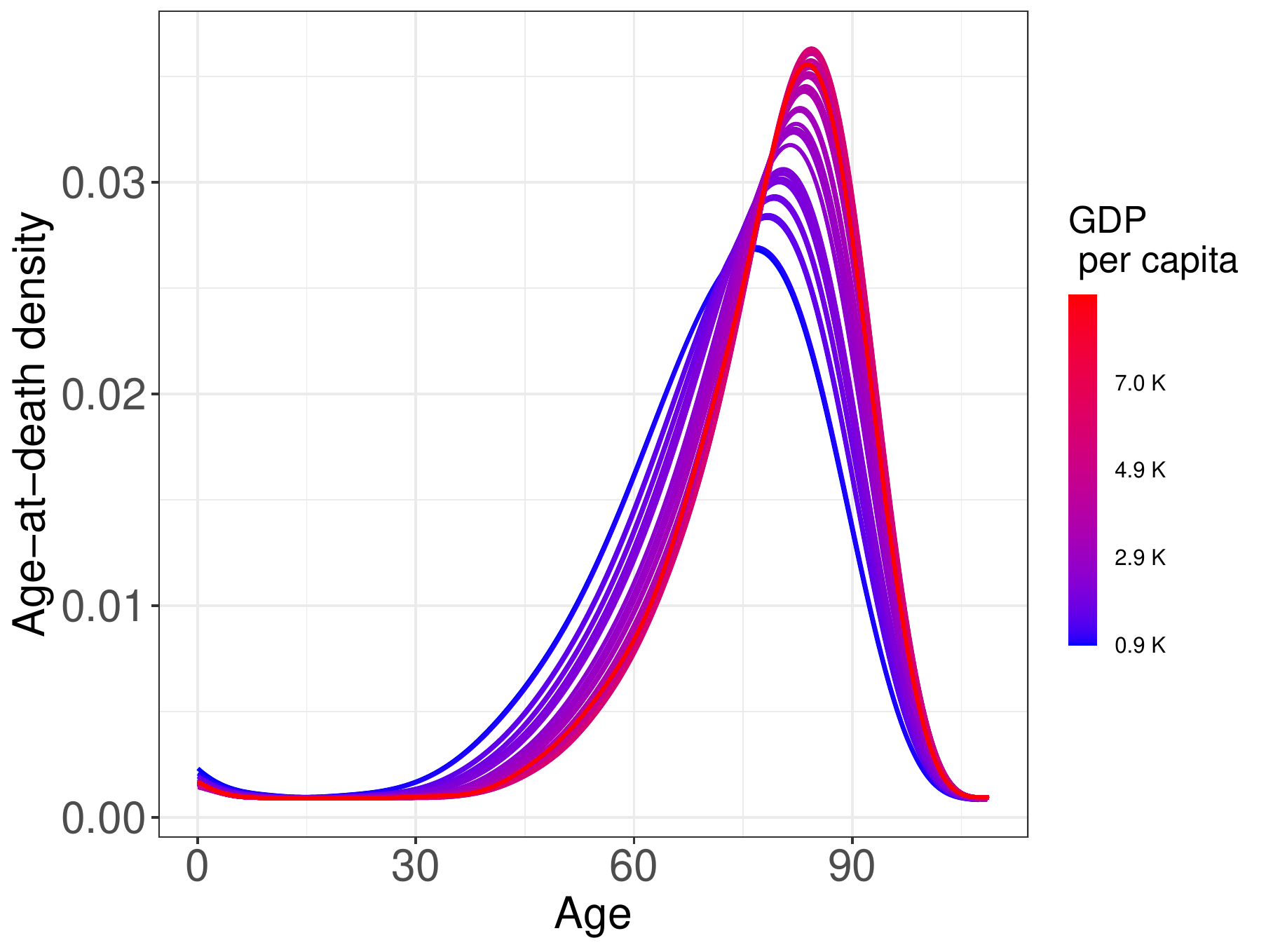}
\end{subfigure}
\hfill
\begin{subfigure}[b]{0.32\textwidth}
\centering
\includegraphics[width = \textwidth]{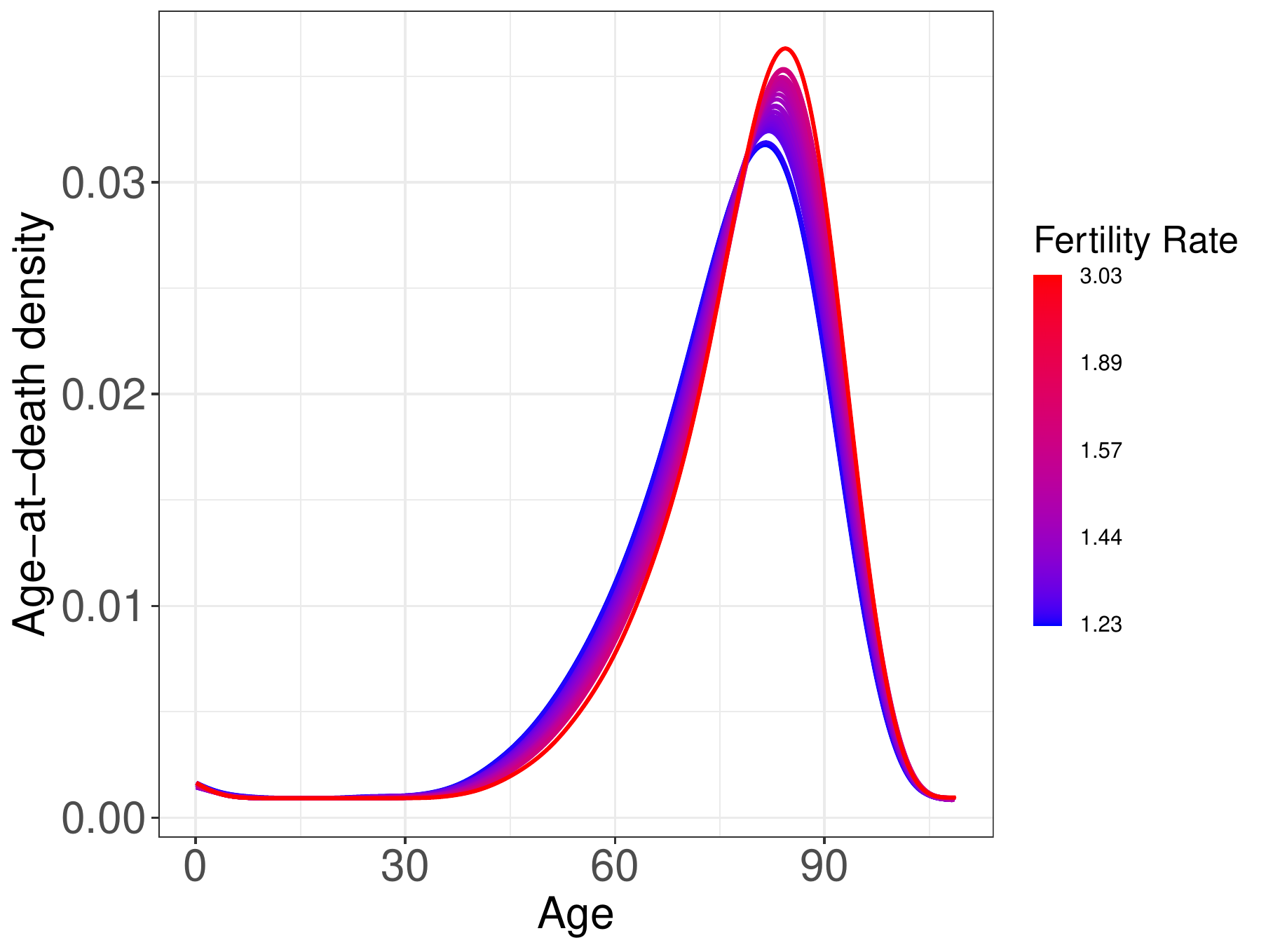}
\end{subfigure}
\hfill
\begin{subfigure}[b]{0.32\textwidth}
\centering
\includegraphics[width = \textwidth]{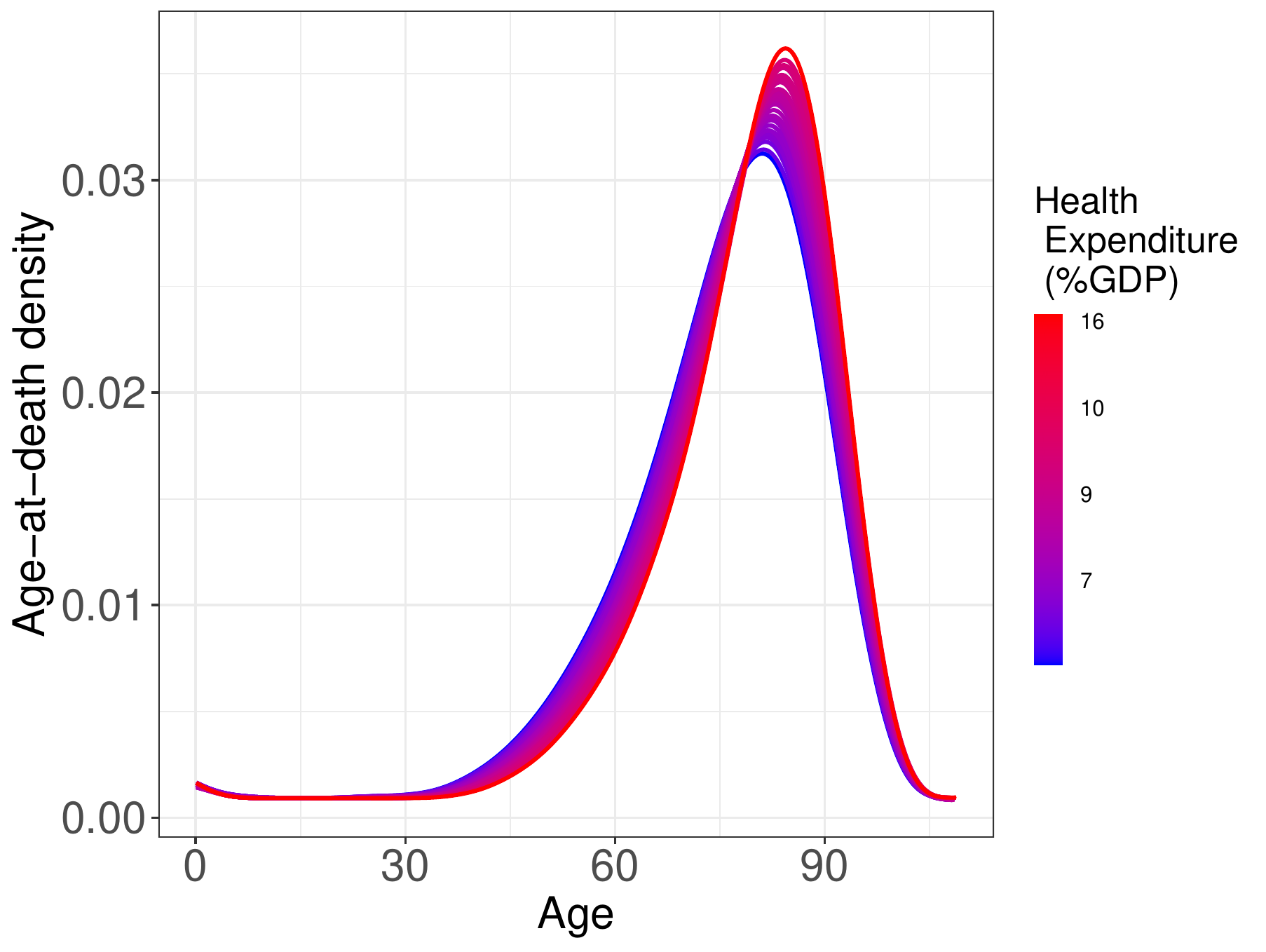}
\end{subfigure}
\caption{\small{Figure showing the effects of the significant predictors $X_3 =$ GDP per capita, $X_2 =$ Fertility rate,  and $X_5 =$ Current health expenditure. The left panel shows the change in density with changing value of $X_3$ from low (blue) to high (red), when $X_2$ and $X_5$ are fixed at their mean level, and analogously for middle and right panels.}}
\label{fig:mort:effect:pred}
\end{figure}
For example, the left-most panel of Figure~\ref{fig:mort:effect:pred} illustrates  how the age-at-death density changes with  increasing levels of GDP per capita, while the other two predictors are kept fixed.
The fitted densities are color coded such that blue to red indicates a smaller to a larger value of GDP.
We find  that smaller values of GDP are associated with left-shifted age-at-death distributions for the population. 
For increasing levels of health expenditure per capita and fertility rates, the age-at-death densities also shift rightwards, but to a lesser extent.

Finally, to illustrate the out-of-sample prediction performance of the proposed IFR model, we 
randomly split the dataset into a training set with sample size $n_{\text{train}} = 20$ and a test set with the remaining $n_{\text{test}}=20$ subjects. The IFR method was implemented as follows: For any given unit direction $\bpara \in \bar{\Theta},$ we partition the domain of the projections into $M$ equal-width non-overlapping bins and consider the representative observations $\Xl$ and $\Yl$ for the data points belonging to the $l-$th bin. The ``true'' index parameter is estimated as $\bhtrue$ as per equation (2.11). 
We then take the fitted objects obtained  from the training set and predict the responses in the test set using the covariates present in the test set. As a measure of the efficacy of the fitted model, we  compute the root mean squared prediction error (RMPE) as
\begin{align}
\label{rmpe:wass}
\text{RMPE} &= \left[\frac{1}{M_{n_{\text{test}}}}\sum_{i=1}^{M_{n_{\text{test}}}}  d_W^2\left(\Yl^{\text{test}},  \hmop{\tilde{\mathbf{X}}_l^{\text{test}\intercal}  \bhtrue, \bhtrue} \right)  \right]^{1/2},
\end{align}
where $\Yl^{\text{test}}$ and $\hmop{\tilde{\mathbf{X}}_l^{\text{test}\intercal}  \bhtrue, \bhtrue} $ denote, respectively, the $l^{\text{th}}$ observed and predicted responses in the test set, evaluated at the binned observation $\tilde{\mathbf{X}}_l^{\text{test}}.$ For any two distribution objects $F,G \in (\Omega,d_W)$, the  Wasserstein-2 distance is given by 
\begin{align*}
\label{wass:metric}
d_W(F,G) = \int_0^1 (F^{-1}(s) - G^{-1}(s))^2 ds,
\end{align*}
where $F^{-1}$ and $G^{-1}$ are the quantile functions corresponding to $F$ and $G$ respectively. 
We repeat this process $500$ times, and compute RMPE for each split for the subjects separately. The mean and sd of the RMPE  over the repetitions are shown  in Table~\ref{tab:rmpe:mort} for the IFR method, as well as for the GFR and individual LFR fits.
\begin{table}[h!]
\centering
\caption{Mean and sd (in parenthesis) of the RMPE as given in~\eqref{rmpe:wass} comparing the performance of various Fr\'echet regression models: Index Fr\'chet Regression (IFR), Global Fr\'echet  Regression (GFR), Local Fr\'echet  Regression (LFR).  The predictors used for the three individual LFR fits are Fertility Rate, GDP per capita at PPP, and Health Expenditure, respectively, as indicated in parentheses.}
\label{tab:rmpe:mort}
\centering
\begin{tabular}{c|c|c|c|c}
\hline
IFR &
GFR &
\begin{tabular}[c]{@{}c@{}}LFR1\\ (on Fertility Rate)\end{tabular} &
\begin{tabular}[c]{@{}c@{}}LFR2\\ (on GDP\\ per Capita-PPP)\end{tabular} &
\begin{tabular}[c]{@{}c@{}}LFR3\\ (on Health\\ Expenditure)\end{tabular} \\ \hline
0.178 (0.0552) &
0.287 (0.0671) &
0.491 (0.0605) &
0.603 (0.0654) &
0.339 (0.0565) \\ \hline
\end{tabular}
\end{table}

Using out-of-sample performance, the IFR model emerges as the best model, as the
average RMPE of $0.178$ is much lower than that of any of the other models. 

\subsection*{S.3.2. Emotional well-being for unemployed workers: Compositional data as random object responses}
\label{suppl:mood}
\hspace*{\fill} \\
We demonstrate the proposed IFR method for the analysis of mood compositional data. 
Compositional data are random vectors with non-negative  components,  where the components of these vectors sum to 1.  With a square-root transformation of the components, compositional  vectors  can be  transformed to unit vectors that lie on the positive segment of a sphere $S^{p-1}$ if the compositional vectors are $p-$dimensional~\citep{scea:11, scea:14}. Thus one can represent compositional data as manifold-valued objects that lie on the surface of a sphere. The data used for this application were collected in the Survey of Unemployed Workers in New Jersey~\citep{krue:11} conducted in the fall of 2009 and the beginning of 2010, during which the unemployment rate in the US peaked at 10\% after the financial crisis of 2007 -- 2008;  similar data were  used to illustrate longitudinal compositional methods  in \cite{dai:21}.  We note that here the object-valued responses lie on a manifold (sphere) with positive curvature. Thus the sufficient (but not necessary) condition for assumption (A5) 
that the underlying metric space behaves like a CAT(0) space is not satisfied. This example thus provides a check on the behavior of IFR when the random objects are situated in a positively curved space.  

Unemployed workers belonging to a  stratified random sample were surveyed at entry into the study, where we analyzed the data for $n=3301$ workers with
complete measurements.  A key variable  in the survey was the  proportion of time the workers spent in each of the four moods: bad, low/irritable, mildly pleasant, and very good while at home; we use this 4-dimensional compositional vector as the response. 
Formally,  the composition measurement of interest is $Z = (Z_1, Z_2, Z_3, Z_4)^\intercal$,  where $Z_j$ is the proportion of time a worker spent in the $j$-th mood when at home, $j = 1, \dots, 4.$ The square-root transformed compositional data 
\begin{align*}
Y = (Y_1, Y_2, Y_3, Y_4)^\intercal = (\sqrt{Z}_1, \sqrt{Z}_2, \sqrt{Z}_3, \sqrt{Z}_4)\t,
\end{align*}
lie on the sphere $\mathcal{S}^3$. We adopt the geodesic metric on this sphere
$d_g(y,y^\ast) = \arccos(y\t y^\ast).$ 

These square root transformed compositional data are treated as the object responses in a regression model with the following  10  baseline predictors obtained from the questionnaire, reflecting various socio-economic and demographic information:  (1) life satisfaction (discrete with levels 0-3, 3 meaning most satisfied)  (2) highest education level (discrete with levels 0-5, indicating high school or less, high school diploma or equivalent, college education, college diploma, graduate school, and graduate degree, respectively),
(3) marital status (discrete with levels 0-5, indicating
single (never married),
married,
separated,
divorced,
widowed, and
domestic partnership (living together but not married), respectively),
(4) number of children (discrete), (5) the number of people in the household (discrete), (6) total annual household income (continuous), (7) hours per week working at the last job (continuous), (8) how the last job ended (discrete with levels 0-2 lost job, quit job, and temporary job ended, respectively), (9) weeks spent looking for work  (continuous), and  (10) credit card balance (continuous). 

\begin{figure}[!htb]
\centering
\includegraphics[width = .7\textwidth]{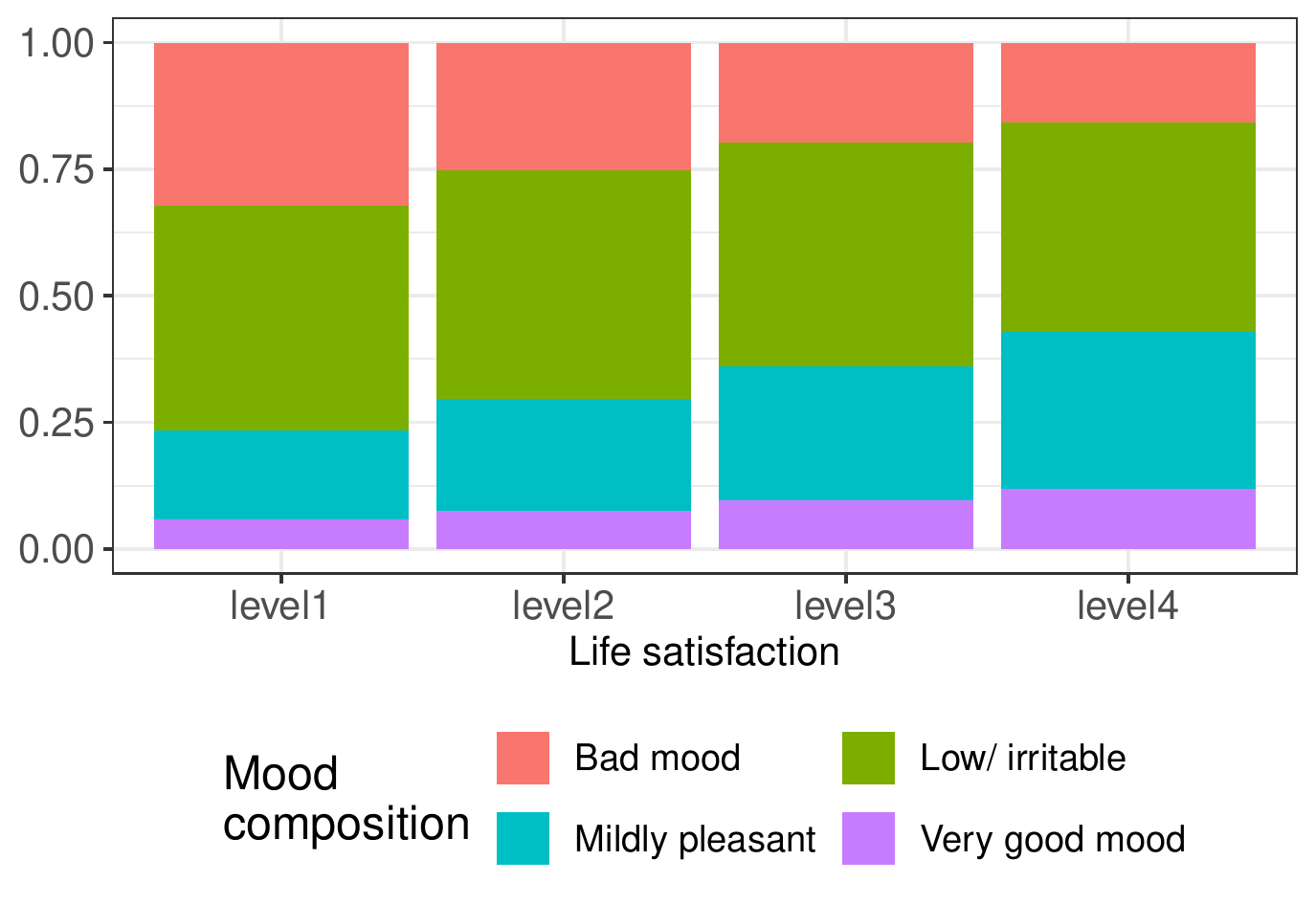}
\centering
\caption{A stacked barplot showing the effect of life satisfaction, from Level 1 (0) to Level 4 (3), on the mood composition,  when all the other predictor levels are kept fixed. A higher life satisfaction level is associated with a larger proportion of good mood.}
\label{fig:mood:barplot}
\end{figure}

For these data, the IFR model produces the  coefficient estimates
\begin{align*}
\bhtrue = (0.483, 0.134, -0.166, -0.190, 0.042, 0.303, 0.075, 0.230, 0.662, -0.307)^\intercal.
\end{align*}
The estimated coefficients can be used to obtain interpretable visualizations of the effect of the individual predictors on the compositional response through the (estimated) single  index link function, which can further lead to effective inference  for the proposed IFR model. For example, we illustrate below (Figure~\ref{fig:mood:barplot}) the effect of the predictor ``life satisfaction'' on the mood compositional data. To this end, the IFR model is fitted over varying levels of life satisfaction, from low (0) to high (3), while the other predictors are fixed at their median levels. We observe an association between a lower life satisfaction level with a higher proportion of bad mood, while a higher value of life satisfaction is associated with a better mood when all of the other predictors are fixed.

The predictive performance of the model is computed based on the root mean prediction error (RMPE) as 
\begin{align*}
\text{RMPE} = \left[\frac{1}{M_{n_{\text{test}}}}\sum_{i=1}^{M_{n_{\text{test}}}}  d_g^2\left(\Yl^{\text{test}},  \hmop{\Xl \t \bhtrue, \bhtrue} \right)  \right]^{1/2},
\end{align*}
where $\Yl^{\text{test}}$ and $\hmop{\Xl \t \bhtrue}$ denote, respectively, the $l^{\text{th}}$ observed and predicted responses in the test set, evaluated at the binned average $\Xl.$  We repeat this process $200$ times, and compute RMPE for each split for the subjects separately. For comparison purposes, we fit the data with the other applicable object regression methods, namely, the global Fr\'echet regression (GFR) method with the four-dimensional mood-compositional data as the response residing on the surface of the sphere $S^3\subset \real^4$, coupled with the 10-dimensional predictors; and individual local linear Fr\'echet regression (LFR) methods accommodating the afore-mentioned object response, while incorporating the continuous predictors  total annual household income, hours per week working at the last job, weeks spent looking for work and credit card balance as univariate predictors. Like nonparametric regression, the LFR method does not work for discrete/ categorical predictors. We denote the results from the four individual univariate local regression by $\text{LFR}_j$,  $j=1,2,3,4$, respectively. Table~\ref{tab:mood:rmpe} summarizes the results.

\begin{table}[]
\caption{Mean and sd (in parenthesis) of root mean prediction error (RMPE) over $200$ repetitions, as obtained from the local fits of the index Fr\'echet regression (IFR) model, the global Fr\'echet regression (GFR) model, and four individual local linear Fr\'echet regression (LFR) models incorporating univariate continuous predictors. Here, $n_{\text{train}}$ and $n_{\text{test}}$ denote the sample sizes for the split training and testing datasets respectively.}
\label{tab:mood:rmpe}
\begin{tabular}{cccccccc}
\hline
$n_{\text{train}}$ & {$n_{\text{test}}$} & IFR & GFR & $\text{LFR}_1$ & $\text{LFR}_2$ & $\text{LFR}_3$ & $\text{LFR}_4$ \\ \hline
$2201$ & $1100$ & \begin{tabular}[c]{@{}c@{}}$0.4779$ \\ $(0.0720)$\end{tabular} & \begin{tabular}[c]{@{}c@{}}$0.7661$\\ $(0.0418)$\end{tabular} & \begin{tabular}[c]{@{}c@{}}$0.6771$\\ $(0.0021)$\end{tabular} & \begin{tabular}[c]{@{}c@{}}$0.7220$\\ $(0.0450)$\end{tabular} & \begin{tabular}[c]{@{}c@{}}$1.1127$\\ $(0.0910)$\end{tabular} & \begin{tabular}[c]{@{}c@{}}$1.0122$\\ $(0.0810)$\end{tabular} \\ \hline
\end{tabular}
\end{table}

We observe that the out-of-sample prediction error is quite low. In fact, it is very close to the average fitting error $(0.351)$, calculated as the average distance between the observed training sample and the predicted objects based on the covariates in the training sets, which supports the validity of the proposed IFR models.

Since  in this example the object-valued responses lie on a manifold (sphere) with positive curvature,  the sufficient (but not necessary) condition for assumption (A5) 
that the underlying metric space behaves like a CAT(0) space is not satisfied. However, the numerical performance of the IFR method is quite good, suggesting a certain degree of model  robustness of the IFR method.

\subsection*{S.3.3. Additional results for the analysis of ADNI neuroimaging data}
\label{suppl:sec:adni}
\hspace*{\fill} \\
The individual effect of the significant predictors- stages of the disease, age, and total score, is illustrated. To this end, the IFR model is fitted over varying values of one predictor, while keeping the other two fixed at their mean levels. 

\begin{figure}[!htb]
\centering
\begin{subfigure}[b]{0.49\textwidth}
\centering
\includegraphics[width=.9\textwidth]{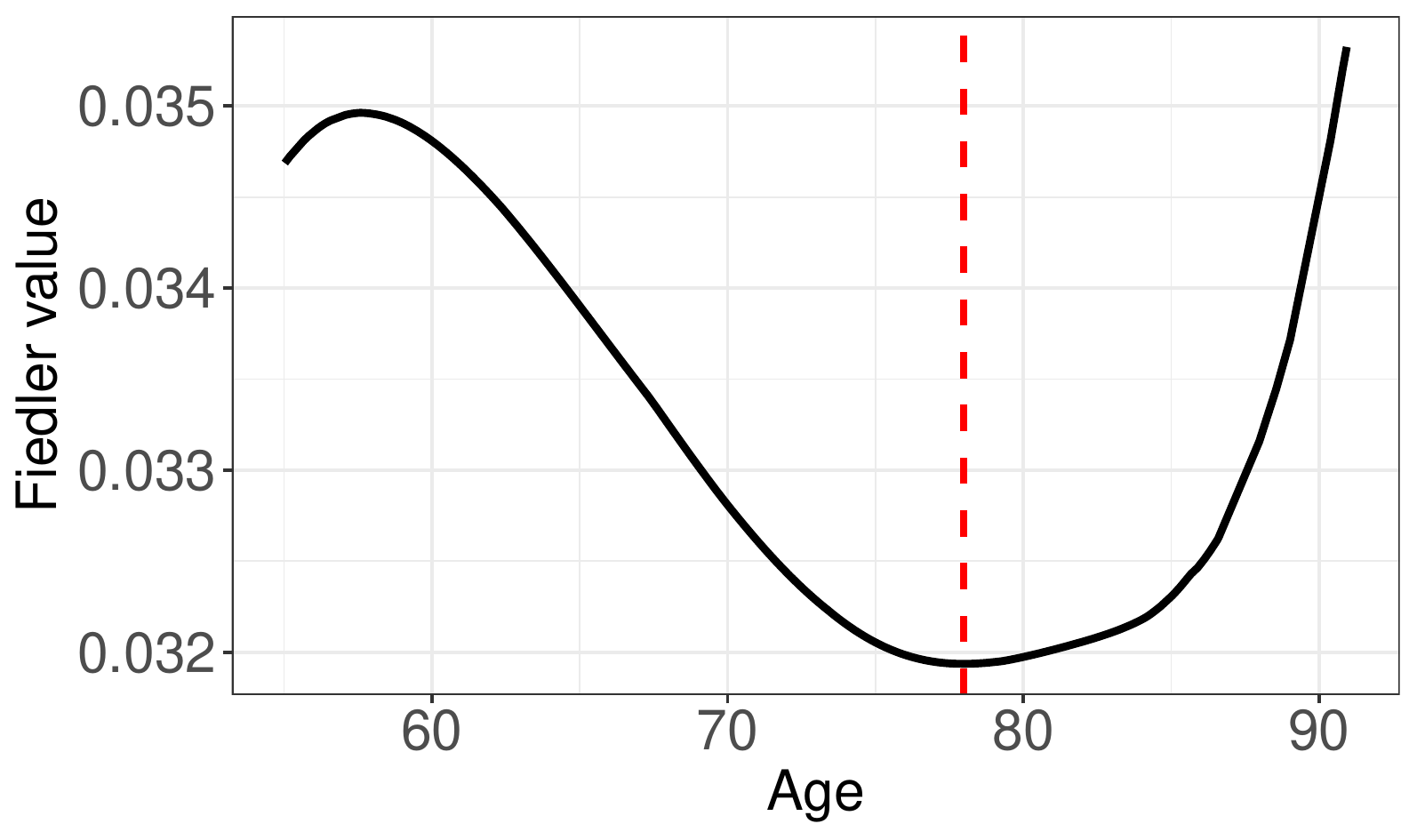}
\end{subfigure}
\hfill
\begin{subfigure}[b]{0.49\textwidth}
\centering
\includegraphics[width=.9\textwidth]{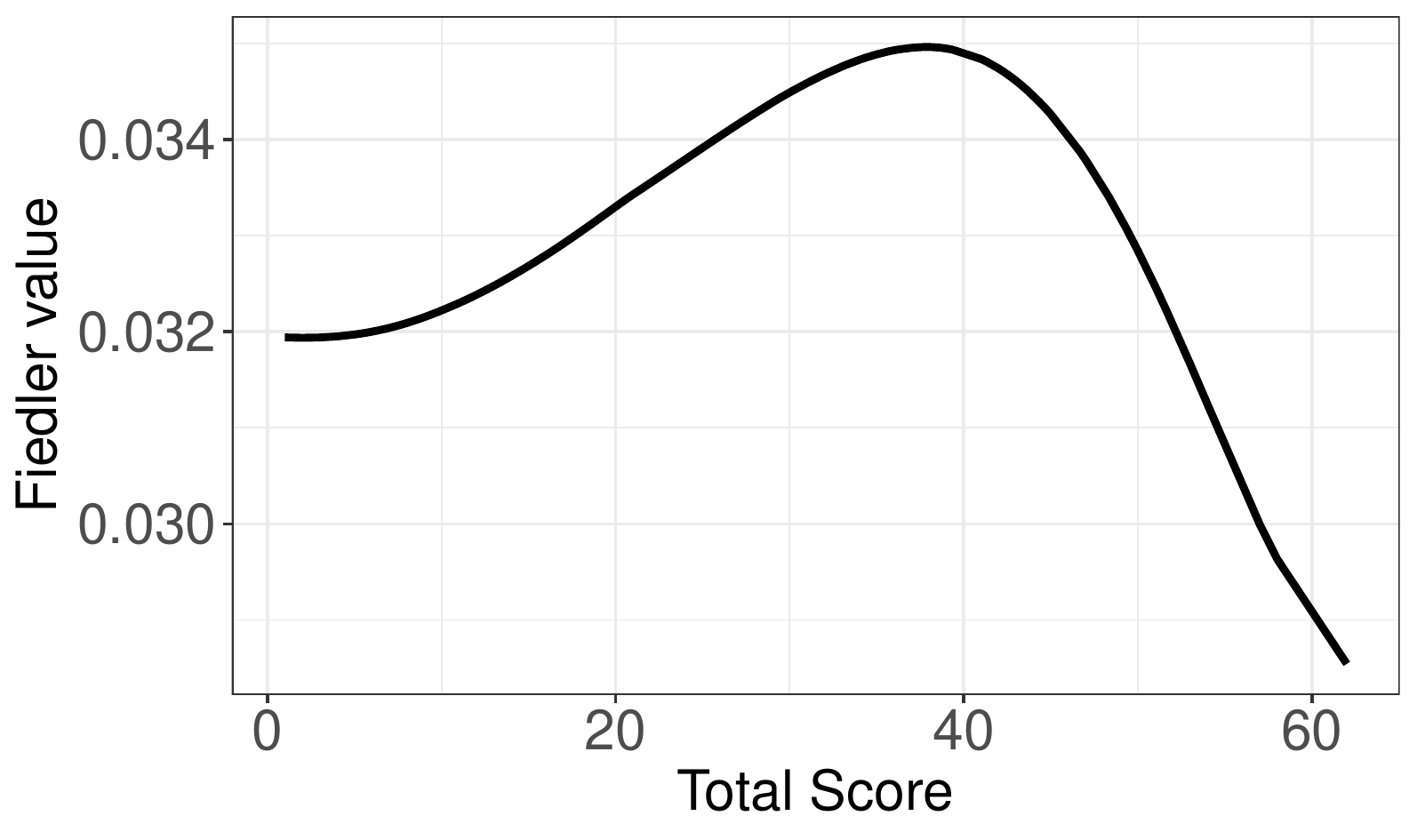}
\end{subfigure}
\caption{Fiedler values as a function of age and total score, corresponding to the index Fréchet regression (IFR) fits for the correlation matrix-valued object responses. The left panel shows the change in the estimated Fiedler value for varying values of age ($X_2$) from low to high when $X_4$ (total score) is fixed at its mean level, with the minimum
attained at 78 years of age marked by a red dashed line. The right panel shows the change in the total score for a fixed mean level of age.}
\label{fig:adni:fied}
\end{figure}

For any $r\times r$ correlation matrix $Y$, the Fiedler value is the second smallest eigenvalue of the corresponding graph Laplacian matrix
\begin{align*}
L(Y) = D(Y) - A(Y).
\end{align*}
Here $A(Y) = (Y - I_r)_{+}$ is the adjacency matrix obtained by applying a threshold and setting the diagonal elements to zero, and $D(Y) = \diag{A(Y)\mathbf{1}_r}$ is the degree matrix, where $I_r = \diag{\mathbf{1}_r}$, $\mathbf{1}_r = (1,\dots,1)\intercal \in \real^r$, and $H_{+} = (\max\{H_{kl},0\})_{k,l=1\dots,r}$ for any matrix $H \in \real^{r\times r}.$ The Fiedler value corresponding to $Y$ is then given by the map
$\lambda_{r-1}(L(Y))$, which produces the $(r-1)$th largest, i.e., second smallest eigen value of $L(Y)$.
After fitting the proposed IFR model, the Fiedler values are calculated over varying values of age and total score. The left panel of Figure~\ref{fig:adni:fied} shows how the Fiedler value changes with increasing age, while the total score is kept fixed at its mean level, while the right panel shows the Fiedler values over the varying total scores for the fixed mean level of age.
In the age-varying Fielder value figure, a convex pattern can be seen around the minimum, which is attained at 78 years of age. This agrees with most studies that have found that functional connectivity decreases during normal aging processes before 80 years of age. Further, we observe that the decrease is reversed for older ages above $80.$ On the other hand, for a higher value of the total score, the Fielder values show a steep decreasing pattern.

\begin{figure}[!htb]
\centering
\includegraphics[width=.6\textwidth]{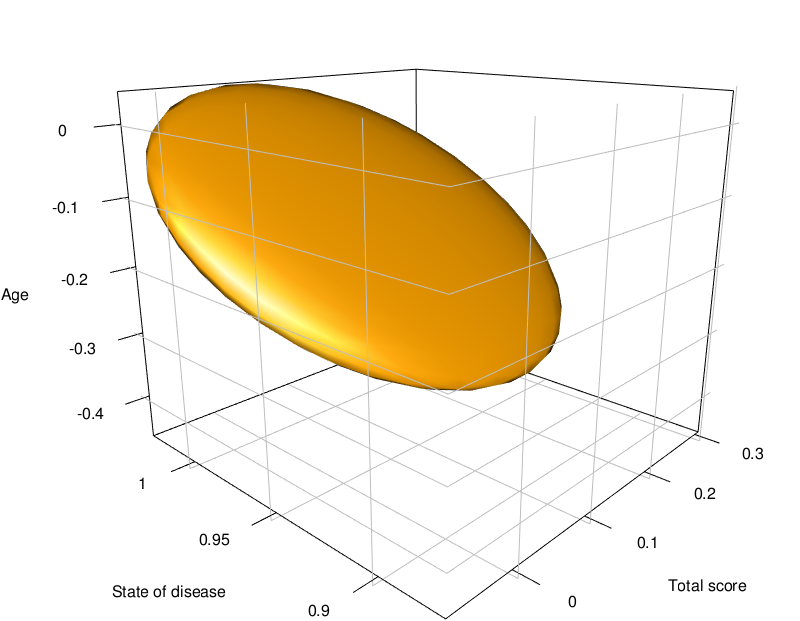}
\centering
\caption{The figure shows the 3-dimensional plot for the $95\%$
confidence region of $(\theta_1,\theta_2,\theta_4)$: the coefficients of the effects of the predictors- age, total score, and stage of the disease, respectively.}
\label{fig:ADNI:CR2}
\end{figure}

Further, continuing from Section 5.1 in the main manuscript, 
we illustrate the $95\%$  confidence region for the coefficients $(\theta_1,\theta_2,\theta_4)$ of the predictors: stages of the disease, age, and total score in a 3-dimensional plot in Figure~\ref{fig:ADNI:CR2}.

\subsection*{S.3.4. Additional simulations for Euclidean responses} \label{supp:sec:sim} 
\begin{figure}[!htb]
\centering
\includegraphics[width=.6\textwidth]{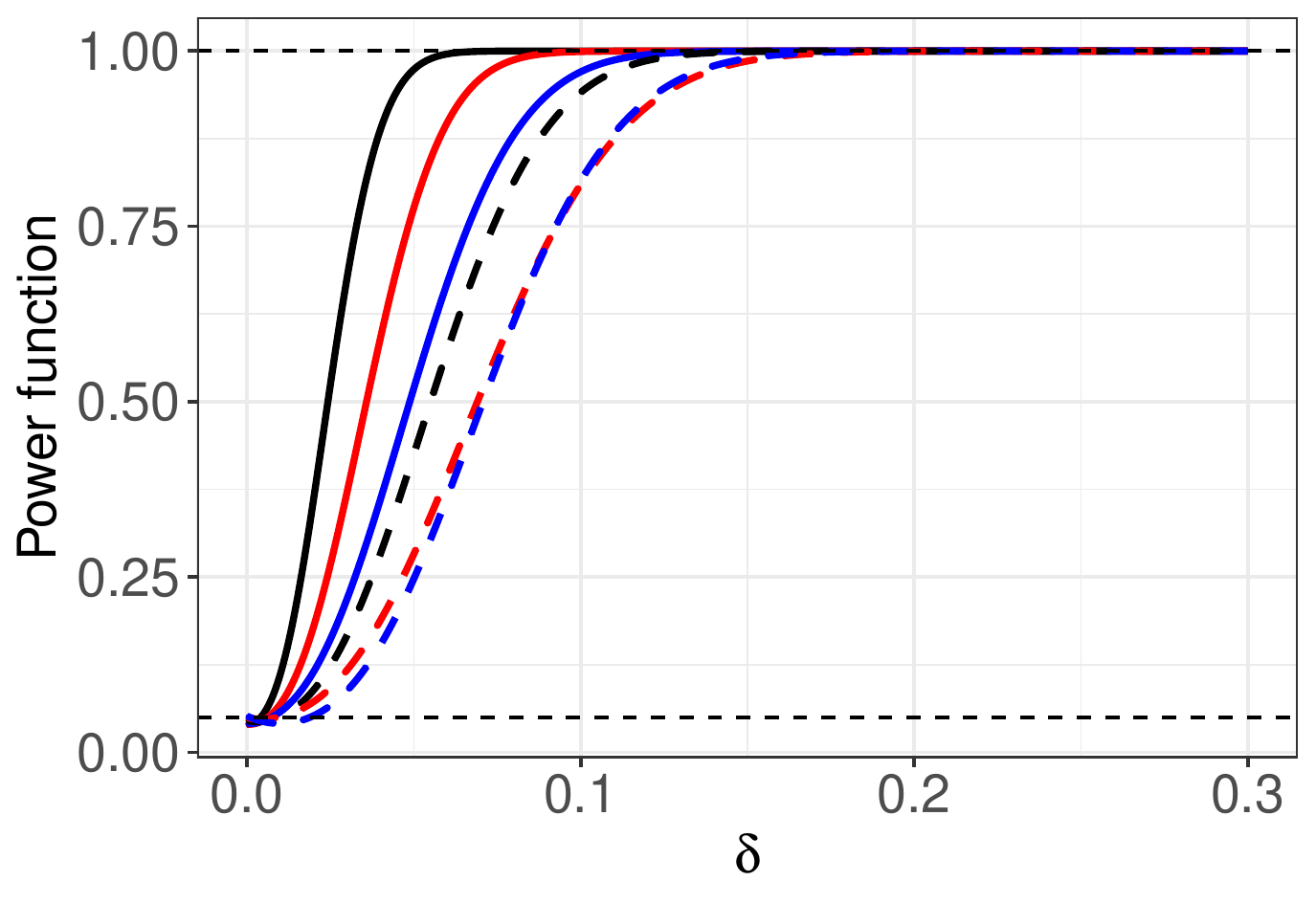}
\centering
\caption{Simulation for Euclidean response using different link functions showing the empirical power function for Euclidean responses. The black, red, and blue curves correspond to the identity, square, and exponential link functions used in the data-generating mechanism, respectively, while the dashed and solid patterns correspond to the varying sample sizes $n=100$ and $n=1000$, respectively. The level of the tests is $\alpha= 0.05$ and is indicated by the dashed line parallel to the x-axis.}
\label{fig:eucl:pow}
\end{figure}

\hspace*{\fill} \\
Here the object response of interest is assumed to lie in the Euclidean space. For generating the predictor vectors we consider a $5-$dimensional vector distributed as truncated multivariate normal distributions, where each of the components is truncated to lie between $[-10,10].$ The components are assumed to be correlated such that $X_1$ correlates with $X_2$ and $X_3$ with $r = 0.5$, and $X_2$ and $X_3$ correlate with $r = 0.25.$ The variances for each of the five components are $0.1.$ 
The empirical power against the sequence of alternatives in equation (3.10) 
increases steeply (see Figure~\ref{fig:eucl:pow}) as we deviate from the null hypothesis in equation (3.9) in Section 3 of  the main manuscript, 
especially corresponding to higher sample size and under identity link. 

The empirical power function, as we deviate from the null hypothesis in equation (3.9) 
is computed and illustrated in the left panel in Figure~\ref{fig:eucl:pow}. Empirical evidence suggests that the proposed test is consistent for a higher sample size of $n= 1000$, and leads to the correct nominal level of the test. 

\begin{figure}[!htb]
\centering
\includegraphics[width =.6\textwidth]{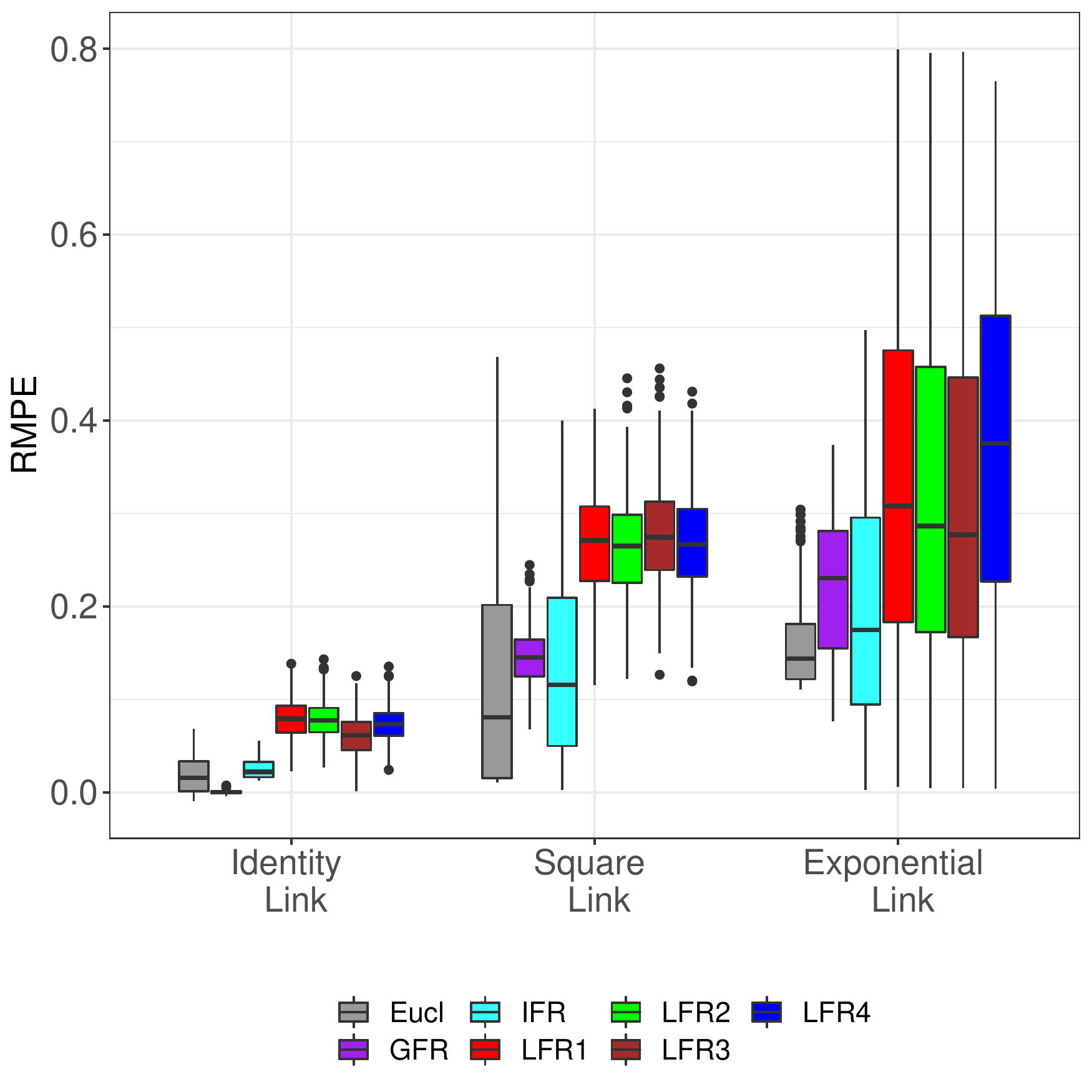}
\centering
\caption{Figure showing boxplot of RMPEs  for various regression methods for simulated Euclidean responses. The methods compared are index Fr\'echet regression (IFR), classical Euclidean single index regression using the R package ``np'' (NP), global Fr\'echet Regression (GFR) with the 4-dimensional predictor, and four individual local linear Fr\'echet regression (LFR) models that treat each predictor components as a univariate predictor. The sample size is fixed at $n = 1000$ and the RMPE are computed over $200$ Monte Carlo simulation runs. }
\label{fig:boxplot:rmpe:eucl}
\end{figure}

The consistency of the estimates is illustrated in Table~\ref{tab:euclidean:set1} based on $500$ replications of the simulation scenario.
\begin{table}[!htb]
\centering
\caption{Table showing bias and variance of $\bhtrue$ (measured in radians) based on $500$ replications for a Euclidean vector response. The predictors $X_1,\dots, X_5$ are generated from a truncated multivariate normal distribution.}
\begin{center}
\begin{tabular}{c|c|c|c|c|c|c}
\hline
& \multicolumn{2}{c|}{link1 ($x \mapsto x$)} & \multicolumn{2}{c|}{link2 ($x \mapsto x^2$)} & \multicolumn{2}{c}{link3 ($x \mapsto e^x$)} \\ \hline
& bias       & dev     & bias         & dev         & bias         & dev         \\ \hline
$n = 100$  & 0.013      & 0.061         & 0.025        & 0.048       & 0.037        & 0.029       \\ \hline
$n = 1000$ & 0.006      & 0.021         & 0.014        & 0.019       & 0.013        & 0.009      \\ \hline
\end{tabular}
\end{center}
\label{tab:euclidean:set1}
\end{table}
Further, the performance of the proposed method is compared to the classical Euclidean single index model fits. To this end, the R package \emph{np}  was called from Julia, for fitting the classical single index regression to the simulated Euclidean responses. The prediction performance of the classical single index fits, denoted by NP, is compared with that of the IFR method, as well as with a Global Fr\'echet Regression (GFR) method and four separate Local Fr\'ecet Regression (LFR) fits. The GFR method utilizes the multi-variate predictors while the four LFR methods treat each of the four-dimensional predictor components as a univariate predictor individually. Note that  in all of the methods- NP, GFR, LFR - binning is not required. The mean and sd of the root mean prediction error (RMPE) over $200$ Monte Carlo simulation runs are reported in Table~\ref{tab:rmpe:eucl}. 
\begin{table}[!htb]
\centering
\caption{Table showing the mean (sd in parenthesis) RMPE for various regression methods for simulated Euclidean responses. The methods compared are index Fr\'echet regression (IFR), classical Euclidean single index regression using the R package ``np'' (NP), global Fr\'echet Regression (GFR) with the 4-dimensional predictor, and four individual local linear Fr\'echet regression (LFR) models that treat each predictor components as a univariate predictor. The sample size is fixed at $n = 1000$ and the RMPE are computed over $200$ Monte Carlo simulation runs. }
\label{tab:rmpe:eucl}
\begin{tabular}{c|c|c|c}
\hline
& Identity link   & Square link     & Exponential link \\ \hline
IFR  & 0.0255 (0.0110) & 0.1383 (0.1031) & 0.1972 (0.1205)  \\ \hline
NP   & 0.0187 (0.0201) & 0.1117 (0.1077) & 0.1578 (0.0442)  \\ \hline
GFR  & 0.0003 (0.0018) & 0.1465 (0.0299) & 0.2181 (0.0748)  \\ \hline
LFR1 & 0.0788 (0.0208) & 0.2686 (0.0558) & 0.3342 (0.1882)  \\ \hline
LFR2 & 0.0784 (0.0204) & 0.2627 (0.0540) & 0.3237 (0.1912)  \\ \hline
LFR3 & 0.0617 (0.0209) & 0.2774 (0.0555) & 0.3162 (0.1892)  \\ \hline
LFR4 & 0.0730 (0.0197) & 0.2694 (0.0561) & 0.3664 (0.1888)  \\ \hline
\end{tabular}
\end{table}	
The data is simulated using three different generating mechanisms - the identity, squared, and exponential link functions, and the sample size $n = 1000$ is considered. For the identity link function, i.e., when the simulated data is generated according to a linear model, the GFR method gives the lowest prediction error. This is indeed expected since the GFR boils down to a linear regression model when the object data are Euclidean. For other situations the NP method for the classical single index model outperforms the other methods, however, the proposed IFR method proves competitive with a comparable magnitude of the prediction error. The boxplot of the RMPEs for the above situations is shown in Figure~\ref{fig:boxplot:rmpe:eucl}.

\subsection*{S.3.5. Simulation results for adjacency matrix as random object responses}
Here we consider responses that are  adjacency matrices obtained for  weighted networks and  equipped with the Frobenius norm. We generated samples of networks with $m= 10$ nodes, as one might encounter in brain networks, represented as weighted adjacency matrices. The predictors were sampled from a $4-$dimensional zero mean multivariate normal distribution with covariance matrix defined by $\corr(X_1,X_2) = \corr(X_1,X_3) =  \corr(X_2,X_3) = 0.3,$ and  $\corr(X_1,X_4) = \corr(X_2,X_4) = -0.4.$ and variances of all components equal to $0.25.$  Subsequently each of the components was truncated to lie between $[-5,5]$.  The elements of the weighted adjacency matrices $Y=(Y_{qr})$ were then obtained as 
\begin{align*}
	Y_{qr} = \zeta(\tbfx\t \btrue) + \epsilon_{qr}, \ q,r = 1,\dots,m,
\end{align*}
where $\epsilon_{qr}$ are independently sampled errors and the link function $\zeta(\cdot)$ 
was taken as the expit function, i.e.,  $\zeta(\tbfx\t \btrue) = 1/(1 + \exp(- \tbfx\t \btrue)).$ For a given index $\tbfx\t \btrue,$ $\epsilon_{qr}$ was sampled from a uniform distribution on $[\max\{0,-\zeta(\tbfx\t \btrue)\}, \min\{1,1-\zeta(\tbfx\t \btrue)\}]$. The matrix responses of interest were thus generated as
$
Y = \zeta(\tbfx\t\btrue)I_m + \bs\eps,
$
where $\bs\eps = \left(\left( \epsilon_{qr}\right)\right)_{q,r = 1,\dots,m}$ as generated above and $I_m$ is the $m \times m$ identity matrix.
\begin{table}[!htb]
	\centering
	\caption{Table showing bias and deviance of $\hat{\theta}$ (measured in radians, as per~eqref{simul:bias:var}) based on $500$ replications for weighted adjacency matrix responses.}
	\label{tab:adjacency:set1}
	\centering
	\begin{tabular}{c|c|c|c}
		\hline
		& \multicolumn{3}{c}{link $(x \mapsto 1/(1+\exp(-x))$} \\ \hline
		& bias & dev & avg. MSD \\ \hline
		$n = 100$ & 0.044 & 0.052 & 0.672 \\ \hline
		$n = 1000$ & 0.021 & 0.019 & 0.041 \\ \hline
	\end{tabular}
\end{table}

Table~\ref{tab:adjacency:set1} presents the bias and variance of the estimator computed based on $500$ replication of the data generating process. 
The mean squared deviation (MSD) was computed as the average distance between the true and estimated adjacency matrices, similar to~\eqref{simu:dens:mse:ifr}. The average mean squared deviation (MSD) over $500$ simulation runs is quite low. With a higher sample size, the estimates seem to perform better consistently. We also note here that the non-zero correlation among the components of the predictor vector does not influence the performance of the nonparametric regression fit negatively. 

\begin{figure}[!htb]
	\centering
	\includegraphics[width=.45\textwidth]{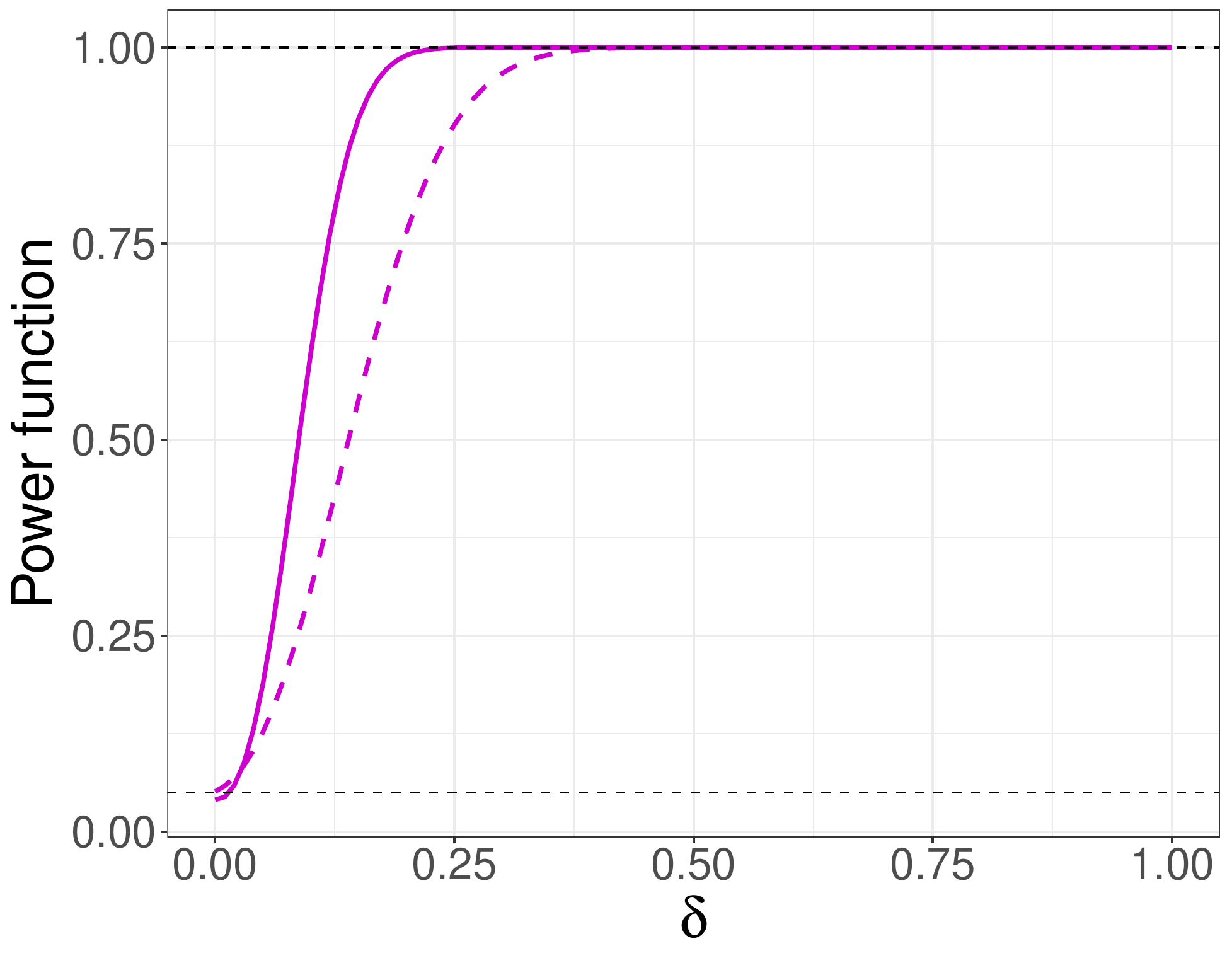}
	\centering
	\caption{Simulation for adjacency matrix response for different sample sizes. The figure displays the empirical power as function of $\delta$ for weighted adjacency matrix responses based on sample sizes $n = 100$ and $n= 1000,$ in dashed and solid lines, respectively. The magenta color corresponds to the expit link function used to generate the data, while the dashed and solid pattern correspond to the varying sample sizes $n=100$ and $n=1000$, respectively. The level of the tests is $\alpha= 0.05$ and is indicated by the dashed line parallel to the x-axis.}
	\label{fig:spd:pow}
\end{figure}

Figure~\ref{fig:spd:pow} shows the empirical power function as we deviate from the null hypothesis in ~\eqref{hyp:null} for two different sample sizes. Empirical evidence suggests that the proposed test is consistent for a higher sample size of $n= 1000$, and leads to the correct empirical level of the test. 

\end{document}